\documentclass[10pt,aps,prb,footinbib,superscriptaddress,twocolumn,showpacs,floatfix]{revtex4-1}
\usepackage{amsbsy}
\usepackage{amsmath}
\usepackage{amssymb}
\usepackage{amsfonts}
\usepackage{graphicx}
\usepackage{pstricks}
\usepackage{pstricks-add}
\usepackage{pst-3dplot}
\usepackage[english]{babel}
\usepackage{color}
\usepackage[normalem]{ulem}
\usepackage{bm}
\usepackage{yfonts}
\usepackage{enumerate}

\begin{document}

\def\myitem#1#2{
%  \begin{minipage}{3in}
  \begin{itemize}
  \item{{\bf[#1]}#2}
  \end{itemize}
%\end{minipage}
}

\date{\today}

\title{Straightening Out the Frobenius-Schur Indicator}

\author{Steven H. Simon}
\affiliation{Rudolf Peierls Centre for Theoretical Physics, Clarendon Laboratory, Oxford, OX1 3PU, UK}

\author{Joost K. Slingerland}
\affiliation{Department of Mathematical Physics, National University of Ireland, Maynooth, Ireland} 
\affiliation{Dublin Institute for Advanced Studies, School of Theoretical Physics, 10 Burlington Rd, Dublin, Ireland.}

\begin{abstract}

  Amplitudes for processes in topological quantum field theory (TQFT)
  are calculated directly from spacetime diagrams depicting the
  motion, creation, annihilation, fusion and splitting of any
  particles involved. One might imagine these amplitudes to be
  invariant under any deformation of the spacetime diagram, this being
  almost the meaning of the "topological" in TQFT.  However, this is
  not always the case and we explore this here, paying particular
  attention to the Frobenius-Schur indicators of particles and
  vertices.  The Frobenius-Schur indicator is a parameter $\kappa_a=\pm 1$ assigned to each
  self-dual particle $a$ in a TQFT, or more generally in any tensor
  category.  If $\kappa_a$ is negative then straightening out a
  timelike zig-zag in the worldline of a particle of type $a$ can
  incur a minus sign and in this case the amplitude associated with
  the diagram is not invariant under deformation.  Negative
  Frobenius-Schur indicators occur even in some of the simplest TQFTs
  such as the $SU(2)_1$ Chern-Simons theory, which describes
  semions. This has caused some confusion about the topological
  invariance of even such a simple theory to space-time deformations.
  In this paper, we clarify that, given a TQFT with negative
  Frobenius-Schur indicators, there are two distinct conventions
  commonly used to interpret a spacetime diagram as a physical
  amplitude, only one of which is isotopy invariant --- the
  non-isotopy invariant interpretation is used more often in the
  physics literature.  We clarify in what sense TQFTs based on
  Chern-Simons theory with negative Frobenius-Schur indicators are
  isotopy invariant, and we explain how the Frobenius-Schur indicator
  is intimately linked with the need to frame world-lines in
  Chern-Simons theory.  Further, in the non-isotopy-invariant
  interpretation of the diagram algebra we show how a trick of
  bookkeeping can usually be invoked to push minus signs onto the
  diagrammatic value of a loop (the ``loop weight''), such that most
  of the evaluation of a diagram does not incur minus signs from
  straightening zig-zags, and only at the last step minus signs are
  added.  We explain the conditions required for this to be possible.
  This bookkeeping trick is particularly useful in the construction of
  string-net wavefunctions, where it can be interpreted as simply a
  well-chosen gauge transformation.  We then further examine what is
  required in order for a theory to have full isotopy invariance of
  planar spacetime diagrams, and discover that, if we have
  successfully pushed the signs from zig-zags onto the loop weight,
  the only possible obstruction to this is given by an object related to
  vertices, known as the ``third Frobenius-Schur indicator''.  We
  finally discuss the extent to which this gives us full isotopy
  invariance for braided theories.
 
\end{abstract}

\pacs{PACS}

\maketitle

%\tableofcontents

%%%
\section{Introduction}

Topology is one of the most prominent themes in modern quantum condensed
matter physics\cite{nayakreview,TopInsulators,MyBook}.  Among the
prevalent topological ideas is that of the topological quantum field
theory (TQFT) which can be used to describe systems with exotic
particles such as anyons.  At the same time there have been substantial
mathematical advances\cite{witten,Kirillov,Turaev,AtiyahBook} in the field of topology based on
the physical ideas of TQFTs.

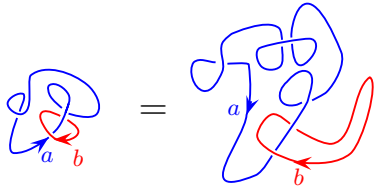
\begin{figure}[h!]
	\begin{center}
		\vspace*{10pt}
		\scalebox{1}{%LaTeX with PSTricks extensions
%%Creator: inkscape 0.91
%%Please note this file requires PSTricks extensions
\psset{xunit=.72pt,yunit=.72pt,runit=.72pt}
\begin{pspicture}(0,950)(744.09448819,1030.36220472)
{
\newrgbcolor{curcolor}{0 0 1}
\pscustom[linewidth=1,linecolor=curcolor]
{
\newpath
\moveto(46.142293,965.70812972)
\curveto(46.142293,965.70812972)(38.918323,947.35459472)(44.627063,947.24810472)
\curveto(51.396103,947.12183472)(58.301753,952.10773472)(62.783783,955.53109572)
\moveto(64.930363,957.80393572)
\curveto(69.412393,961.22729572)(71.370663,967.75967372)(73.288863,973.30901472)
\moveto(71.521093,968.51078972)
\curveto(72.570423,971.54647372)(74.611443,975.13913072)(73.162593,978.00571372)
\curveto(71.481503,981.33179572)(66.332783,984.96499772)(63.162593,983.00571372)
\curveto(61.744843,982.12949472)(62.553613,979.55713872)(63.162593,978.00571372)
\curveto(64.329123,975.03388772)(69.677823,970.98540872)(69.677823,970.98540872)
}
}

\psline[linewidth=1,linecolor=curcolor,arrowsize=7]{->}(58.301753,952.10773472)(62.783783,955.53109572)
\rput(62,945){$\color{blue} a$}

{
\newrgbcolor{curcolor}{0 0 1}
\pscustom[linewidth=1,linecolor=curcolor]
{
\newpath
\moveto(73.162593,968.00571372)
\curveto(73.162593,968.00571372)(85.239043,963.94949572)(88.162593,968.00571372)
\curveto(91.244283,972.28133572)(87.169463,979.58186472)(83.162593,983.00571372)
\curveto(75.455213,989.59161872)(60.898043,994.55862172)(53.162593,988.00571372)
\curveto(51.129303,983.68349572)(55.419423,969.68933872)(48.667673,967.75317572)
\curveto(45.795093,967.09061272)(41.036753,970.43841272)(41.142283,973.38452072)
\curveto(41.242583,976.18431372)(47.085333,979.58741272)(49.551553,978.25825172)
\curveto(50.826503,975.87223172)(47.967693,971.79735972)(47.531243,969.09137272)
}
}
{
\newrgbcolor{curcolor}{1 0 0}
\pscustom[linewidth=1,linecolor=curcolor]
{
\newpath
\moveto(72.531253,964.64720872)
\curveto(72.531253,964.64720872)(78.431193,961.45304572)(78.339633,958.43528572)
\curveto(78.248033,955.41751572)(72.678433,953.46964572)(69.299023,953.76332572)
\curveto(65.864093,954.06183572)(62.898373,957.10485572)(61.066793,960.02601572)
\curveto(59.488303,962.54352572)(56.800113,966.83685172)(59.299023,968.76332872)
\curveto(61.797933,970.68980572)(68.793943,966.74301872)(68.793943,966.74301872)
}
}

\psline[linewidth=1,linecolor=curcolor,arrowsize=7]{->}(71.678433,954.46964572)(65.299023,954.36332572)
\rput(78,945){$\color{red} b$}

\put(110,965){\Large $=$}
%{
%\newrgbcolor{curcolor}{0 1 1}
%\pscustom[linestyle=none,fillstyle=solid,fillcolor=curcolor]
%{
%\newpath
%\moveto(99.57218933,977.15391462)
%\lineto(124.61125183,977.15391462)
%\lineto(124.61125183,973.87266462)
%\lineto(99.57218933,973.87266462)
%\lineto(99.57218933,977.15391462)
%\closepath
%\moveto(99.57218933,969.18516462)
%\lineto(124.61125183,969.18516462)
%\lineto(124.61125183,965.86485212)
%\lineto(99.57218933,965.86485212)
%\lineto(99.57218933,969.18516462)
%\closepath
%}
%}
{
\newrgbcolor{curcolor}{0 0 1}
\pscustom[linewidth=1,linecolor=curcolor]
{
\newpath
\moveto(181.90117,948.68654472)
\curveto(181.90117,948.68654472)(198.0405,971.34817172)(198.89674,974.97525672)
\curveto(200.0455,979.84150172)(203.33854,987.67952672)(198.89674,989.97525672)
\curveto(193.98344,992.51467772)(188.01286,984.79822472)(188.01286,984.79822472)
\curveto(181.54839,979.39467372)(183.01013,970.14616372)(193.26539,972.70241372)
\lineto(195.79078,973.83883472)
}
}

{
\newrgbcolor{curcolor}{0 0 1}
\pscustom[linewidth=1,linecolor=curcolor]
{
\newpath
\moveto(201.11882,974.97525672)
\curveto(201.11882,974.97525672)(207.46569,978.08989372)(208.6442,979.26840472)
\curveto(213.35825,983.98244972)(218.84781,994.35141472)(216.34662,1000.53109572)
\curveto(196.0061,1050.78626622)(189.19103,1011.23692772)(190.31044,1008.00571472)
\moveto(190.56298,1004.72271872)
\curveto(191.47222,989.76012572)(196.3961,992.12183072)(200.78676,997.66673472)
\curveto(204.40409,1002.23501472)(203.95689,1007.29519072)(199.64837,1007.39304372)
\curveto(195.65348,1007.48377372)(178.63971,1004.66777472)(173.77047,1002.57614472)
\curveto(170.70775,1001.26052372)(172.0448,995.18570572)(174.40182,992.82868272)
\curveto(176.75884,990.47165972)(182.0448,990.47165972)(184.40182,992.82868272)
\curveto(186.19337,994.62022872)(185.53824,999.54568672)(185.53824,999.54568672)
\lineto(185.2857,1003.25825272)
}
}
{
\newrgbcolor{curcolor}{0 0 1}
\pscustom[linewidth=1,linecolor=curcolor]
{
\newpath
\moveto(185.2857,1006.81852872)
\curveto(185.2857,1006.81852872)(185.18265,1011.76180572)(183.69496,1013.00571472)
\curveto(177.30635,1018.34743972)(158.89674,1009.97525672)(158.89674,1009.97525672)
\curveto(158.89674,1009.97525672)(154.79873,1007.15286172)(153.89674,1004.97525672)
\curveto(151.98332,1000.35585872)(155.81016,994.59465472)(153.89674,989.97525672)
\curveto(152.09275,985.62004672)(148.48468,978.89219072)(143.89674,979.97525672)
\curveto(138.76727,981.18616072)(135.81505,990.69963472)(138.89674,994.97525672)
\curveto(142.26895,997.76515672)(146.57787,994.97525672)(150.41197,994.97525672)
}
}

{
\newrgbcolor{curcolor}{0 0 1}
\pscustom[linewidth=1,linecolor=curcolor]
{
\newpath
\moveto(156.6239,994.21764272)
\curveto(156.7003,994.19854272)(167.3777,995.08048172)(167.81108,993.83883472)
\curveto(168.98577,984.95487072)(169.17402,974.52094372)(166.67466,967.37436872)
\curveto(166.0247,954.74556572)(136.06476,921.06145472)(171.6239,935.68211472)
\curveto(175.59017,937.66235472)(180.03316,945.98541472)(180.03316,945.98541472)
}
}

\psline[linewidth=1,linecolor=blue,arrowsize=7]{->}(169.17402,974.52094372)(166.67466,967.37436872)
\rput(160,970){$\color{blue} a$}

{
\newrgbcolor{curcolor}{1 0 0}
\pscustom[linewidth=1,linecolor=curcolor]
{
\newpath
\moveto(189.50206,962.62691572)
\curveto(189.50206,962.62691572)(183.50384,968.44351772)(180.25968,967.67767472)
\curveto(176.63259,966.82143572)(172.87695,960.67422572)(172.27998,956.99556572)
\curveto(171.52488,952.34239472)(181.3961,947.12183472)(181.3961,947.12183472)
\curveto(181.3961,947.12183472)(191.13582,942.44932472)(196.3961,942.12183472)
\curveto(204.87804,941.59376472)(214.8626,941.68716472)(221.3961,947.12183472)
\curveto(228.07419,953.46661572)(236.47515,984.60359172)(231.3961,987.12183072)
\curveto(221.74648,980.23302472)(224.10903,953.92906572)(211.3961,952.12183472)
\curveto(205.41609,951.52969472)(191.97668,960.48033572)(192.97668,959.48033572)
}
}

\psline[linewidth=1,linecolor=red,arrowsize=7]{<-}(191.13582,942.44932472)(196.3961,942.12183472)
\rput(194,935){$\color{red} b$}

\end{pspicture}}
	\end{center}
	\caption{The two diagrams are (regular) isotopy equivalent to each other, i.e., they can be smoothly deformed into each other without cutting, treating each line as a thickened ribbon.  In a TQFT we expect that these two diagrams should have the same amplitude.}
	\label{fig:fig1}
\end{figure}

While there are several, essentially equivalent, definitions of a TQFT\cite{nayakreview,MyBook,Kirillov,AtiyahBook}, for our purposes we will think of a TQFT as a set of rules that
 takes as an input a space-time diagram of particles
tracing out world-lines and gives a complex number, or amplitude, as
an output.  The input space-time diagrams can include simple
space-time motion of particles; creation or annihilation of
particle-antiparticle pairs; fusion of particles when two particles
meet to form a combined ``bound state" particle; and the reverse
process, splitting, where one particle divides to form two others (See
Fig. \ref{fig:fig1} or \ref{fig:SU22} for example).  The essence of the TQFT is nothing
more than a diagrammatic calculus that maps from the input diagram to
the output amplitude.

One of the key applications of TQFTs is in the construction of knot
and link invariants.  The input space-time diagram can be a labeled
knot or link (representing world lines of particle types) and the
output amplitude is the knot invariant.  Knot invariants of this type
were famously constructed by Witten\cite{witten} using Chern-Simons
TQFTs.  For such TQFTs the diagrammatic rules are constructed so as to
be (regular) ``isotopy invariant" meaning that any smooth deformation
of the diagram leaves the output amplitude unchanged as shown in
Fig.~\ref{fig:fig1} (the word ``regular'', here means we should treat
strands as thickened ribbons so as to keep track of self twists).

%By choosing normalizations of vertices (when particle are created, annihilated, fuse, or split) appropriately and choosing a convenient gauge  the diagrammatic calculus is usually arranged\cite{Kitaev20062,bondersonthesis} to be so-called ``space-time isotopy invariant".   This means that any space-time diagram that can be smoothly deformed into another space-time diagram will have the same amplitude (See right versus left of Fig \ref{fig:fig1} for example).      Isotopy invariance has great advantages in that it makes many calculations extremely easy as simple geometric manipulations of a diagram.   Further this makes a direct connection between the diagrammatic calculus of a TQFT and certain knot invariants\cite{witten}.  

However, in TQFTs which contain a particle which happens to be its own
antiparticle (we say the particle is ``self-dual''), the usually
applied rules of diagrammatic evaluation can have an obstruction to
isotopy invariance.  Each self-dual particle has a property known as
its Frobenius-Schur indicator.  The Frobenius-Schur indicator is
simply a sign $\kappa = \pm 1$.  Cases where $\kappa=-1$ are cases
where there is an obstruction to isotopy invariance.  In particular,
as shown in Fig.~\ref{fig:fig2}, straightening out a ``zig-zag" in the
space-time path of a particle with $\kappa = -1$ incurs a minus sign
in the amplitude corresponding to the space-time diagram compared to
the diagram without the space-time zig-zag.  (A zig-zag, as shown in
Fig.~\ref{fig:fig2} involves one particle pair creation event and one
particle pair annihilation event.)  This minus sign from this zig-zag
cannot be removed by gauge choices and normalization choices, and must
be treated very carefully.  A good discussion of this issue is given,
for example, in Ref.~\onlinecite{WangBook}. These signs have often been
treated incorrectly or omitted in the physics literature.   Such nontrivial signs occur
even in theories, such as Chern-Simons theories, which we think of as
being diffeomorphism invariant.  The purpose of this article is to
clarify the physics of the Frobenius-Schur indicator.

\begin{figure}[h!]
\begin{pspicture}(0,0)(3,1.5)
\psset{linewidth=.03,linecolor=black}
\pscurve(0,0)(.25,1)(-.25,.5)(0,1.5)
\rput(1.25,.75){\scalebox{1.25}{$= \kappa_a$}}
\psline(2,0)(2.25,1.5)
\rput(.3,.2){$a$}
\rput(2.2,.2){$a$}
\end{pspicture}
%	\end{center}
%	\vspace*{-15pt}
\caption{A space-time path (time directed upwards) of a particle which
  is its own antiparticle (hence there is no directed arrow on the
  world-line).  The Frobenius-Schur indicator $\kappa_a=\pm 1$ for
  particle type $a$ is included for removing the ``zig-zag" in
  space-time.  (A zig-zag is defined to involve one particle pair
  creation event and one particle pair annihilation event.)  If
  $\kappa_a = -1$ for a particle, then diagrams do not have isotopy
  invariance as the zig-zag on the left incurs a minus sign when it is
  straightened.  Note, in section \ref{sec:flags} we will clarify that
  this diagram is to be interpreted via ``Convention 1'' where flags
  are assigned to all be pointing to the right.}
	\label{fig:fig2}
\end{figure}

Even if we are not considering fully-fledged 2+1 dimensional TQFTs,
many tensor categories can be thought of as diagrammatic algebras ---
planar algebras without necessarily having a notion of over- and
under-crossings.  Self-dual objects in these theories can also be
assigned Frobenius-Schur indicators.  One of the main applications of
such planar diagrammatic algebra is in building toy model 2+1
dimensional Hamiltonians that have topological properties --- so
called string-net Hamiltonians\cite{LevinWen,Burnell}.  We will comment
briefly on how, in the context of string nets, the Frobenius-Schur
indicator can be handled.

The realization that theories we think of as being isotopy (or
diffeomorphism) invariant may incur signs when a zig-zag is
straightened leads us to reconsider how invariant these theories
actually are --- and whether TQFTs have other constraints
in deforming diagrams.

The outline of this paper is as follows:

In section \ref{sec:basics}
we introduce the basic definitions associated with the Frobenius-Schur
indicator.  In section \ref{sec:analogy}, to make this physics a bit
more familiar, we draw an analogy with spin 1/2 particles.

In section \ref{sec:flags} we describe the use of so-called ``flags''
for keeping track of signs of Frobenius-Schur indicators --- whereby
we associate one of two possible flag directions with particle
creation or annihilation.  This technique is used commonly in the
literature (See for example
Refs.~\onlinecite{Kitaev20062,bondersonthesis,MyBook}).  We then emphasize
that there are two different conventions commonly used in how one
assigns the direction of these flags to a diagram, and these different
conventions have different physical interpretations.  The first
convention (``Convention 1'') gives an output from any diagram that
explicity incorporates signs associated with straightening zig-zags as
shown in Fig.~\ref{fig:fig2}.  However, the second convention
(``Convention 2'') is constructed so as to not give such signs and
thereby give an output which is fully isotopy invariant.

In section \ref{sec:ChernSimons} we explain how the knot invariants
famously derived by Witten\cite{witten} from Chern-Simons theory
correspond to our ``Convention 2'' for assigning these flags so as to
obtain an isotopy invariant output from any input knot diagram. We
explain how this convention arises from the need to give Wilson loops
(or particle world lines) a framing.

In section \ref{sec:su21} we consider the very simplest anyon theory
(or modular tensor category) having a nontrivial Frobenius-Schur
indicator, the so-called semion theory or $SU(2)_1$.  We consider in
detail (in sections \ref{sub:SU2Con2} and \ref{sub:nonisocon1}) the
above mentioned two conventions for assigning flags to a diagram and
discuss the differing outputs from diagrammatic evaluations.  We then
introduce a third convention in section \ref{sub:capcounting} which is
physically equivalent to ``Convention 1'' but uses a bookkeeping trick
called ``Cap Counting'' to take care of minus signs by pushing 
signs onto the diagrammatic value of a loop (the ``loop weight'').  As a result,
the evaluation of the diagram {\it seems} isotopy invariant, until the
last step where certain signs are added and the isotopy invariance is
broken again.

We then turn to a very brief discussion of string nets in section
\ref{sec:stringnets}.  We explain in section \ref{sub:stringnetsemion}
how the cap counting technique is particularly convenient in the case
of building a string net from $SU(2)_1$ as it makes the ground state
wavefunction look isotopy invariant.  We explain how the cap counting
technique applied to a string net can be interpreted as simply a gauge
transform.  Further we discuss the effect of the interpretation of the
Frobenius-Schur indicator in the TQFT that emerges from this string
net.  In section \ref{sec:SU22} we briefly outline how this cap
counting would apply to another simple theory $SU(2)_2$, suggesting
that the idea is more general.

In section \ref{sec:generalizedcapcounting} we then explain how the
cap counting technique (and pushing minus signs onto the loop weight)
can be generalized to many other theories so long as they admit a so-called ``${\mathbb{Z}_2}$ Frobenius-Schur
grading''.  Existence of such a grading is extremely common, particularly among braided theories. It is fairly hard to find cases which do not admit such a grading --- we give some examples in the appendices.

Having discussed isotopy invariance applied to evaluation of
knots, in section \ref{sec:fullisotopy} we discuss whether there can
be further impediments to isotopy invariance of diagram algebras
coming from behavior at vertices (fusions and splittings).  The
particular worry is that there may be a phase incurred if we try to
mutate a vertex with two legs going down and one going up to a vertex
with two legs going up and one leg going down.  Assuming
  that we have handled the signs from removing zig-zags using the
  cap-counting techniques, we find that there is only a single
  possible obstruction to turning legs up and down, and this is
  related to a closely related quantity known as the {\it third}
  Frobenius-Schur indicator.  If this indicator is nontrivial, there
  is an obstruction to turning legs up and down (i.e., a phase that
  cannot be removed).  This obstruction can occur, even for braided
  (or modular) theories --- although such situations appear to be
extremely unusual.

Assuming no such third Frobenius-Schur indicator obstruction occurs,
and assuming we have a $\mathbb{Z}_2$ Frobenius-Schur grading, then
for planar diagram algebra we have full isotopy --- i.e., diagrams can
be deformed smoothly within the plane and the value of the diagram remains
unchanged.  For 2+1 dimensional theories, i.e., theories with well
defined braidings and which give rise to invariants of links and knots, we might wonder if we also have full isotopy invariance of diagrams with vertices.
The answer is a bit subtle and we discuss this in more detail in
section \ref{sub:howmuch}.  In particular one is allowed to deform
diagrams --- but only in limited ways. If we think of the diagrams as
being made of ribbons, we are only allowed to consider configurations
with one side of the ribbon facing forwards at both the beginning and
end of the deformation.  One way to describe this is to say that if we
have a diagram in the shape of a tetrahedron, we are allowed to rotate
the tetrahedron arbitrarily, but not invert it.  There are some
theories where inversion of the tetrahedron is also allowed, which we
call ``full tetrahedral symmetry'', but these are not generic.

Finally in section \ref{sec:moregeneral} we briefly discuss more
general transformations that might be possible beyond our ``cap
counting'' scheme, and in section \ref{sec:conclusions} we give some brief
conclusions.

There are several major appendices to this work.  Appendices
\ref{sec:many} are devoted to exploring how common it is to have
theories that have $\mathbb{Z}_2$ Frobenius-Schur gradings.  We claim,
particularly among braided theories, that it is extremely common and it is
rather hard to find exceptions (although exceptions do exist), and we
detail this claim in the appendix.

In Appendix \ref{app:turning} we detail the proof that the only
obstruction to turning up and down legs of vertices is given by the
third Frobenius-Schur indicator.  We further prove that for braided (ribbon)
theories, such obstruction cannot occur unless there are fusion
multiplicities.

Finally Appendix \ref{app:examples} discusses a number of unusual
examples.  Section \ref{sub:exceptions} considers theories which do
not admit a $\mathbb{Z}_2$ Frobenius-Schur grading.  For braided
theories, these are fairly hard to find.  Section \ref{app:modularZ3}
gives an example of a modular (and therefore braided) theory which has
a nontrivial third Frobenius-Schur indicator.  Section
\ref{app:modularinv} discusses a modular (and therefore braided)
theory which has isotopy invariance that allows rotation of
tetrahedral diagrams, but not inversion.

\section{Some basics regarding the definition of the Frobenius-Schur Indicator}
\label{sec:basics}

\subsection{Review of Diagrams, F-moves, Gauge Choice, and Isotopy Invariance}

Let us first remind ourselves of the idea of $F$-moves and the
structure of Hilbert space in topologically ordered systems.  We will
always assume we are describing a unitary topological system (a
unitary TQFT) --- i.e., a quantum mechanical system that might be
physically realized.  We draw diagrams of lines labeled with particle
types ($a,b,c, \ldots$) and arrows. Each particle has an antiparticle
which we label with an overline such as $\bar a$.  Reversing an arrow
changes a particle to its anti-particle.  If $a=\bar a$ we say the
particle is self-dual and we do not draw an arrow on the corresponding
line.

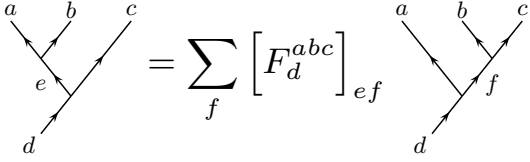
\begin{figure}[h!]
\newcommand{\texttauhere}{2}		
	\scalebox{.5}{\begin{pspicture}(2,0)(10,3) 
		%% tree
		\psset{linewidth=.04,linecolor=black,arrowsize=.2,xunit=.8cm,yunit=1cm}
		\psline[ArrowInside={->}](1,2)(2,3)
		\psline[ArrowInside={->}](1,2)(0,3)
		\psline[ArrowInside={->}](2,1)(1,2)
		\psline[ArrowInside={->}](2,1)(4,3)
		\psline[ArrowInside={->}](1,0)(2,1)
		\rput[cB](0,3.3){\scalebox{\texttauhere}{\color{black}{$a$}}}
		\rput[cB](2,3.3){\scalebox{\texttauhere}{\color{black}{$b$}}}
		\rput[cB](4,3.3){\scalebox{\texttauhere}{\color{black}{$c$}}}
		\rput[cB](1.0,1.3){\scalebox{\texttauhere}{\color{black}{$e$}}}
		\rput[cB](0.6,-0.3){\scalebox{\texttauhere}{\color{black}{$d$}}}
		
\rput(8.5,1.5){\scalebox{3}{$=\displaystyle{\sum_f} \left[F^{abc}_d\right]_{ef}$}}

\rput(13,0){
		%% tree
		\psset{linewidth=.04,linecolor=black,arrowsize=.2,xunit=.8cm,yunit=1cm}
		\psline[ArrowInside={->}](3,2)(2,3)
		\psline[ArrowInside={->}](2,1)(0,3)
		\psline[ArrowInside={->}](2,1)(3,2)
		\psline[ArrowInside={->}](3,2)(4,3)
		\psline[ArrowInside={->}](1,0)(2,1)
		\rput[cB](0,3.3){\scalebox{\texttauhere}{\color{black}{$a$}}}
		\rput[cB](2,3.3){\scalebox{\texttauhere}{\color{black}{$b$}}}
		\rput[cB](4,3.3){\scalebox{\texttauhere}{\color{black}{$c$}}}
		\rput[cB](3.0,1.3){\scalebox{\texttauhere}{\color{black}{$f$}}}
		\rput[cB](0.6,-0.3){\scalebox{\texttauhere}{\color{black}{$d$}}}

		%			\rput[cB](.7,-.3){\scalebox{\texttauhere}{\color{blue}{$\tau$}}}			
	}
		\end{pspicture}}

%          \vspace*{10pt}
%		\includegraphics[width=3in]{Fmatrix.png}
%\scalebox{.13}{\input{Diagram3.tex}}
%	\vspace*{5pt}
              \caption{Fusion trees describing the structure of
                Hilbert space.  The labels $a,b,c$ are the physical
                particle types we are describing. Here, $d$, $e$ and
                $f$ are the results of fusing together these particles
                in various orders as shown in the diagrams.  The basis
                of states shown in the left figure ($a$ and $b$ fusing
                to $e$) is described in terms of the basis of states
                described by the right figure ($b$ and $c$ fusing to
                $f$).  The two different bases are related by the
                $F$-matrix as shown\cite{Note1}.  }
	\label{fig:Fmatrix}
\end{figure}

For a system with multiple anyons at a fixed moment in time, the
Hilbert space dimension depends on the fusion rules of the theory.
The (generically) multi-dimensional space can be described in several
different bases, which we typically draw as trees as shown in, say,
the left of Fig.~\ref{fig:Fmatrix}.  For example, the diagram on the
left of the figure represents a particular state in the Hilbert space of
three particles of types $a$, $b$ and $c$.  For this particular state
(reading upwards) particle $e$ splits into $a$ and $b$ showing that the
quantum number of $e$ is the same as that of $a$ and $b$ put
together.  We may also say that $a$ and $b$ fuse to $e$. 
Similarly the quantum number of all three particles $a$,
$b$, $c$ put together is the same as that of particle type $d$.  It is
important to note that diagrams having different values of $e$ are
orthogonal to each other.

On the other hand, we can also describe the same states via the tree
on the right hand side, where $f$ splits into $b$ and $c$ (and again,
all three particles have the same quantum number as $d$).  These two
possible bases on the left and right of the figure are related by the
$F$-matrix $F^{abc}_d$ with elements $[F^{abc}_d]_{ef}$, the $F$-symbols, as shown in
Fig.~\ref{fig:Fmatrix}\footnote{It is assumed here for simplicity of
  presentation that there are no fusion multiplicities greater than 1,
  otherwise vertices would contain an additional index. See
  Refs. \onlinecite{Kitaev20062,bondersonthesis,MyBook} }.

\begin{figure}[h!]
	\begin{center}
		\vspace*{0pt}
\scalebox{.15}{\input{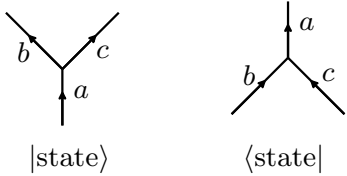}}
	\end{center}
	\vspace*{-15pt}
	\caption{A ket (left) and its corresponding bra (right)\cite{Note1}.  }
	\label{fig:braandket}
\end{figure}

For each possible vertex (intersection of three lines) in a diagram,
there is a choice of gauge.  If we change this choice of gauge we
multiply each vertex by an arbitrary phase factor\footnote{In the case
  where there is fusion multiplicity at the vertex the factors
  $u^{bc}_a$ would instead be a unitary matrix. See
  Refs. \onlinecite{Kitaev20062,bondersonthesis,MyBook}.} 
$u^{bc}_a$ for a diagram where $a$ splits into $b$ and $c$ like
the one on the left of Fig.~\ref{fig:braandket}.  Such gauge changes
result in a change in the $F$-matrix via
\begin{equation}
  [F^{abc}_d]_{ef}^{{\rm new}}  = \frac{u^{bc}_f u^{af}_d}{u^{ab}_e u^{ec}_d} [F^{abc}_d]_{ef}^{{\rm old}}
  \label{eq:gaugechange}
\end{equation}

So far we have described splitting diagrams, where, going upwards in
the diagram (forward in time) a particle splits into others.  We
should think of such diagrams as being kets.  The corresponding bras
are obtained by turning the diagrams upside down and reversing all
arrows -- so for example, the diagram on the left of
Fig.~\ref{fig:braandket} is the ket corresponding to the bra on the
right.

\begin{figure}[h!]
	\begin{center}
		\vspace*{-5pt}
\scalebox{.1}{\input{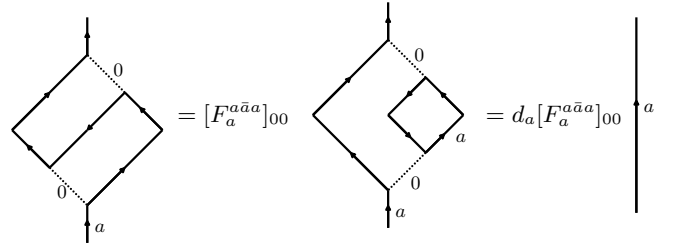}}
	\end{center}
	\vspace*{-15pt}
	\caption{A zig-zag diagram.   Here $0$ represents the identity or vacuum particle.  }
	\label{fig:Wig}
\end{figure}

Now consider the diagram shown in Fig.~\ref{fig:Wig} which is the same
as the left of Fig.~\ref{fig:fig2}, but with some identity lines drawn
(dotted and labeled with 0).  Here we view the up-down zig-zag as an
inner product between a bra (upper half) and a ket (lower half) when
cut horizontally at mid-height.  In the middle of Fig.~\ref{fig:Wig}
we have evaluated the left diagram using the $F$-move from
Fig.~\ref{fig:Fmatrix} (as well as using the fact that different
values of the particle $f$ in the ket on the right of
Fig.~\ref{fig:Fmatrix} are orthogonal to each other --- so we need
only $[F]_{00}$ not $[F]_{0j}$).

We would like our diagrams to obey isotopy invariance, so that we can
straighten out wiggly lines as in Fig.~\ref{fig:fig2} or the right of
Fig.~\ref{fig:Wig}.  Examining Fig.~\ref{fig:Wig}, we can achieve this
invariance at least up to a phase factor by defining the diagrammatic loop weight $d_a$ of particle
$a$ as
$$
 d_a = \left|  [F^{a \bar{a} a}_{a}]_{00} \right|^{-1}.
$$
where $0$ indicates the identity, or vacuum particle.  With this
definition, this quantity, $d_a > 0$,  is also known as the
``quantum dimension'' of the particle $a$ which describes how the
dimension of the Hilbert space increases asymptotically\cite{Kitaev20062} with the number $N_a$ of
particles of type $a$, as $N_a$ grows large,  via ${\rm Dim} \sim d_a^{N_a}$.

We assign this factor $d_a$ to a closed loop of the particle $a$
(i.e., creation of $\bar a$ and $a$ followed by reannihilation of the
same), as in Fig.~\ref{fig:Wig}.  For consistency we must include
factors of $d_a^{1/4}$ at any vertex including particle type $a$. In
particular, a vertex like those shown in Fig.~\ref{fig:braandket}
should be associated with a factor
\begin{equation}
\label{eq:vertexfactor} \mbox{vertex factor} =  \left(\frac{d_b d_c}{d_a}\right)^{1/4}
\end{equation}
This then gives us the modified orthonormality shown in
Fig.~\ref{fig:bubblecollapse} and the modified completeness relation
shown in Fig.~\ref{fig:completeness}.  In particular, in
Fig.~\ref{fig:bubblecollapse} with $a$ being replaced by the identity,
we must have $b=\bar c$, resulting in a closed loop of type $b$ which
is properly assigned the value $d_b$ (since $d_0 = 1$ and $d_c=d_{\bar c}$).  Note that
since this closed particle loop is the inner product between a bra and
its corresponding ket, we expect its inner product, and hence the
loop weight $d_b$ is positive definite.

\begin{figure}[h]
\hspace*{-2cm}\scalebox{.7}{\begin{pspicture}(6,.5)(0,3.5)
\psline[ArrowInside=->,linewidth=.04,arrowsize=.2](1,.5)(1,1.25)
\psline[ArrowInside=->,linewidth=.04,arrowsize=.2](1,2.75)(1,3.5)
\psline[ArrowInside=->,linewidth=.04,arrowsize=.2](1,1.25)(1.5,2)
\psline[ArrowInside=->,linewidth=.04,arrowsize=.2](1,1.25)(.5,2)
\psline[ArrowInside=->,linewidth=.04,arrowsize=.2](1.5,2)(1,2.75)
\psline[ArrowInside=->,linewidth=.04,arrowsize=.2](.5,2)(1,2.75)

\rput(1.3,.75){\scalebox{2}{\color{black}{$a$}}}
\rput(1.3,3.25){\scalebox{2}{\color{black}{$d$}}}
\rput(1.55,1.5){\scalebox{2}{\color{black}{$c$}}}
\rput(.5,1.5){\scalebox{2}{\color{black}{$b$}}}
\rput(1.55,2.5){\scalebox{2}{\color{black}{$c$}}}
\rput(.5,2.5){\scalebox{2}{\color{black}{$b$}}}

\rput(3,0){
  \psline[ArrowInside=->,linewidth=.04,arrowsize=.2](1,.5)(1,1.25)
  \psline[ArrowInside=->,linewidth=.04,arrowsize=.2](1,2.75)(1,3.5)
  \psbezier[ArrowInside=->,linewidth=.04,arrowsize=.2,ArrowInsidePos=.7](1,1.25)(.5,2)(.5,2)(1,2.75)
  \psbezier[ArrowInside=->,linewidth=.04,arrowsize=.2,ArrowInsidePos=.7](1,1.25)(1.5,2)(1.5,2)(1,2.75)

  \rput(1.3,.75){\scalebox{2}{\color{black}{$a$}}}
  \rput(1.3,3.25){\scalebox{2}{\color{black}{$d$}}}
\rput(1.55,2.5){\scalebox{2}{\color{black}{$c$}}}
\rput(.5,2.5){\scalebox{2}{\color{black}{$b$}}}

}

\rput(2.5,1.8){\scalebox{2}{\color{black}{$\displaystyle{=}$}}}

\rput(2.5,0){
\rput(4.25,2){\scalebox{2}{\color{black}{$\displaystyle{= \delta_{ad} \sqrt{\frac{d_b d_c}{d_a}}}$}}}

\psline[ArrowInside=->,linewidth=.04,arrowsize=.2](7.0,.5)(7.0,3.5)
\rput(7.5,2){\scalebox{2}{\color{black}{$a$}}}
}

\end{pspicture}}
\caption{Modified Orthonormality (Bubble-Collapsing).}
\label{fig:bubblecollapse}
\end{figure}
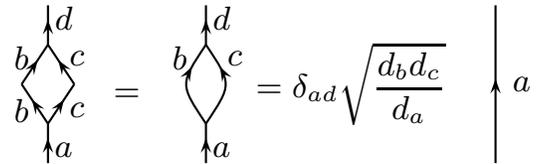  

\begin{figure}[h]
\hspace*{-4cm}\scalebox{.7}{\begin{pspicture}(4,.5)(0,2.5)
	\psline[ArrowInside=->,linewidth=.04,arrowsize=.2](1,.5)(1,2.75)
	\rput(.7,1.5){\scalebox{2}{\color{black}{$a$}}}
	\psline[ArrowInside=->,linewidth=.04,arrowsize=.2](1.8,.5)(1.8,2.75)	
	\rput(2.1,1.5){\scalebox{2}{\color{black}{$b$}}}
	
	\rput(4.5,1.5){\scalebox{2}{\color{black}{$\displaystyle{=\sum_{\color{black}c} \sqrt{\frac{d_c}{d_a d_b}}}$}}}

	\rput(3,0){

	\psline[ArrowInside=->,linewidth=.04,arrowsize=.2](5,.5)(5.5,1.25)
	\psline[ArrowInside=->,linewidth=.04,arrowsize=.2](6,.5)(5.5,1.25)
	\psline[ArrowInside=->,linewidth=.04,arrowsize=.2](5.5,1.25)(5.5,2.0)

	\psline[ArrowInside=->,linewidth=.04,arrowsize=.2](5.5,2.0)(5,2.75)
	\psline[ArrowInside=->,linewidth=.04,arrowsize=.2](5.5,2.)(6,2.75)

	\rput(5.8,1.5){\scalebox{2}{\color{black}{$c$}}}
	
	\rput(4.75,.5){\scalebox{2}{\color{black}{$a$}}}
	\rput(6.25,.5){\scalebox{2}{\color{black}{$b$}}}
	
	\rput(4.75,2.5){\scalebox{2}{\color{black}{$a$}}}
	\rput(6.25,2.5){\scalebox{2}{\color{black}{$b$}}}}
    \end{pspicture}}
  \caption{Modified Completeness Relation}
  \label{fig:completeness}
\end{figure}
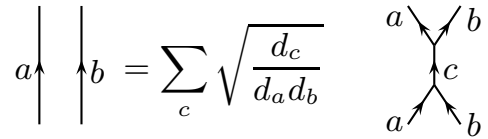

With the three diagrammatic rules ($F$-moves (Fig.~\ref{fig:Fmatrix}),
completeness (Fig.~\ref{fig:completeness}), and bubble-collapsing
(Fig.~\ref{fig:bubblecollapse}) we can evaluate any planar diagram.

In order for this modified vertex normalization to allow us to freely
straighten out a wiggly line as in Fig.~\ref{fig:Wig} without
incurring any phase, we only require that $[F^{a \bar{a} a}_{a}]_{00}$
be real and positive, so that the prefactor of the diagram on the right hand side of Fig.~\ref{fig:Wig}
cancels completely.  Since we are free to choose the gauge of the two
vertices which create $a$ and $\bar a$  from the vacuum, we can use this
freedom (See Eq.~\ref{eq:gaugechange}) to fix $[F^{a \bar{a} a}_{a}]_{00}$ to be positive if $a \neq \bar a$, by choosing the gauge transform constant $u^{a \bar a}_0/u^{\bar a a}_0$.  This then achieves the desired
isotopy invariance whenever $a \neq \bar a$.

\subsection{Frobenius-Schur Indicator}

Unfortunately, in the case where a particle is self-dual $a = \bar a$,
the quantity $[F^{a\bar a a}_a]_{00} = [F^{aaa}_a]_{00}$ is gauge invariant, and gauge
  transformation is not possible.  Hence there may be an obstruction
  to having a diagrammatic algebra that enjoys isotopy invariance.

  For a self-dual particle $a$ the quantity we are interested in is
  $[F_a^{aaa}]_{00}$.  It is fairly easy to show that this quantity
  must be real\cite{Kitaev20062}.  We thus define a quantity known as
  the Frobenius-Schur indicator\footnote{In the language of Ref.~\onlinecite{Ng} the Frobenius-Schur indicator that we are discussing for most of this paper would be called the {\it second} Frobenius-Schur indicator.} for a self-dual particle type
  ($a= \bar a$) to be given by
$$
 \kappa_a = {\rm{sign}} \left[F_a^{a a a}\right]_{{}_{00}}~~~.
$$
%Strictly speaking this is the {\it second} Frobenius-Schur indicator, but we will not
%speak of other indicators until section \ref{sec:fullisotopy}, so until that point this is what we will mean when we say ``Frobenius-Schur Indicator''. 
For a particle which is not its own antiparticle, there are various
differing definitions of Frobenius-Schur
indicator\cite{Bantay,Kitaev20062}.  We will thus avoid making any
statements about Frobenius-Schur indicators for non-self-dual
particles.

In the case where $\kappa_a = -1$, one might think that we can make
some redefinition to get rid of this sign. In particular we can
declare that the  loop weight $d_a$ (the diagrammatic value of the closed
loop) to be precisely $[F^{aaa}_a]_{00}^{-1}$ rather than its absolute
value.  However, this is problematic because then the inner product of
a state with itself (the lower half of the loop with the upper half of
the loop) would give a negative result --- thus sacrificing
positivity of norms, and resulting in a nonunitary theory that should
not be used for quantum mechanical systems (although it may be
perfectly good for defining knot invariants).

Thus when there is a negative Frobenius-Schur indicator, we find
ourselves in a situation where we must either accept a negative
quantum dimension (and negative normed states) or we must accept that
removing zig-zags (as in Fig.~\ref{fig:fig2} or Fig.~\ref{fig:Wig})
accumulates a minus sign.

This may seem quite disturbing.  Particularly if we consider, for
example, Chern-Simons theories, such as $SU(2)_1$, which are supposed
to be diffeomorphism invariant and unitary.  Yet, many particles in Chern-Simons theory (such as 
the only nontrivial
particle in $SU(2)_1$) do indeed have a negative Frobenius-Schur
indicator.  In section \ref{sec:ChernSimons} below we discuss how
the idea of Frobenius-Schur indicators fits into Chern-Simons theory.

\section{Analogy with Spin-1/2 particles}
\label{sec:analogy}

While the physics of the Frobenius-Schur indicator might appear a bit
unfamiliar it turns out that there is a familiar analog in angular
momentum addition --- where the particle type (the label $a,b,c$ etc)
corresponds to the eigenvalue of $J^2$.

Consider three spin-1/2 particles which all taken together are in an
eigenstate of $J=1/2$.  We can describe the possible states of the
system with fusion diagrams analogous to Fig.~\ref{fig:Fmatrix} --- in
this case where $a,b,c$ and $e$ are all labeled as individually having
$J=1/2$.  In Fig.~\ref{fig:Fmatrix} we can (on the left of the figure)
consider either the fusion of the leftmost two particles to some
angular momentum $d=0$ or $d=1$, or we can (on the right of the
figure) consider fusion of the rightmost two particles to either
$f=0$ or $f=1$.  (Here, 0 and 1 refer to singlet and triplet).  The
$F$-matrix that relates these two descriptions of the same space is
given by $[F^{\frac{1}{2} \frac{1}{2}\frac{1}{2}}_\frac{1}{2}]_{df}$
which is often known as a $6j$ symbol in the theory of angular
momentum addition.

Given that the total spin is 1/2 we can focus on the case where the
total $z$-component of angular momentum is $J_z = 1/2$ as well.  The
state where the leftmost two particles fuse to the identity (or
singlet $J=d=0$) can then be written explicitly as
\begin{equation}
\label{eq:psi}
|\psi\rangle =  \frac{1}{\sqrt{2}} \left(|\uparrow_1 \downarrow_2 \rangle -   |\downarrow_1 \uparrow_2\rangle\right)  \otimes |\uparrow_3\rangle  
\end{equation}
where the subscripts are the particle labels given in left to right order.  This wavefunction is precisely analogous to the lower half (the ``ket") of the left hand side of Fig.~\ref{fig:Wig}.

On the other hand, we could use a basis where we instead fuse the rightmost two particles together first, as in the righthand side of Fig.~\ref{fig:Fmatrix}.   We can write the state where the right two fuse to $J=f=0$ analogously as 
\begin{equation}
\label{eq:psi'}
|\psi'\rangle =  |\uparrow_1\rangle  \otimes   \left(|\uparrow_2 \downarrow_3 \rangle -   |\downarrow_2 \uparrow_3\rangle\right) \frac{1}{\sqrt{2}}
\end{equation}
which is precisly analogous to  (but the conjugate of) the top half (the ``bra") of the left hand side of Fig.~\ref{fig:Wig}.   

It is easy to check that the inner product of these two states
$|\psi\rangle$ and $|\psi'\rangle$ (corresponding to the middle of
Fig.\ref{fig:Wig} is\footnote{This result of $-1/2$ is precisely the
  $6j$ symbol
  $$\left\{\begin{array}{ccc} 1/2 & 1/2 & 0 \\ 1/2 & 1/2
                                        &0 \end{array} \right\}=-1/2$$}
$$
 \langle \psi' | \psi \rangle = -1/2 
$$
        
By redefining the normalization of these states, we can arrange that this overlap have unit magnitude.    However, the sign cannot be removed.  The situation is the same for any two half-odd-integer spins fused to a singlet.

It is worth thinking for a moment about why the sign is hard to remove by any sort of redefinitions.  
The convention we have used above
in Eq.~\ref{eq:psi} and \ref{eq:psi'} is that the fusion of two
particles to the identity (to a singlet) should give
$ (|\uparrow_a \downarrow_b \rangle - |\downarrow_a\uparrow_b\rangle)/\sqrt{2}$ where $a$
is always to the left of $b$.  Whatever convention we choose should be
something that we can evaluate locally --- i.e., the result of fusing
$a$ with $b$ should be the same whether we fuse the result with some
$c$ on its left later, or some $d$ on its right later.  This sort of
locality requirement prevents us from finding a way to insert a minus
sign into the definition of $|\psi\rangle$ or $|\psi'\rangle$.

\section{Flags}
\label{sec:flags}
%
%The conventional method of evauating diagrams for a TQFT is discussed in detail, for example, by Refs.~
%\onlinecite{Kitaev20062,bondersonthesis}.  In the case of self-dual particles with nontrivial Frobenius-Schur indicators, one introduces a so-called "flag" on the world line of particles to aid with the bookkeeping.   When a self-dual particle pair is created from the vacuum it is assigned a (conventionally) right  pointing flag.   Similarly when a self-dual particle pair annihilates it is assigned a right-pointing flag.  We can switch the direction of a flag at the price of multiplying by a the Frobenius-Schur indicator, as shown

Let us now be a bit more precise with the rules for evaluating
diagrams.  Here we will use the so-called ``flag" method.  This is the method developed
by Ref.~\onlinecite{Kitaev20062} and also used in
Ref.~\onlinecite{bondersonthesis}.  It is equivalent to discussions in
Refs~\onlinecite{Turaev,Kirillov,Kassel}.

%In order to keep track of the signs that are incurred in this theory (and for any theory with nontrivial Frobenius-Schur indicators), various strategies are taken

%In references \onlinecite{Turaev,Kirillov,Kassel} fictitious degrees of freedom are introduced to distinguish between a particle going forwards and the antiparticle going backwards (although both of these are actually the same).  

%  We will then develop yet another bookkeeping scheme which has some advantages and some disadvantages which we will explain. 

We first work in a gauge (as discussed above) where we can remove
zig-zags freely for any non-self-dual particle, and also we can remove
zig-zags freely for any self-dual particle with Frobenius-Schur
indicator $\kappa=+1$.  The difficult part is in keeping track of
factors of the $-1$ for particles with the Frobenius-Schur indicator
$\kappa=-1$.  For these particles we introduce large arrows on the
particle lines in the diagram, known as flags, whenever a pair of such
self-dual particles is either created or destroyed\footnote{In
  category theory these flags correspond to choosing between $ev$ and
  $\tilde {ev}$ or $coev$ and $\tilde{coev}$.  In fact in category
  theory it is more natural to keep such flags for all particles
  whether or not they are self dual, and for both $\kappa_a = \pm$.
  However, here we are concerned mainly with the case where we cannot
  gauge away complications.} as shown in Fig.~\ref{fig:flags1}.  We
emphasize that these flag arrows are not the same arrows we put on
lines to distinguish $a$ from $\bar a$ (which we draw as smaller
arrows as in Fig.~\ref{fig:Fmatrix} for example).  Here the particles
are self-dual, and for self-dual particles we do not draw the smaller
arrows since $a=\bar a$.

We call the creation diagram a cup (right diagrams of top two lines of
Fig.~\ref{fig:flags1}) and the annihilation diagrams are called caps
(left diagrams of top two lines of Fig.~\ref{fig:flags1}).  The reason
we introduce the flags is because we wish to use two different cup
states and two different cap states, their conjugates. The two cup
(and cap) states differ by a factor $\kappa_a$, which allows us to
remove any explicit factors of $\kappa_a$ by choosing which of the cup
and cap states to use in the diagrams. We then need the flags to
distinguish the two cup (and cap) states.  Hermitian conjugation of a
cup gives a cap (and vice versa) but does not reverse the direction of
the flag.  As shown in the third line of Fig.~\ref{fig:flags1}, a cup
and a cap may be assembled to form a positive definite inner product
assuming that their flags are aligned.

When evaluating a diagram, a flag can be reversed at the price of a
single factor of $\kappa_a$ as shown in Fig.~\ref{fig:flags2}.  Two
 flags in successive cup-cap pairs can be cancelled with each other to
remove a zig-zag (as shown in Fig \ref{fig:flags3}) if the arrows are
oppositely directed.

Note that under twisting of lines, flags are rotated with the lines as in Fig.~\ref{fig:Rtwist}.
This is because the Frobenius-Schur indicator enters into the relationship between $R$
matrix element $R^{aa}_0$ for the self-dual particle $a$ and the twist factor $\theta_a$ via
$$
  R^{aa}_0 = \theta^*_a \kappa_a ~~~~~ (a = \bar a),
$$
as illustrated in Fig.~\ref{fig:Rmatrix},\ref{fig:Rtwist},\ref{fig:thetatwist}.

It turns out the that the flags are not just an accounting trick but they represent a genuine
degree of freedom of self-dual particles, connected to framing.  We will return to discuss the
meaning of this degree of freedom in section \ref{sec:ChernSimons}.

\begin{figure}[h!]
	\begin{center}
          \begin{pspicture}(0,-3)(4,2)
  \psset{linewidth=.04,linecolor=black,arrowsize=.3}

\rput(0,0){  
\rput(1,0){
  \psarc(0,1){.5}{0}{180}
  \rput(0,1.5){\scalebox{1.2}{\color{white}$\blacktriangleleft$}}
  \rput(0,1.46){\scalebox{2}{\color{black}$\triangleleft$}}
   
\rput(.4,1.6){\scalebox{1.2}{\color{black}{$a$}}}

   }

\rput(2.1,1.2){\scalebox{1.4}{\color{black}$=$}}

\rput(3.3,1.25){\scalebox{2}{\color{black}$[~~~~~]$}}
\rput(4.1,1.7){\scalebox{1.2}{\color{black}$\dagger$}}
\rput(3.25,2.5){ \psscalebox{1 -1}{
  \psarc(0,1){.5}{0}{180}
  \rput(0,1.5){\scalebox{1.2}{\color{white}$\blacktriangleleft$}}
  \rput(0,1.46){\scalebox{2}{\color{black}$\triangleleft$}}
  
\rput(.4,1.6){\psscalebox{1.2 -1.2}{\color{black}{$a$}}}
  
  }

}
}

\rput(0,-1.5){  
\rput(1,0){
  \psarc(0,1){.5}{0}{180}
  \rput(0,1.5){\scalebox{1.2}{\color{white}$\blacktriangleright$}}
  \rput(0,1.46){\scalebox{2}{\color{black}$\triangleright$}}
  \rput(.4,1.6){\scalebox{1.2}{\color{black}{$a$}}}

}

\rput(2.1,1.2){\scalebox{1.4}{\color{black}$=$}}

\rput(3.3,1.25){\scalebox{2}{\color{black}$[~~~~~]$}}
\rput(4.1,1.7){\scalebox{1.2}{\color{black}$\dagger$}}
\rput(3.25,2.5){ \psscalebox{1 -1}{
  \psarc(0,1){.5}{0}{180}
  \rput(0,1.5){\scalebox{1.2}{\color{white}$\blacktriangleright$}}
  \rput(0,1.46){\scalebox{2}{\color{black}$\triangleright$}}
  \rput(.4,1.6){\psscalebox{1.2 -1.2}{\color{black}{$a$}}}
  }
}
}

\rput(0,-3){

\rput(1,0){
  \pscircle(0,1){.5}
  \rput(0,1.5){\scalebox{1.2}{\color{white}$\blacktriangleleft$}}
  \rput(0,1.46){\scalebox{2}{\color{black}$\triangleleft$}}

  \rput(.4,.4){\scalebox{1.2}{\color{black}{$a$}}}
  
  \rput(0,.5){\scalebox{1.2}{\color{white}$\blacktriangleleft$}}
  \rput(0,.52){\scalebox{2}{\color{black}$\triangleleft$}}
  
}

\rput(2.1,1){\scalebox{1.4}{\color{black}$=$}}

\rput(3,0){
  \pscircle(0,1){.5}
  \rput(0,1.5){\scalebox{1.2}{\color{white}$\blacktriangleright$}}
  \rput(0,1.46){\scalebox{2}{\color{black}$\triangleright$}}
  \rput(.4,.4){\scalebox{1.2}{\color{black}{$a$}}}
  
  \rput(0,.5){\scalebox{1.2}{\color{white}$\blacktriangleright$}}
  \rput(0,.52){\scalebox{2}{\color{black}$\triangleright$}}
}

\rput(4.8,1.05){\scalebox{1.4}{\color{black}$=|d_a|$}}
}

\end{pspicture}
	\end{center}
	\vspace*{-15pt}
	\caption{Some of the flag-bookkeeping rules for keeping track of signs given by Frobenius-Schur indicators used by Ref.~\onlinecite{Kitaev20062}.   Large arrows represent flags on self-dual particle lines.     The inner product of a cup and cap is positive definite if they have aligned flags. 
}
	\label{fig:flags1}
\end{figure}

\begin{figure}[h!]
\begin{center}
\begin{pspicture}(1,1)(4,1.5)
\psset{linewidth=.04,linecolor=black,arrowsize=.3}

\rput(1,0){
  \psarc(0,1){.5}{0}{180}
  \rput(0,1.5){\scalebox{1.2}{\color{white}$\blacktriangleleft$}}
  \rput(0,1.46){\scalebox{2}{\color{black}$\triangleleft$}}
  \rput(.7,1.2){\scalebox{1.2}{\color{black}{$a$}}}
}

\rput(3,1.2){\scalebox{1.4}{\color{black}$= \kappa_a$}}

\rput(4.5,0){
  \psarc(0,1){.5}{0}{180}
  \rput(0,1.5){\scalebox{1.2}{\color{white}$\blacktriangleright$}}
  \rput(0,1.46){\scalebox{2}{\color{black}$\triangleright$}}
  \rput(.7,1.2){\scalebox{1.2}{\color{black}{$a$}}}
}
\end{pspicture}  
\end{center}
\caption{Reversing a flag incurs a factor of the Frobenius-Schur indicator $\kappa_a$.}
\label{fig:flags2}
\end{figure}

\begin{figure}[h!]
\begin{pspicture}(3.5,-2.75)(4,2)
  \psset{linewidth=.04,linecolor=black,arrowsize=.3}

\rput(0,0){

  \rput(0,0){  
\rput(1,0){
  \psarc(0,1){.5}{0}{180}
  \rput(0,1.5){\scalebox{1.2}{\color{white}$\blacktriangleleft$}}
  \rput(0,1.46){\scalebox{2}{\color{black}$\triangleleft$}}
}
\psline(.5,1)(.5,0)
\psline(2.5,1)(2.5,2)
  \rput(.3,.2){\scalebox{1.2}{\color{black}{$a$}}}

\rput(2,2){ \psscalebox{1 -1}{
  \psarc(0,1){.5}{0}{180}
  \rput(0,1.5){\scalebox{1.2}{\color{white}$\blacktriangleright$}}
  \rput(0,1.46){\scalebox{2}{\color{black}$\triangleright$}}
  }
}
}
\rput(3.0,1){\scalebox{1.4}{\color{black}$=$}}

\rput(3.0,0){  
\rput(1,0){
  \psarc(0,1){.5}{0}{180}
  \rput(0,1.5){\scalebox{1.2}{\color{white}$\blacktriangleright$}}
  \rput(0,1.46){\scalebox{2}{\color{black}$\triangleright$}}
}

\psline(.5,1)(.5,0)
\psline(2.5,1)(2.5,2)
  \rput(.3,.2){\scalebox{1.2}{\color{black}{$a$}}}

\rput(2,2){ \psscalebox{1 -1}{
  \psarc(0,1){.5}{0}{180}
  \rput(0,1.5){\scalebox{1.2}{\color{white}$\blacktriangleleft$}}
  \rput(0,1.46){\scalebox{2}{\color{black}$\triangleleft$}}
  }
}
}

\rput(6,1){\scalebox{1.4}{\color{black}$=$}}

\rput(6.0,0){
\psline(.5,0)(.5,2)
  \rput(.3,.2){\scalebox{1.2}{\color{black}{$a$}}}
}
}

\rput(0,-2.5){

  \rput(0,0){

\rput(3,0){   \psscalebox{-1 1}{ 
\rput(1,0){
  \psarc(0,1){.5}{0}{180}
  \rput(0,1.5){\scalebox{1.2}{\color{white}$\blacktriangleleft$}}
  \rput(0,1.46){\scalebox{2}{\color{black}$\triangleleft$}}
}
\psline(.5,1)(.5,0)
\psline(2.5,1)(2.5,2)
  \rput(.3,.2){\psscalebox{-1.2 1.2}{\color{black}{$a$}}}

\rput(2,2){ \psscalebox{1 -1}{
  \psarc(0,1){.5}{0}{180}
  \rput(0,1.5){\scalebox{1.2}{\color{white}$\blacktriangleright$}}
  \rput(0,1.46){\scalebox{2}{\color{black}$\triangleright$}}
  }
}
}
}
}

\rput(3.0,1){\scalebox{1.4}{\color{black}$=$}}

\rput(3.0,0){  

 \rput(3,0){ \psscalebox{-1 1}{
  \rput(1,0){
  \psarc(0,1){.5}{0}{180}
  \rput(0,1.5){\scalebox{1.2}{\color{white}$\blacktriangleright$}}
  \rput(0,1.46){\scalebox{2}{\color{black}$\triangleright$}}
}

\psline(.5,1)(.5,0)
\psline(2.5,1)(2.5,2)
  \rput(.3,.2){\psscalebox{-1.2 1.2}{\color{black}{$a$}}}

\rput(2,2){ \psscalebox{1 -1}{
  \psarc(0,1){.5}{0}{180}
  \rput(0,1.5){\scalebox{1.2}{\color{white}$\blacktriangleleft$}}
  \rput(0,1.46){\scalebox{2}{\color{black}$\triangleleft$}}
  }
}
}
}
}

\rput(6,1){\scalebox{1.4}{\color{black}$=$}}

\rput(6.0,0){
\psline(.5,0)(.5,2)
  \rput(.3,.2){\scalebox{1.2}{\color{black}{$a$}}}
}
}

\end{pspicture}
\caption{Opposite directed flags may be cancelled and zig-zags may be removed.}
\label{fig:flags3}
\end{figure}

\begin{figure}[h!]
	\begin{center}
\scalebox{.15}{\input{Diagram7.tex}}
	\end{center}
	\vspace*{-15pt}
	\caption{The general definition of the $R$ matrix.\cite{Note1}  }
	\label{fig:Rmatrix}
\end{figure}

\begin{figure}[h!]
	\begin{center}
\scalebox{.18}{\input{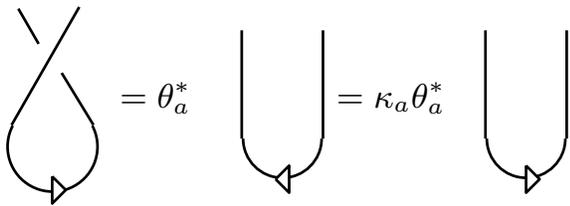}}
	\end{center}
	\vspace*{-15pt}
	\caption{When untwisting, the direction of the flag follows
          the change in direction of the string (as if the flag is
          attached to the string).  In order to get the flag pointing
          to the right again, one multiplies by the Frobenius-Schur
          indicator.  Thus for a self-dual particle
          $R_0^{aa} = \theta_a^* \kappa_a$.}
	\label{fig:Rtwist}
\end{figure}

\begin{figure}[h!]
	\begin{center}
\scalebox{.15}{\input{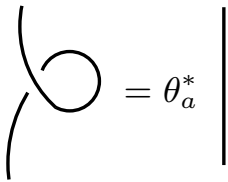}}
	\end{center}
	\vspace*{-15pt}
	\caption{In comparison to the $R$-matrix, a twist of this sort does not accumulate a factor of the Frobenius-Schur indicator.  This is true in either convention 1 or convention 2 for assigning flag directions.}
	\label{fig:thetatwist}
\end{figure}

We note that the mathematical purist may want to decorate all cups and
caps with flags whether or not the particle is self-dual and whether
or not $\kappa_a = \pm 1$.  While there is a certain appeal to
treating all particles on equal footing it is only for the self-dual
particles with $\kappa_a = -1$ that we cannot live without the flags,
and it is usually easier not to draw them in all other cases. 

Suppose now we want to turn a labeled knot or link  diagram
into an output (our interpretation of a TQFT is a prescription for
turning a diagram into an output).  It is important to emphasize that
if one is given such a diagram one must first decorate cups and caps
with flags in order to fully define the value of the diagram.  There
are two conventions we may choose to use, so as to determine how
flags are meant to point if they are not drawn (and the two different
conventions give physically different results).

\vspace*{10pt}

\noindent {\bf Convention 1:  All Flags Point Right}

In this convention we simply declare that at the beginning of a
calculation all flags are right-pointing.  Further, if we ever
generate a left-pointing flag (say by the left equality of
Fig.~\ref{fig:Rtwist}) we will immediately apply $\kappa_a$ to
reorient the flag again to point right (as in the right of
Fig.~\ref{fig:Rtwist}).

With this convention of all flags pointing right, note that
straightening out a zig-zag necessarily incurs a factor of $\kappa_a$
(this can be seen by combining Fig. \ref{fig:flags2} and
Fig.~\ref{fig:flags3}).  As a result this means that our diagrams are
not isotopy invariant.

Note that we have implicitly used this right-pointing convention in
defining our $F$-matrices in Fig.~\ref{fig:Fmatrix}.  In cases where
there is a cup (in either diagram) we assume the corresponding flag
points right. Further, Fig.~\ref{fig:fig2} has implicitly used this
convention as well.  Most diagrams drawn in the 
condensed-matter literature use this convention even if it is
not stated.

 This convention is most natural when one is describing
  physics in a Hamiltonian formalism. In such a case one might be
  presented with a particular (two-dimensional) wavefunction of a
  planar system.  Give such a ket, one can label the positions and
  types of each particle that is found in the system.  When two such
  particles (possibly self-dual) are created (annihilated), each such
  creation (annihilation) event looks identical to every other
  one. Thus it makes sense to label them all the same way and make
  sure the flags all point the same way.

  With this convention, it is important to note that even though we do
  not have full isotopy invariance (straightening a zig-zag can incur
  a minus sign) there is still some degree of deformation of diagrams
  that is allowed.  In particular any deformation of a diagram is
  allowed that does not change the time-direction of motion of any
  particles, and does not add or remove particle creation/annihilation
  events.  See the related discussion in the Conclusions.

\vspace*{10pt}

\noindent {\bf Convention 2:  Alternating Flags}

A different convention we might choose is that (starting with a knot
or link diagram) we should decorate cups and caps with flags that
alternate direction as we walk along the length of a strand (meaning
that the first flag points in the direction of walking, the second
points opposite the direction of walking and so forth).  With such a
convention, zig-zags can be freely straightened out (as in
Fig.~\ref{fig:flags3}).  The resulting diagrammatic evaluations are
then (regular) isotopy invariant, thus giving us a way to construct
true knot invariants.

The rule of decorating cups and caps with alternating flags may seem
like a disturbingly nonlocal rule: if one locally examines a cup or a
cap, one may have to walk a very long distance along the strand to
find out whether this cup or cap should be decorated with a
left-pointing or right-pointing flag.  Nonetheless, as we shall see in
section \ref{sec:ChernSimons}, there is a sensible physical interpretation to this rule,
and it is precisely this rule that needs to be applied for turning
Chern-Simons theories into a knot invariant\cite{witten}.

Note that flipping all of the flags in a knot or link leaves the final
evaluation unchanged.  This suggests that we have not actually
distinguished between $a$ and $\bar a$ which are meant to be the same.

This convention (or a convention equivalent to this in some language)
is used commonly by mathematicians (category theorists, knot
theorists, etc) and, as we will see in the next section, it arises
naturally in an action formalism such as Chern-Simons theory where the
knots are Wilson loop operators.  Here, it is much more natural to
think in terms of space-time histories and the particular ``state'' of
a system at some time-slice (which we discussed in ``Convention 1''
above) is not really a feature of the model.

\section{Isotopy Invariant Chern-Simons Theory}
\label{sec:ChernSimons}

Let us review some basic notions of constructing knot invariants from
Chern-Simons theory as pioneered by Witten\cite{witten}.  We will be
brief here, referring to the literature for further
details\cite{nayakreview,MyBook,witten}.

For a given reference manifold ${\cal M}$ (often taken to be $S^3$)
the Chern-Simons partition function for that manifold is given by the
functional integral
$$
Z({\cal M}) = \int_{{\cal M}} {\cal D} a_\mu(x) \,\, e^{i S_{CS}[a_\mu(x)]}
$$
where $S_{CS}$ is the Chern-Simons action and $a_\mu(x)$ is the
Chern-Simons vector potential, which is a Lie algebra valued vector
for some given Lie algebra.

We now consider a Wilson loop operator $\hat W_L$ defined along a closed
curve $L$
$$
 \hat W_L^R = {\rm Tr}^R\left[  P e^{\oint_L dl^\mu a_\mu} \right]
 $$
 where $P$ means that the integral should be path-ordered, and the
 superscript ${}^R$ means that the trace is taken in a representation
 $R$ of the Lie algebra.

For simplicity, we assume that the closed curve $L$ is embedded in a
reference manifold ${\cal M} = S^3$.  The famous result by
Witten\cite{witten} is that one can define a knot invariant as the
expectation of the Wilson loop operator
\begin{eqnarray} \label{eq:knotwitten}
  \mbox{Knot Invariant}^R_L &=& \langle \hat W_L^R \rangle
                              \\ &=& \frac{
  \int_{S^3} {\cal D}a_\mu(x) \,\, \hat W_L^R \,\, e^{i
    S_{CS}[a_\mu(x)]}}{\int_{{S^3}} {\cal D} a_\mu(x) \,\, e^{i
    S_{CS}[a_\mu(x)]}} \nonumber
 \end{eqnarray}
These knot invariants correspond to a diagram with particle type $R$ traveling along the closed loop $L$.

This prescription, while simple sounding, actually has a number of
subtleties.  The main subtlety we will focus on here is that the
expression for the knot invariant Eq.~\ref{eq:knotwitten} is actually
not well defined as it is written.  One must specify not only the
oriented path of the Wilson line, but also a {\it framing} of the
knot.  By framing here we mean that we not only specify a loop $L$ but
we attach a narrow ribbon to $L$ such that $L$ is one of the
boundaries of the ribbon.  The reason for this requirement is that in
order to evaluate integrals such as that in Eq.~\ref{eq:knotwitten}
one needs to regularize the Wilson loop integral by point-splitting,
where the two edges of the ribbon correspond to the splitting of a
single point into two very close points as shown in
Fig.~\ref{fig:framing1} (See the discussion in
Ref.~\onlinecite{witten}).  Since our theory is
  topological we expect that any physical results we may calculate
  should depend only on the topological properties of the framing.

If the ribbon twists around the particle path as shown in the middle
of Fig.~\ref{fig:framing1}, the twist in framing can be removed at the
cost of a factor of $\theta_a^*$ or $\theta_a$ depending on the
chirality of the twist.

For simplicity of presentation, and in the spirit of
  developing a planar diagram algebra, one often uses so-called {\it
    blackboard framing} in drawing knots and links.  In this
  representation the ribbon is assumed to lie flat on the page.  A
  twist such as the middle of Fig.~\ref{fig:framing1} is represented
  instead with a blackboard framed curl such as
  Fig.~\ref{fig:thetatwist}.

  With this convention we can define a reference frame in the
  following way.  The $z$-axis of our coordinate system points along
  the direction of the arrow.  The $y$-axis points from the solid to
  the dashed line (the two edges of the ribbon), and the $x$-axis of
  our coordinate frame always points out of the plane of the page
  towards the reader as in Fig.~\ref{fig:framing2}.    

  If the dashed line lies to the right of the solid line while walking
  along the direction of the arrow, we have specified a right-handed
  coordinate system, whereas if the dashed line lies to the left of
  the solid line, then we have a left-handed coordinate system.  Note
  that particle $a$ being framed right handed is equivalent to
  particle $\bar a$ being framed left-handed as shown in the left of
  Fig.~\ref{fig:framing2}.  This can be interpreted as the CPT
  theorem: charge, parity, and direction of motion are all reversed.  A right-handed $a$ can annihilate with a right-handed $\bar a$ as shown in the right of Fig.~\ref{fig:framing2}.

 \begin{figure}
    \begin{pspicture}(.5,0)(4,4)
    \psset{arrowsize=.2}
    \psline{->}(0,0)(0,1)
    \psline(0,1)(0,3)
    \rput(-.4,1){\scalebox{1.5}{$a$}}

    \rput(-.1,3.5){\scalebox{1}{unframed}}

    \rput(2,0){
    \psset{arrowsize=.2}
    \psline{->}(0,0)(0,1)
    \psline(0,0)(0,3)
    \rput(-.4,1){\scalebox{1.5}{$a$}}

    \psbezier[linestyle=dashed,border=.1,bordercolor=white](-.2,1.8)(-.2,2.3)(.2,2.3)(.2,3)
    
    \psline[linestyle=dashed](.2,0)(.2,.8)
    \psbezier[linestyle=dashed](.2,.6)(.2,1.6)(-.2,1.3)(-.2,1.8)
    \psline[border=.1,bordercolor=white](0,1)(0,2)

    \rput(.55,0){\scalebox{1.5}{\pstThreeDCoor[linewidth=.5pt,arrowsize=3pt,linecolor=blue,Alpha=80,Beta=30,xMax=.4,yMax=.4,zMax=.4,xMin=0,yMin=0,zMin=0,nameX={},nameY={},nameZ={}]
      \rput(0,.4){\scalebox{.6}{\color{blue}$z$}}
      \rput(-.1,-.25){\scalebox{.6}{\color{blue}$x$}}
      \rput(.3,-.15){\scalebox{.6}{\color{blue}$y$}}
    }}

      \rput(.55,2.6){\scalebox{1.5}{\pstThreeDCoor[linewidth=.5pt,arrowsize=3pt,linecolor=blue,Alpha=80,Beta=30,xMax=.4,yMax=.4,zMax=.4,xMin=0,yMin=0,zMin=0,nameX={},nameY={},nameZ={}]
      \rput(0,.4){\scalebox{.6}{\color{blue}$z$}}
      \rput(-.1,-.25){\scalebox{.6}{\color{blue}$x$}}
      \rput(.3,-.15){\scalebox{.6}{\color{blue}$y$}}
    }}

        \rput(-.7,1.6){\scalebox{1.5}{\pstThreeDCoor[linewidth=.5pt,arrowsize=3pt,linecolor=blue,Alpha=-110,Beta=30,xMax=.4,yMax=.4,zMax=.4,xMin=0,yMin=0,zMin=0,nameX={},nameY={},nameZ={}]
      \rput(0,.4){\scalebox{.6}{\color{blue}$z$}}
      \rput(.15,.25){\scalebox{.6}{\color{blue}$x$}}
      \rput(-.25,-.1){\scalebox{.6}{\color{blue}$y$}}
    }}

    \rput(-.1,3.5){\scalebox{1}{framed}}
  }

    \rput(5,0){\psset{arrowsize=.2}
    \psline{->}(0,0)(0,1)
    \psline(0,1)(0,3)
    \psline[linestyle=dashed](.2,0)(.2,3)
    \rput(-.4,1){\scalebox{1.5}{$a$}}
    \rput(-1.25,1.5){\scalebox{1.5}{$=\theta_a^*$}}

    \rput(.55,1.4){\scalebox{1.5}{\pstThreeDCoor[linewidth=.5pt,arrowsize=3pt,linecolor=blue,Alpha=80,Beta=30,xMax=.4,yMax=.4,zMax=.4,xMin=0,yMin=0,zMin=0,nameX={},nameY={},nameZ={}]
      \rput(0,.4){\scalebox{.6}{\color{blue}$z$}}
      \rput(-.1,-.25){\scalebox{.6}{\color{blue}$x$}}
      \rput(.3,-.15){\scalebox{.6}{\color{blue}$y$}}
    }}

  }

   \end{pspicture}
  \caption{The unframed figure on the left is ambiguous in
    Chern-Simons theory.  A framing is given for the diagrams on the
    right.}
  \label{fig:framing1}
\end{figure}
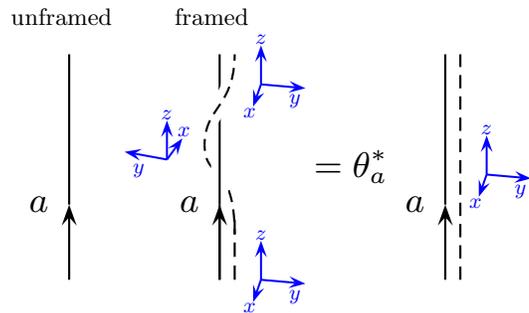

\begin{figure}
    \begin{pspicture}(2,-.25)(4,2.9)

 \rput(-1,0){   
    \rput(2.0,0){\psset{arrowsize=.2}
    \psline{->}(0,0)(0,1)
    \psline(0,1)(0,3)
    \psline[linestyle=dashed](.2,0)(.2,3)
    \rput(-.4,1){\scalebox{1.5}{$a$}}

    \rput(.55,0){\scalebox{1.5}{\pstThreeDCoor[linewidth=.5pt,arrowsize=3pt,linecolor=blue,Alpha=80,Beta=30,xMax=.4,yMax=.4,zMax=.4,xMin=0,yMin=0,zMin=0,nameX={},nameY={},nameZ={}]
      \rput(0,.4){\scalebox{.6}{\color{blue}$z$}}
      \rput(-.1,-.25){\scalebox{.6}{\color{blue}$x$}}
      \rput(.3,-.15){\scalebox{.6}{\color{blue}$y$}}
    }}

}  
  
    \rput(4,0){\psset{arrowsize=.2}
    \psline(0,1)(0,0)
    \psline{<-}(0,1)(0,3)
    \psline[linestyle=dashed](.2,0)(.2,3)
    \rput(-.4,1){\scalebox{1.5}{$\bar a$}}
    \rput(-1,1.5){\scalebox{1.5}{$=$}}

    % \rput(1.2,.5){\scalebox{1.5}{\pstThreeDCoor[linewidth=.5pt,arrowsize=3pt,linecolor=blue,Alpha=345,Beta=150,xMax=.4,yMax=.4,zMax=.5,xMin=0,yMin=0,zMin=0,nameX={},nameY={},nameZ={}]
    %   \rput(0,-.5){\scalebox{.6}{\color{blue}$z$}}
    %   \rput(-.15,-.25){\scalebox{.6}{\color{blue}$x$}}
    %   \rput(-.3,-.1){\scalebox{.6}{\color{blue}$y$}}
    % }}
    % this points down (z) out (x) left (y)

   \rput(.7,.5){\scalebox{1.5}{\pstThreeDCoor[linewidth=.5pt,arrowsize=3pt,linecolor=blue,Alpha=75,Beta=150,xMax=.45,yMax=.3,zMax=.45,xMin=0,yMin=0,zMin=0,nameX={},nameY={},nameZ={}]
       \rput(0,-.45){\scalebox{.6}{\color{blue}$z$}}
      \rput(-.2,-.2){\scalebox{.6}{\color{blue}$x$}}
       \rput(.25,-.14){\scalebox{.6}{\color{blue}$y$}}
     }}

  }}
    \end{pspicture}
  \begin{pspicture}(1,-.7)(4,1.7)
  \psset{linecolor=black,arrowsize=.3}

 %top 
\rput(0,0){

\rput(4.25,0){
  \psarc(0,1){.5}{0}{180}
  \rput(-.4,-.2){\scalebox{1.4 }{\color{black}{$a$}}}
  \psarc[linestyle=dashed](0,1){.3}{0}{180}
  \psline[linestyle=dashed](-.3,1)(-.3,0)
  \psline[linestyle=dashed](.3,1)(.3,0)

  \psline[ArrowInside=->,arrowsize=.2](-.5,0)(-.5,1)
    \psline[ArrowInside=->,arrowsize=.2](.5,1)(.5,0)
  
   }

\rput(0,-.8){

    \rput(2.8,1.2){\scalebox{1.5}{\pstThreeDCoor[linewidth=.5pt,arrowsize=3pt,linecolor=blue,Alpha=80,Beta=30,xMax=.4,yMax=.4,zMax=.4,xMin=0,yMin=0,zMin=0,nameX={},nameY={},nameZ={}]
      \rput(0,.4){\scalebox{.6}{\color{blue}$z$}}
      \rput(-.1,-.25){\scalebox{.6}{\color{blue}$x$}}
      \rput(.3,-.15){\scalebox{.6}{\color{blue}$y$}}
    }}

     \rput(5.6,1.5){\scalebox{1.5}{\pstThreeDCoor[linewidth=.5pt,arrowsize=3pt,linecolor=blue,Alpha=345,Beta=150,xMax=.4,yMax=.4,zMax=.5,xMin=0,yMin=0,zMin=0,nameX={},nameY={},nameZ={}]
       \rput(0,-.5){\scalebox{.6}{\color{blue}$z$}}
       \rput(-.15,-.25){\scalebox{.6}{\color{blue}$x$}}
       \rput(-.3,-.1){\scalebox{.6}{\color{blue}$y$}}
     }}
    % this points down (z) out (x) left (y)
}}  
\end{pspicture}                    
  \caption{Left: Particle $a$ being framed right-handed is equivalent to
    particle $\bar a$ being framed left-handed.  Right: a right-handed particle $a$ can annihilate a right-handed particle $\bar a$.    }
  \label{fig:framing2}
\end{figure}
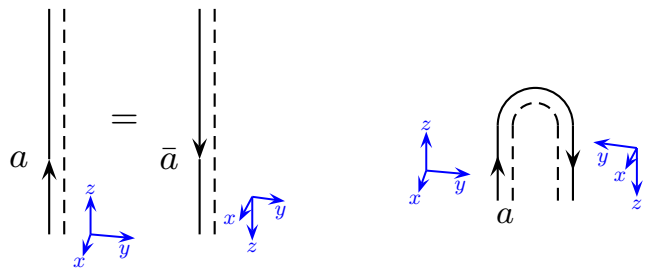

\begin{figure}
 \hspace*{-1.5cm} \begin{pspicture}(1,-3.5)(4,1.7)
  \psset{linewidth=.04,linecolor=black,arrowsize=.3}

 %top 
\rput(0,0){  
\rput(1,0){
  \psarc(0,1){.5}{0}{180}
  \rput(0,1.5){\scalebox{1.2}{\color{white}$\blacktriangleright$}}
  \rput(0,1.46){\scalebox{2}{\color{black}$\triangleright$}}
  \rput(.4,1.6){\scalebox{1.2}{\color{black}{$a$}}}
   
}

\rput(2.1,1.2){\scalebox{1.4}{\color{black}$=$}}

\rput(4.25,0){
  \psarc(0,1){.5}{0}{180}
  \rput(.4,1.6){\scalebox{1.2}{\color{black}{$a$}}}
  \psarc[linestyle=dashed](0,1){.3}{0}{180}
   }

    \rput(2.8,1.2){\scalebox{1.5}{\pstThreeDCoor[linewidth=.5pt,arrowsize=3pt,linecolor=blue,Alpha=80,Beta=30,xMax=.4,yMax=.4,zMax=.4,xMin=0,yMin=0,zMin=0,nameX={},nameY={},nameZ={}]
      \rput(0,.4){\scalebox{.6}{\color{blue}$z$}}
      \rput(-.1,-.25){\scalebox{.6}{\color{blue}$x$}}
      \rput(.3,-.15){\scalebox{.6}{\color{blue}$y$}}
    }}

     \rput(5.6,1.5){\scalebox{1.5}{\pstThreeDCoor[linewidth=.5pt,arrowsize=3pt,linecolor=blue,Alpha=345,Beta=150,xMax=.4,yMax=.4,zMax=.5,xMin=0,yMin=0,zMin=0,nameX={},nameY={},nameZ={}]
       \rput(0,-.5){\scalebox{.6}{\color{blue}$z$}}
       \rput(-.15,-.25){\scalebox{.6}{\color{blue}$x$}}
       \rput(-.3,-.1){\scalebox{.6}{\color{blue}$y$}}
     }}
    % this points down (z) out (x) left (y)

  %   \rput(5.5,1){\scalebox{1.5}{\pstThreeDCoor[linewidth=.5pt,arrowsize=3pt,linecolor=blue,Alpha=-110,Beta=30,xMax=.4,yMax=.3,zMax=.4,xMin=0,yMin=0,zMin=0,nameX={},nameY={},nameZ={}]
  %     \rput(0,.4){\scalebox{.6}{\color{blue}$z$}}
  %     \rput(.15,.25){\scalebox{.6}{\color{blue}$x$}}
  %     \rput(-.25,-.1){\scalebox{.6}{\color{blue}$y$}}
  %   }}
% this has into plane, pointing left, pointing up.

%    \rput(5.5,1){\scalebox{1.5}{\pstThreeDCoor[linewidth=.5pt,arrowsize=3pt,linecolor=blue,Alpha=-10,Beta=30,xMax=.3,yMax=.45,zMax=.4,xMin=0,yMin=0,zMin=0,nameX={},nameY={},nameZ={}]
%       \rput(0,.4){\scalebox{.6}{\color{blue}$z$}}
%       \rput(-.1,-.25){\scalebox{.6}{\color{blue}$x$}}
%       \rput(-.25,-.1){\scalebox{.6}{\color{blue}$y$}}
%     }}
% up z left y x out

  % \rput(3,1.5){\scalebox{1.5}{\pstThreeDCoor[linewidth=.5pt,arrowsize=3pt,linecolor=blue,Alpha=170,Beta=150,xMax=.3,yMax=.4,zMax=.4,xMin=0,yMin=0,zMin=0,nameX={},nameY={},nameZ={}]
  %     \rput(0,-.4){\scalebox{.6}{\color{blue}$z$}}
  %    \rput(.1,.25){\scalebox{.6}{\color{blue}$x$}}
  %     \rput(.25,-.14){\scalebox{.6}{\color{blue}$y$}}
  %   }}
  % this has into plane pointing right pointing down

%   \rput(3,1.5){\scalebox{1.5}{\pstThreeDCoor[linewidth=.5pt,arrowsize=3pt,linecolor=blue,Alpha=70,Beta=150,xMax=.4,yMax=.3,zMax=.4,xMin=0,yMin=0,zMin=0,nameX={},nameY={},nameZ={}]
%       \rput(0,-.4){\scalebox{.6}{\color{blue}$z$}}
%      \rput(-.2,-.2){\scalebox{.6}{\color{blue}$x$}}
%       \rput(.25,-.14){\scalebox{.6}{\color{blue}$y$}}
%     }}
  % down z  x out y right  

}

% second
\rput(0,-1.5){  
\rput(1,0){
  \psarc(0,1){.5}{0}{180}
  \rput(0,1.5){\scalebox{1.2}{\color{white}$\blacktriangleleft$}}
  \rput(0,1.46){\scalebox{2}{\color{black}$\triangleleft$}}
  \rput(.4,1.6){\scalebox{1.2}{\color{black}{$a$}}}
}

\rput(2.1,1.2){\scalebox{1.4}{\color{black}$=$}}

\rput(4.25,0){
  \psarc(0,1){.5}{0}{180}
  \rput(.2,1.2){\scalebox{1.2}{\color{black}{$a$}}}
  \psarc[linestyle=dashed](0,1){.7}{0}{180}
}

    \rput(5.4,1.2){\scalebox{1.5}{\pstThreeDCoor[linewidth=.5pt,arrowsize=3pt,linecolor=blue,Alpha=80,Beta=30,xMax=.4,yMax=.4,zMax=.4,xMin=0,yMin=0,zMin=0,nameX={},nameY={},nameZ={}]
      \rput(0,.4){\scalebox{.6}{\color{blue}$z$}}
      \rput(-.1,-.25){\scalebox{.6}{\color{blue}$x$}}
      \rput(.3,-.15){\scalebox{.6}{\color{blue}$y$}}
    }}

     \rput(3.2,1.5){\scalebox{1.5}{\pstThreeDCoor[linewidth=.5pt,arrowsize=3pt,linecolor=blue,Alpha=345,Beta=150,xMax=.4,yMax=.4,zMax=.5,xMin=0,yMin=0,zMin=0,nameX={},nameY={},nameZ={}]
       \rput(0,-.5){\scalebox{.6}{\color{blue}$z$}}
       \rput(-.15,-.25){\scalebox{.6}{\color{blue}$x$}}
       \rput(-.3,-.1){\scalebox{.6}{\color{blue}$y$}}
     }}

}

%    \rput(5.25,-.5){\scalebox{1.5}{\pstThreeDCoor[linewidth=.5pt,arrowsize=3pt,linecolor=blue,Alpha=80,Beta=30,xMax=.4,yMax=.4,zMax=.4,xMin=0,yMin=0,zMin=0,nameX={},nameY={},nameZ={}]
%      \rput(0,.4){\scalebox{.6}{\color{blue}$z$}}
%      \rput(-.1,-.25){\scalebox{.6}{\color{blue}$x$}}
%      \rput(.3,-.15){\scalebox{.6}{\color{blue}$y$}}
%    }}
%%%  left y out x down z

%     \rput(3.2,0){\scalebox{1.5}{\pstThreeDCoor[linewidth=.5pt,arrowsize=3pt,linecolor=blue,Alpha=345,Beta=150,xMax=.4,yMax=.4,zMax=.5,xMin=0,yMin=0,zMin=0,nameX={},nameY={},nameZ={}]
%       \rput(0,-.5){\scalebox{.6}{\color{blue}$z$}}
%       \rput(-.15,-.25){\scalebox{.6}{\color{blue}$x$}}
%       \rput(-.3,-.1){\scalebox{.6}{\color{blue}$y$}}
%     }}
%  right y out x up z

% third
\rput(0,-2.25){  
  \rput(1,0){
   \psarc(0,1){.5}{180}{360}
 \rput(0,-.975){ \rput(0,1.5){\scalebox{1.2}{\color{white}$\blacktriangleright$}}
  \rput(0,1.46){\scalebox{2}{\color{black}$\triangleright$}}
  \rput(.4,1.4){\scalebox{1.2}{\color{black}{$a$}}}
   }
}

\rput(2.1,.7){\scalebox{1.4}{\color{black}$=$}}

\rput(4.25,0){
  \psarc(0,1){.5}{180}{360}
  \rput(.4,.45){\scalebox{1.2}{\color{black}{$a$}}}
  \psarc[linestyle=dashed](0,1){.3}{180}{360}
   }

    \rput(2.8,.7){\scalebox{1.5}{\pstThreeDCoor[linewidth=.5pt,arrowsize=3pt,linecolor=blue,Alpha=80,Beta=30,xMax=.4,yMax=.4,zMax=.4,xMin=0,yMin=0,zMin=0,nameX={},nameY={},nameZ={}]
      \rput(0,.4){\scalebox{.6}{\color{blue}$z$}}
      \rput(-.1,-.25){\scalebox{.6}{\color{blue}$x$}}
      \rput(.3,-.15){\scalebox{.6}{\color{blue}$y$}}
    }}

     \rput(5.6,1.0){\scalebox{1.5}{\pstThreeDCoor[linewidth=.5pt,arrowsize=3pt,linecolor=blue,Alpha=345,Beta=150,xMax=.4,yMax=.4,zMax=.5,xMin=0,yMin=0,zMin=0,nameX={},nameY={},nameZ={}]
       \rput(0,-.5){\scalebox{.6}{\color{blue}$z$}}
       \rput(-.15,-.25){\scalebox{.6}{\color{blue}$x$}}
       \rput(-.3,-.1){\scalebox{.6}{\color{blue}$y$}}
     }}
    % this points down (z) out (x) left (y)

}

% last
\rput(0,-3.75){  
\rput(1,0){
  \psarc(0,1){.5}{180}{360}
  \rput(0,-.975){
  \rput(0,1.5){\scalebox{1.2}{\color{white}$\blacktriangleleft$}}
  \rput(0,1.46){\scalebox{2}{\color{black}$\triangleleft$}}
  \rput(.4,1.4){\scalebox{1.2}{\color{black}{$a$}}}
}
}

\rput(2.1,.7){\scalebox{1.4}{\color{black}$=$}}

\rput(4.25,0){
  \psarc(0,1){.5}{180}{360}
  \rput(.2,.7){\scalebox{1.2}{\color{black}{$a$}}}
  \psarc[linestyle=dashed](0,1){.7}{180}{360}
}

    \rput(5.4,0.7){\scalebox{1.5}{\pstThreeDCoor[linewidth=.5pt,arrowsize=3pt,linecolor=blue,Alpha=80,Beta=30,xMax=.4,yMax=.4,zMax=.4,xMin=0,yMin=0,zMin=0,nameX={},nameY={},nameZ={}]
      \rput(0,.4){\scalebox{.6}{\color{blue}$z$}}
      \rput(-.1,-.25){\scalebox{.6}{\color{blue}$x$}}
      \rput(.3,-.15){\scalebox{.6}{\color{blue}$y$}}
    }}

     \rput(3.2,1.0){\scalebox{1.5}{\pstThreeDCoor[linewidth=.5pt,arrowsize=3pt,linecolor=blue,Alpha=345,Beta=150,xMax=.4,yMax=.4,zMax=.5,xMin=0,yMin=0,zMin=0,nameX={},nameY={},nameZ={}]
       \rput(0,-.5){\scalebox{.6}{\color{blue}$z$}}
       \rput(-.15,-.25){\scalebox{.6}{\color{blue}$x$}}
       \rput(-.3,-.1){\scalebox{.6}{\color{blue}$y$}}
     }}

}

\end{pspicture}
  \caption{A dictionary between flags and framing for self-dual particles.
The orientation of the local frames on the left and right of the ribbon is indicated. The $y$-axis always points from the solid edge to the dashed edge of the ribbon, the $x$-axis always points out of the page and the $z$-axis follows the solid edge in such a way that the frame is right handed.}
 \label{fig:flagframing}
\end{figure}
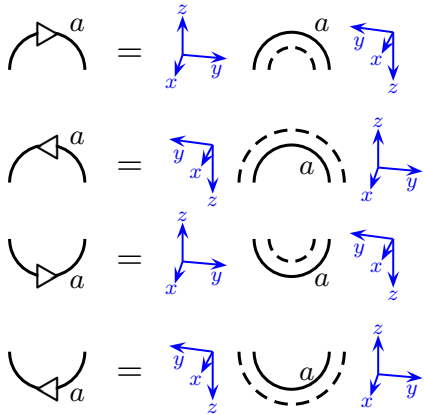

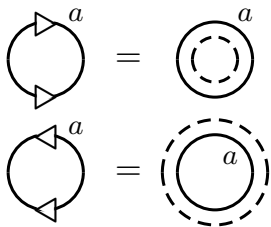
\begin{figure}
 \begin{pspicture}(1,-1.5)(4,1.7)
  \psset{linewidth=.04,linecolor=black,arrowsize=.3}

 %top 
\rput(0,0){  
\rput(1,0){
  \psarc(0,1){.5}{0}{180}
  \rput(0,1.5){\scalebox{1.2}{\color{white}$\blacktriangleright$}}
  \rput(0,1.46){\scalebox{2}{\color{black}$\triangleright$}}
  \rput(.4,1.6){\scalebox{1.2}{\color{black}{$a$}}}

   \psarc(0,1){.5}{180}{360}
 \rput(0,-.975){ \rput(0,1.5){\scalebox{1.2}{\color{white}$\blacktriangleright$}}
  \rput(0,1.46){\scalebox{2}{\color{black}$\triangleright$}}
%  \rput(.4,1.4){\scalebox{1.2}{\color{black}{$a$}}}

}
}

 \rput(2.1,1){\scalebox{1.4}{\color{black}$=$}}

  \rput(3.25,0){
    \psarc(0,1){.5}{0}{180}
    \rput(.4,1.6){\scalebox{1.2}{\color{black}{$a$}}}
    \psarc[linestyle=dashed](0,1){.3}{0}{180}
    \psarc(0,1){.5}{180}{360}
 %  \rput(.4,.45){\scalebox{1.2}{\color{black}{$a$}}}
   \psarc[linestyle=dashed](0,1){.3}{180}{360}

 }
 }

% second
 \rput(0,-1.5){  
 \rput(1,0){
   \psarc(0,1){.5}{0}{180}
   \rput(0,1.5){\scalebox{1.2}{\color{white}$\blacktriangleleft$}}
   \rput(0,1.46){\scalebox{2}{\color{black}$\triangleleft$}}
   \rput(.4,1.6){\scalebox{1.2}{\color{black}{$a$}}}
   \psarc(0,1){.5}{180}{360}
   \rput(0,-.975){
   \rput(0,1.5){\scalebox{1.2}{\color{white}$\blacktriangleleft$}}
   \rput(0,1.46){\scalebox{2}{\color{black}$\triangleleft$}}
  % \rput(.4,1.4){\scalebox{1.2}{\color{black}{$a$}}}
 }
 }

 \rput(2.1,1.0){\scalebox{1.4}{\color{black}$=$}}

 \rput(3.25,0){
   \psarc(0,1){.5}{0}{180}
   \rput(.2,1.2){\scalebox{1.2}{\color{black}{$a$}}}
   \psarc[linestyle=dashed](0,1){.7}{0}{180}
   \psarc(0,1){.5}{180}{360}
  % \rput(.2,.7){\scalebox{1.2}{\color{black}{$a$}}}
   \psarc[linestyle=dashed](0,1){.7}{180}{360}
 
 }
 }

\end{pspicture}
  \caption{Taking the inner product of a ket and its Hermitian
    conjugate bra results in a properly framed diagram.}
  \label{fig:framedloops}
\end{figure}

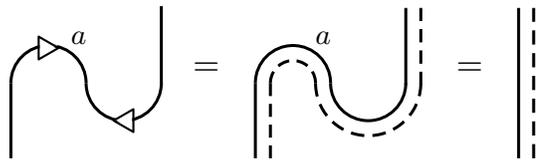
\begin{figure}
  \begin{pspicture}(3.5,0)(4,2)
  \psset{linewidth=.04,linecolor=black,arrowsize=.3}

 %top 
\rput(0,0){  
\rput(1,0){
  \psarc(0,1){.5}{0}{180}
  \rput(0,1.5){\scalebox{1.2}{\color{white}$\blacktriangleright$}}
  \rput(0,1.46){\scalebox{2}{\color{black}$\triangleright$}}
  \rput(.4,1.6){\scalebox{1.2}{\color{black}{$a$}}}

  \psline(-.5,1)(-.5,0)
  
}

\rput(3.1,1.2){\scalebox{1.4}{\color{black}$=$}}

\rput(6.6,1.2){\scalebox{1.4}{\color{black}$=$}}

\rput(4.25,0){
  \psarc(0,1){.5}{0}{180}
  \rput(.4,1.6){\scalebox{1.2}{\color{black}{$a$}}}
  \psarc[linestyle=dashed](0,1){.3}{0}{180}

  \psline(-.5,1)(-.5,0)
  \psline[linestyle=dashed](-.3,1)(-.3,0)

   }

% last
\rput(1,0){  
\rput(1,0){
  \psarc(0,1){.5}{180}{360}
  \rput(0,-.975){
  \rput(0,1.5){\scalebox{1.2}{\color{white}$\blacktriangleleft$}}
  \rput(0,1.46){\scalebox{2}{\color{black}$\triangleleft$}}
%  \rput(.4,1.4){\scalebox{1.2}{\color{black}{$a$}}}

    \psline(.5,1.95)(.5,3)
  
}
}

%\rput(3.1,.5){\scalebox{1.4}{\color{black}$=$}}

\rput(4.25,0){
  \psarc(0,1){.5}{180}{360}
%  \rput(.2,.7){\scalebox{1.2}{\color{black}{$a$}}}
  \psarc[linestyle=dashed](0,1){.7}{180}{360}

  \psline(.5,.975)(.5,2)
  \psline[linestyle=dashed](.7,.975)(.7,2)
  
   }
}

\psline(7.25,0)(7.25,2)
\psline[linestyle=dashed](7.45,0)(7.45,2)

}

% \rput(0,-2){

% % second
% \rput(0,-1.5){  
% \rput(1,0){
%   \psarc(0,1){.5}{0}{180}
%   \rput(0,1.5){\scalebox{1.2}{\color{white}$\blacktriangleright$}}
%   \rput(0,1.46){\scalebox{2}{\color{black}$\triangleright$}}
%   \rput(.4,1.6){\scalebox{1.2}{\color{black}{$a$}}}

% }

% \rput(3.1,1.2){\scalebox{1.4}{\color{black}$=$}}

% \rput(4.25,0){
%   \psarc(0,1){.5}{0}{180}
%   \rput(.2,1.2){\scalebox{1.2}{\color{black}{$a$}}}
%   \psarc[linestyle=dashed](0,1){.7}{0}{180}
%    }
% }

% % third
% \rput(0,-2.25){  
%   \rput(1,0){
%    \psarc(0,1){.5}{180}{360}
%  \rput(0,-.975){ \rput(0,1.5){\scalebox{1.2}{\color{white}$\blacktriangleleft$}}
%   \rput(0,1.46){\scalebox{2}{\color{black}$\triangleleft$}}
%   \rput(.4,1.4){\scalebox{1.2}{\color{black}{$a$}}}
%    }
% }

% \rput(3.1,.7){\scalebox{1.4}{\color{black}$=$}}

% \rput(4.25,0){
%   \psarc(0,1){.5}{180}{360}
%   \rput(.4,.45){\scalebox{1.2}{\color{black}{$a$}}}
%   \psarc[linestyle=dashed](0,1){.3}{180}{360}
%    }

% }

% }

\end{pspicture}
  \caption{If flags orientations alternate as you walk along a string,
    then the corresponding framing is consistent with the ribbon
    extended on one side of the string.  This allows the string to be
    smoothly deformed, thus giving the zig-zag identity of
    Fig.~\ref{fig:flags3}.}
  \label{fig:flags3a}
\end{figure}

{{\color{blue}} However, we now see that we have an ambiguity if we
  try to establish a similar convention for self-dual particles.
  Since we do not have arrows along lines, we need to invoke flags in
  order to clarify how a line is meant to be framed.  The flags are
  precisely what is needed to define a consistent framing convention.
  Let us use the dictionary shown in Fig.~\ref{fig:flagframing} which
  assigns a right-handed frame with the ribbon on the inside of the
  curved line to a right pointing cap/cup and assigns a right-handed
  frame with ribbon on the outside of the curved line to a left
  pointing cap/cup.  Note this means that the $z$-axis follows
the direction of the flag for caps and opposes the direction of the
flag for cups.  Other consistent conventions are also possible.}
Such a dictionary has a number of appealing features.  First, inner
products between a cup/cap and its Hermitian conjugate results in a
properly framed positive definite diagram, as shown in
Fig.~\ref{fig:framedloops}.  In addition the ribbons connect together
so as to reproduce the zig-zag identity (Fig.~\ref{fig:flags3a}).
I.e., properly framed zig-zags can be pulled straight --- meaning that
properly framed zig-zags correspond to alternating flag directions.

Now that we understand that flags are essentially a stand-in for
framing of the strands, this sheds light on the alternating flag-assignment rule in ``Convention 2'' above. The demand that in any diagram representing an amplitude in Chern-Simons theory all curves must posses a consistent framing is a global requirement on the whole curves. This is why the flag assignment in ``Convention 2'' appears as a non-local procedure.
Nevertheless, so long as we adhere to the rule of alternating flags, we will obtain
a consistent framing of the curve, and Chern-Simons theory will give
us an isotopy invariant result. 

One should realize that Wilson lines in Chern-Simons theory {\it must}
be assigned a framing corresponding to alternating flags.  While one
can twist framings as in Fig.~\ref{fig:framing1}, one cannot flip a
flag as in Fig.~\ref{fig:flags2}.  Thus the most general TQFT
structure which allows arbitrary flag labels on diagrams is actually
beyond what can be achieved with a Chern-Simons theory.

It is worth mentioning that the machinery of specifying a framing and
putting flags on lines is not fundamentally special to self-dual
particles.  However, for non-self dual particles it is easiest to
simply work with conventions that allow us to never think about flags.

Having now understood the meaning of the Frobenius-Schur indicator, it
is useful to work through some examples.  Further, we will introduce a
third convention for evaluating diagrams which will make our
``Convention 1'' at least appear a bit more isotopy invariant.

\section{Simplest Nontrivial Frobenius-Schur Anyon: $SU(2)_1$, the Semion Model}
\label{sec:su21}

The simplest example of a theory with a nontrivial Frobenius-Schur
indicator is $SU(2)_1$, the semion model. This model has the identity
particle and a single nontrivial particle which we will call $a$ which
is self-dual with $\kappa_a = -1$.  Diagrams are simple unlabeled linked loop diagrams (loops with no branching, with over and undercrossings
allowed) with the lines representing the world lines of particle $a$.
It is perhaps underappreciated that even this extremely simple anyon
theory has a nontrivial Frobenius-Schur indicator.

We will focus on explaining the physics of this simple case.  Once this
much is understood, the more general case follows fairly
straightforwardly.

\subsection{Isotopy Invariant Approach: Convention 2}
\label{sub:SU2Con2}

If we want to construct an isotopy invariant (and unitary) theory, we
must follows the rules of alternating flag directions as discussed in
``Convention 2'' above.  For $SU(2)_1$ evaluating any {\it planar}
diagram using this convention simply gives unity as an output.  For
diagrams with over and undercrossings, one can use two rules to turn
diagrams into planar diagrams which then have the value unity.

\vspace*{5pt}

\myitem{Rule A1\,}{ Turning an overcrossing into an undercrossing (or vice-versa) multiplies a diagram by $-1$.}

% \hspace*{.3cm} \begin{minipage}{3in}
%   \begin{itemize}
% \item[Rule A1] Turning an overcrossing into an undercrossing (or vice-versa) multiplies a diagram by $-1$. 

\myitem{Rule A2\,}{ One can use the twist factor $\theta_a^*=-i$ to remove
   twists as in Fig.~\ref{fig:thetatwist} or as in the left equality of
   Fig.~\ref{fig:Rtwist} (Removing the mirror image twist gets the
   complex conjugate phase).}

\myitem{Rule A3\,}{ Any loop (unlinked to any other loop, and without self-twists) has value +1 independent of whether it has
   zig-zags.}

% \end{itemize}
% \end{minipage}

% \vspace*{10pt}

 This set of rules is now isotopy invariant.  Note crucially in Rule
 A2, that if the twist is oriented as in the left of
 Fig.~\ref{fig:Rtwist} one now does not flip the direction of the flag
 after untwisting (one does not implement the last equality of
 Fig.~\ref{fig:Rtwist}).  The reason for this is our rule that flag
 directions should alternate as we walk along a strand.  Once a planar
 diagram is achieved, the value of that planar diagram is always
 unity.

\subsection{Non-Isotopy Invariant: Convention 1}
\label{sub:nonisocon1}

One may also choose to interpret a diagram using ``Convention 1'',
which assumes all flags are right-pointing.  In this convention, one
must keep track of zig-zags.  Each time a zig-zag is straightened, one
incurs a factor of $\kappa_a = -1$.  A loop with no zig-zags has
value of 1.  For planar diagrams the value of the diagram is given by
\begin{equation}
  \mbox{Diagram} = (-1)^{\mbox{number of zig-zags to be straightened}}
  \label{eq:Diazz}
\end{equation}
where here we count the number of zig-zags that must be straightened
in order to obtain a set of loops in the plane each with no zig-zags.

For non-planar diagrams, in addition to this zig-zag rule, Rule A1
still applies.  However, instead of using Rule A2 to remove twists, we now must
consider how the twist is oriented on the page before deciding whether
we need to reverse a flag after removing the twist.  Thus our
evaluation rules for non-planar diagrams are

\vspace*{5pt} 
  \myitem{Rule B1
%\phantom{.1}
\,}{ Turning an overcrossing into an undercrossing (or vice-versa) multiplies a diagram by $-1$.}

  \myitem{Rule B2.1\,}{ One can use the twist factor $\theta_a^*=-i$ to remove twists as in
Fig.~\ref{fig:thetatwist}   (removing the mirror image twist gets the complex
conjugate phase).}

\myitem{Rule B2.2\,}{ One can use the twist factor
  $\kappa_a \theta_a^*= +i$ to remove twists as in
  Fig.~\ref{fig:Rtwist} (removing the mirror image twist gets the
  complex conjugate phase).}

\myitem{Rule B3
%\phantom{.1}
\,}{ A simple loop (unlinked to other loops and with no zig-zags and no self-twists) has value $+1$. Removing a zig-zag multiplies the diagram by $-1$.}

The two cases of Rule B2 make it crucial here to keep track of whether
a twist is oriented vertically or horizontally, as the phase for
removing the twist differs in the two cases (Fig.~\ref{fig:thetatwist}
versus Fig.~\ref{fig:Rtwist}).  These rules are not isotopy invariant.

\subsection{Convention 3:  Cap Counting as an  Alternative to Convention 1}
\label{sub:capcounting}

We now propose a different bookkeeping scheme which ends up equivalent
to Convention 1, but is simpler in some respects.  Here, we propose to
move the minus sign onto the loop weight $d_a$, the diagrammatic value of a
loop.  However, to make up for this, and so that the value of a single
loop is still positive, we evaluate a diagram and at the end, we
multiply by $(-1)$ to the power of the number of ``caps" in the
original diagram.  A cap in this case, as shown in Fig.~\ref{fig:acap}, is a
place where a particle annihilates with another particle to form the
vacuum (or equivalently it ``turns over" in time).  For example, a
simple loop acquires $d=-1$ for being a loop, but since it has a cap,
it gets an additional factor of $-1$.  If we add a zig-zag to a line, as
in the left of Fig.~\ref{fig:fig2}, then we count another cap, and
hence we accumulate another factor of $-1$, thus accounting for the
Frobenius-Schur indicator.

\begin{figure}[h!]
  \begin{center}
    \vspace*{10pt}
\scalebox{.2}{\input{Diagram12.tex}}
	\end{center}
%	\vspace*{-pt}
	\caption{A cap is when a particle going upwards in time annihilates with another particle to form the vacuum.}
	\label{fig:acap}
\end{figure}

% \begin{figure}[h!]
% \hspace*{-2.25in}\scalebox{.2}{\input{Diagram13.tex}}
% 	\caption{Evaluation of a single loop in the semion model give $+1 = (-|d|) \times (-1)^{\mbox{number of caps}}$ since the circle has a single cap and $d=1$.}
% 	\label{fig:acircle}
% \end{figure}

The general evaluation of a planar diagram is given by
\begin{equation}
  {\rm Diagram} = (-1)^{\mbox{number of loops + number of caps}}
  \label{eq:loopspluscaps}
\end{equation}
which can be easily seen to be equivalent to Eq.~\ref{eq:Diazz}.

In evaluating a non-planar diagram, one must be cautious because
removal of a twist as in Fig.~\ref{fig:thetatwist} now also removes a
cap. Thus the net twist factor incurred is $\tilde \theta_a^* = -\theta_a^*$.  Note
that this factor now matches the factor for removing a twist in
Fig.~\ref{fig:Rtwist}, where the twist is simply oriented in a
different direction on the page.    So our rules for evaluating diagrams here are

\myitem{Rule C0\,}{ Before evaluating a diagram count the number of caps, and call it $n$.} 
    
\myitem{Rule C1\,}{ Turning an overcrossing into an undercrossing (or vice-versa) multiplies a diagram by $-1$.}
    
\myitem{Rule C2\,}{ One can use the twist factor $\tilde \theta_a^*=i$ to remove
  twists as in Fig.~\ref{fig:thetatwist} or as in the left equality of
  Fig.~\ref{fig:Rtwist} (Removing the mirror image twist gets the
  complex conjugate phase $\tilde \theta_a=-i$).}

\myitem{Rule C3\,}{ Any loop (unlinked to other loops and with no self-twists) has value -1 independent of whether it has
  zig-zags.}

\myitem{Rule C4\,}{ At the end of the evaluation, multiply the final result by $(-1)^n$.}

The diagrammatic rules C1-C3 are (regular) isotopy invariant.  We call
these steps, the {\it nonunitary evaluation} of the diagram, since
they correspond to an (auxiliary) nonunitary theory with negative loop
weight ($d_a<0$) which has a non-positive definite norm.  (Note that
if one is not interested in building physical models, that need to be
unitary, one can certainly consider such non-unitary diagram rules for
building knot invariants, for example).

While the nonunitary evaluation of the diagram is isotopy invariant,
the application of rule Rule C0,C4 breaks this isotopy invariance, and
in fact gives the same result as the evaluation described in section
\ref{sub:nonisocon1} (Convention 1).  The advantage of this scheme is that for the
main part of the evaluation (Rules C1,C2,C3) one works with isotopy
invariant rules, and only at the beginning and the end does one break
this invariance.

Note that for any planar diagram, one only applies Rules C0,C3,C4 and
one recovers the result stated in Eq.~\ref{eq:loopspluscaps}.

This cap counting scheme has a significant advantage when the diagram
algebra is used to build a string net
model\cite{LevinWen,Freedman2004} which we discuss next.

\section{String Nets}

\label{sec:stringnets}            

\subsection{String Nets in Brief}

String nets\cite{LevinWen,LevinLin,Burnell} are a general construction
which take as an input a planar diagram algebra (a spherical
category).  From this algebra one can build an explicit local Hamiltonian for a 2+1 dimensional system whose ground states are described by a 2+1 dimensional TQFT,
and which correspondingly has anyon excitations.  The TQFT which results is
known as the quantum double (or Drinfel'd double) of the input
spherical category (the input planar diagram algebra).  We will give a
very brief description of string nets here referring to the
literature\cite{LevinWen} for further details.

To build a string net, one usually starts with a 2D honeycomb lattice,
though any type of planar trivalent lattice will do.  One assigns a
particle type of the planar diagram algebra to each directed edge of the
lattice. These particle types then meet at the vertices and low energy
configurations are required to satisfy the fusion rules there.  Each
such assignment is viewed as a fusion diagram in the sense of the
diagram algebra.  The string-net ground state wavefunction is given by
\begin{equation}
  \label{eq:stringnet}
 |\psi\rangle = \sum_{\mbox{\small all planar diagrams}} \!\!\!\!\!\!\!\! W({\rm diagram}) \,\, |\mbox{diagram}\rangle
\end{equation}
where $W({\rm diagram})$ is the amplitude that results from the
evaluation of the planar diagram. (We have ignored here possible ground state degeneracies, which can occur if the lattice models a surface of nontrivial topology rather than a plane, e.g. a torus.)

Interestingly, if the input planar algebra comes from a modular tensor
category (a sufficiently well behaved anyon theory) then the output
TQFT is just the product of the input anyon theory and its mirror
image.  Note that in the construction of the string net model, we
never actually input any of the braiding properties associated with
the planar algebra --- the output braiding properties are emergent.
Somehow in the system ``knows'' when the planar diagram algebra stems
from a 2+1 dimensional theory and the emergent theory correctly
reflects this.

One can also use input planar algebras that are not consistent
with any braiding, and still the string net construction will produce
a well defined 2+1 dimensional TQFT with nontrivial anyons as an
output --- although in such a case the output TQFT is very
nontrivially related to the input.

If the input theory includes a negative Frobenius-Schur indicator,
then the amplitudes $W({\rm diagram})$ in Eq.~\ref{eq:stringnet}
should be evaluated using ``Convention 1''.  This requires us to set
some convention as to what we mean as ``up" on the two dimensional
plane.  ``Up'' has the physical meaning of time in a 2+1 dimensional
theory, but here ``up'' is just a reference direction in the plane.
Note that the use of Convention 1 allows for the desired amplitudes to
be generated by a local and Hermitian Hamiltonian, and the resulting
string net correctly generates a TQFT which is the quantum double of
the input spherical category.  One may be tempted to define a model by
the requirement that the ground states look similar to
Eq.~\ref{eq:stringnet} but with the amplitudes evaluated using
Convention 2, but this does not correctly generate the quantum double.  Since
switching between the two conventions is a non-local procedure, the
resulting topological orders will generally be different.

% The Hamiltonian that gives the wavefunction Eq.~\ref{eq:stringnet} as the
% ground state is built from a set of commuting projectors:
% $$
% H = -\sum_{\mbox{\small vertices } \alpha} V_\alpha -
% \sum_{\mbox{\small plaquettes } \beta} P_\beta
% $$
% where $V_\alpha$ is a projector that acts at vertex $\alpha$ and
% $P_\beta$ is a projector that acts on plaquette $\beta$.  The ground
% state is the simultaneous $+1$ eigenstate of all of these projectors.
% Quasiparticle excitations involve projectors in the $0$ eigenstate.

% Topological excitations (or quasiparticles) can be created and moved
% by so-called ``string-operators.''  These are operators that act along
% a one-dimensional path of edges (and also depend on the state of the
% edges immediately adjacent to this path).  These operators act such
% that plaquette and/or vertex excitations are created at the ends of the
% string, but are not created along the length
% of the string.  Acting on the ground
% state with a string opertor creates a particle type at one end of the
% string operator, and the corresponding antiparticle type at the
% opposite end.  

\subsection{String Nets in the Semion Model}
\label{sub:stringnetsemion}

We now apply the string-net construction to the above discussed
$SU(2)_1$ rules (Convention 1).  Using the evaluation of diagrams from
Eq.~\ref{eq:Diazz}, we can write the string net wavefunction as
\begin{equation} \label{eq:semionstringnet1}
|\psi\rangle = \sum_{\mbox{\small all planar diagrams}}
\!\!\!\!\!\!\!\! (-1)^{\mbox{number of zig-zags}} \,\, |\mbox{diagram}\rangle
\end{equation}
where a diagram here is any planar diagrams of loops (i.e., a loop
gas) now on a honeycomb lattice.  Here we must know what direction is
``up'' in order to know what constitutes a zig-zag.  The topological
content of this string net is (i.e., the resulting output TQFT is)
$SU(2)_1 \times {\overline{SU(2)_1}}$ with the overline meaning the
mirror image theory.

Now we can try to instead use our cap counting scheme (Convention 3)
for constructing a string net.  Here we would equivalently write the
signs as in Eq.~\ref{eq:loopspluscaps}.  However, now we are not
describing space-time world lines, but rather a component of a
wavefunction.  Thus, we can assign the $-1$ factors on every cap as
being simply a gauge transformation on wavefunctions.  To be more
explicit, on a honeycomb lattice, we can choose a local basis for the
Hilbert space of the direct neighborhood of each vertex in the spatial
lattice which has two links pointing down and one link pointing
up. (The links incident on these vertices also cover the neighborhoods
of the remaining vertices).  Such a basis would be labeled by (fusion
respecting) particle type labelings around the vertex. Basis states
for the Hilbert space of the entire lattice can be built as tensor
products of these vertex states.  We can then change the original
choice of basis by multiplying any vertex basis vector representing a
cap by $-1$.  If we make such a transform on our in-plane
wavefunctions
$$
 |\psi\rangle \rightarrow (-1)^{\mbox{number of of caps}} |\psi\rangle
$$
where $\psi$ is a particular in-plane loop configuration, we would
completely remove the need to keep track of these factors in our
bookkeeping.  The remaining effect of the Frobenius-Schur indicator is
that the value of a loop is $d = -1$.  In particular, one obtains a
fully isotropic (and fully isotopic) quantum loop gas on the honeycomb
lattice where the ground state wavefunction is simply of the form
\begin{equation}
  \label{eq:semionstringnet2}
 |\psi\rangle = \sum_{\mbox{\small all planar diagrams}} \!\!\!\!\!\!\!\! (-1)^{\mbox{number of loops}}  \,  |{\mbox{diagram}}\rangle
\end{equation}
The result of this construction is precisely the ``double semion"
wavefunction as it is described in
Refs.~\onlinecite{LevinWen,Freedman2004,CurtSemion}.  An explicit
Hamiltonian is written down by Ref.~\onlinecite{LevinWen} that
generates this wavefunction as a ground state.  

The gauge transformation on the wave functions that we have performed here has an interesting relation to the  gauge transformations $u^{ab}_{c}$ in the tensor category that  we discussed before Eq.~\ref{eq:gaugechange}. If the theory is unitary, then the phase factors will need to satisfy $u_{ab}^{c}=(u^{ab}_{c})^{*}$, so that the inner product is preserved and equations like Eq.~\ref{fig:bubblecollapse}
and Eq.~\ref{fig:completeness} remain valid with the same, positive, coefficients. However, if unitarity is not required then $u_{ab}^{c}$ and $u^{ab}_{c}$ could be chosen independently and this would allow for instance $u_{a\bar{a}}^{1}=-1$ and $u^{a\bar{a}}_{1}=1$, which would change inner products and in particular the sign of the $a$-loop amplitude. We see then that the unitary gauge transform that we perform on the lattice achieves the same effect on lattice wave functions as such a non-unitary gauge at the level of the tensor category.

The interesting consequence here is that, although our string net is based on a unitary theory $SU(2)_1$
(so it gives a nice unitary theory as its output),  it can be presented
in a different gauge so that it appears to stem from a nonunitary
theory (with $d=-1$).  Nonetheless, the two string net wavefunctions
Eqs.~\ref{eq:semionstringnet1} and \ref{eq:semionstringnet2} are
actually just unitary gauge transforms of each other!

As an aside we mention that, as noted, the tensor categories we deal
with are all so called spherical categories and as such they are
furnished with a spherical structure, which helps to define quantum
traces (see
e.g. Refs.~\onlinecite{Barrett1999spherical,Etingof2016tensor} for
details). In particular, if a category allows multiple spherical
structures then changing the spherical structure can change the loop
values (they are the traces of the trivial diagrams with a single
line).  The trick we use here appears to be related to spherical
structures but we must remember that our trick is for diagram
evaluation with the interpretation of Convention 1 and spherical
structures are natural in the context of Convention 2. In particular,
though different choices of spherical structure will give rise to
different loop values in Convention 2, all choices should give rise to
planar isotopy invariance of the diagrammatic calculus.

Note that, were we to build a semion string-net based on the
Convention 2 rules listed in section \ref{sub:SU2Con2}, we would
obtain an equal weighting of all loop diagrams.  This is
the Toric Code ground state, rather than
$SU(2)_1 \times \overline{SU(2)_1}$.

\subsection{String Nets in $SU(2)_2$}
\label{sec:SU22}

As in the case of the doubled-semion model, the cap counting technique
is useful in more general string-net models.  As an example (and a
preview of upcoming sections) let us consider a string net built from
the commonly-discussed theory $SU(2)_2$.  This theory also has a
particle type with negative Frobenius-Schur indicator, whereas its
close relative, the Ising theory does not.  $SU(2)_2$ has two
nontrivial particles, which we will call $\psi$ and $\sigma$ (another
common notation is $1$ and $1/2$, in analogy with $SU(2)$ spins).  The
nontrivial fusion rules are
\begin{eqnarray*}
 \psi \times \psi  &=& I \\
 \psi \times \sigma &=& \sigma \\
 \sigma \times \sigma &=& I + \psi 
\end{eqnarray*}
with $I$ being the identity particle.  The Frobenius-Schur indicators
of $\sigma$ is $\kappa_\sigma=-1$ whereas $\kappa_\psi=+1$.  The
fusion rules of the Ising theory are exactly the same as those of
$SU(2)_2$ although all particles in the Ising theory have +1
Frobenius-Schur indicator.

% \begin{center}
% \begin{tabular}{c|r}
%   particle &  $\kappa$ \\
%   \hline
%  \rule{0pt}{12pt} $I$ &  1  \\
%   $\sigma$ & $-1$ \\
%   $\psi $ & 1
% \end{tabular}
% \end{center}

\begin{figure}
\begin{center}
\scalebox{.5}{%LaTeX with PSTricks extensions
%%Creator: inkscape 0.91
%%Please note this file requires PSTricks extensions
\psset{xunit=.5pt,yunit=.5pt,runit=.5pt}
\newcommand{\mylinewidth}{3}
\begin{pspicture}(744.09448819,1052.36220472)
{
\newrgbcolor{curcolor}{0 0 0}
\pscustom[linewidth=\mylinewidth,linecolor=curcolor]
{
\newpath
\moveto(140,627.14286472)
\curveto(95.714284,738.57142472)(80.000003,969.99999972)(248.57144,888.57142472)
\curveto(575.48692,731.79229472)(141.23129,488.23277472)(298.57143,731.42855472)
\moveto(314.28571,745.71428472)
\curveto(341.46374,769.05264472)(319.65968,796.00580472)(332.85714,821.42857472)
}
}
{
\newrgbcolor{curcolor}{0 0 0}
\pscustom[linewidth=\mylinewidth,linecolor=curcolor]
{
\newpath
\moveto(341.42857,837.14285472)
\curveto(392.85714,952.85714372)(435.71428,937.14285472)(500,798.57142472)
\curveto(564.28571,659.99999472)(394.31625,323.69278472)(350,421.42857472)
\curveto(319.91291,487.78314472)(302.92466,462.59571472)(295.71429,358.57142472)
\curveto(279.52802,125.05133472)(171.42858,538.09523472)(150.00001,607.14285472)
}
}
{
\newrgbcolor{curcolor}{0 0 0}
\pscustom[linewidth=\mylinewidth,linecolor=curcolor]
{
\newpath
\moveto(114.28571,672.85714472)
\curveto(-21.42857,544.28571472)(229.31315,623.43426472)(220,670.00000472)
\curveto(202.85714,755.71428472)(141.42857,688.57142472)(141.42857,688.57142472)
}
}

{
\newrgbcolor{curcolor}{0 0 0}
\pscustom[linewidth=\mylinewidth,linecolor=curcolor]
{
\newpath
\moveto(648.57141876,779.99999159)
\curveto(648.57141876,745.28494988)(630.98260584,717.14284819)(609.28570557,717.14284819)
\curveto(587.58880529,717.14284819)(569.99999237,745.28494988)(569.99999237,779.99999159)
\curveto(569.99999237,814.7150333)(587.58880529,842.85713499)(609.28570557,842.85713499)
\curveto(630.98260584,842.85713499)(648.57141876,814.7150333)(648.57141876,779.99999159)
\closepath
}
}
{
\newrgbcolor{curcolor}{0 0 1}
\pscustom[linewidth=\mylinewidth,linestyle=dashed,linecolor=curcolor]
{
\newpath
\moveto(574.28571,811.42857472)
\curveto(644.28571,748.57142472)(608.57143,715.71428472)(608.57143,715.71428472)
}
}
{
\newrgbcolor{curcolor}{0 0 1}
\pscustom[linewidth=\mylinewidth,linestyle=dashed,linecolor=curcolor]
{
\newpath
\moveto(358.57142,801.42857472)
\curveto(447.14285,1001.42856372)(402.85714,528.57143472)(442.85714,605.71428472)
\curveto(482.85714,682.85714472)(497.14286,801.42857472)(497.14286,801.42857472)
}
}
{
\newrgbcolor{curcolor}{0 0 1}
\pscustom[linewidth=\mylinewidth,linestyle=dashed,linecolor=curcolor]
{
\newpath
\moveto(198.57143,708.57142472)
\curveto(198.57143,784.28571472)(280,872.85714472)(280,872.85714472)
}
}
{
\newrgbcolor{curcolor}{0 0 1}
\pscustom[linewidth=\mylinewidth,linecolor=curcolor,linestyle=dashed]
{
\newpath
\moveto(341.42857,664.28571472)
\curveto(254.28572,951.42856472)(170.24159,466.26473472)(284.28571,508.57142472)
\curveto(372.85714,541.42857472)(321.42857,642.85714472)(321.42857,642.85714472)
}
}
{
\newrgbcolor{curcolor}{0 0 1}
\pscustom[linewidth=\mylinewidth,linestyle=dashed,linecolor=curcolor]
{
\newpath
\moveto(368.57143,565.71428472)
\curveto(392.85714,645.71428472)(389.92015,483.94858472)(402.85714,538.57143472)
\curveto(428.57143,647.14285472)(444.28571,444.28571472)(392.85714,451.42857472)
\curveto(341.42857,458.57142472)(368.57143,565.71428472)(368.57143,565.71428472)
\closepath
}
}
\end{pspicture}}
\end{center}
\vspace*{-1in}
\caption{A space-time diagram for $SU(2)_2$.  The solid lines represent $\sigma$ and the dotted lines represent $\psi$.}
\label{fig:SU22}
\end{figure}
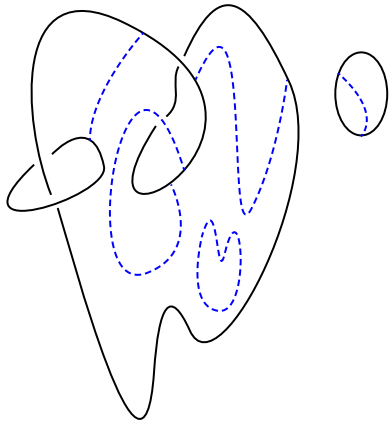

A typical space-time diagram is shown in Fig.~\ref{fig:SU22} where the
solid lines represent $\sigma$ and the dotted lines represent $\psi$.
Notice that the $\sigma$ lines, the particles with the nontrivial
Frobenius-Schur indicator, form closed loops.  Thus, we can handle the
bookkeeping exactly the same as we did for the semion theory in
section \ref{sub:capcounting} above.  We count a closed loop of
$\sigma$ as $-|d_\sigma|$ and each cap for a $\sigma$ particle
accumulates a $-1$.  

When we build a string-net model, as in the case of the semion model,
we pick out a preferred direction for ``up" within the plane.  Our
fusion algebra is given in terms of the $F$-matrices of $SU(2)_2$ in
``Convention 1''.  However, as in in section \ref{sub:stringnetsemion}
we can simplify the theory by using instead ``Convention 3'' where we
push the minus sign onto the loop weight.  Then, as in the case
of the semion model we can gauge transform to remove a factor of $-1$
for each $\sigma$ particle cap. The result is an isotropic string net model with
isotopic weights in the plane and negative loop weight for
the $\sigma$ particle.  Such a string net model has been studied in
depth in Ref.~\onlinecite{Vidal}.

The fact that we can apply the same scheme of generally moving minus
signs from the Frobenius-Schur indicators onto the loop weight
suggests how we can generally use cap counting for more general anyon
theories.

\section{Generalizing Cap Counting to Other Anyon Theories}
\label{sec:generalizedcapcounting}

It is useful to carefully extend the principle of cap-counting
bookkeeping to other categories.  This then gives us a convenient
method of bookkeeping for complicated theories with nontrivial
Frobenius-Schur indicators.

%It further gives us a prescription for
%how to treat Frobenius-Schur indicators in string-net models.

%In theories more complicated than $SU(2)_1$ we have fusion and
%splitting of particle world lines.  The key to our approach, one that
%works for {\it most} theories of interest, is that the union of a
%particular set of particle world lines (those with $\kappa=-1$, and
%perhaps some others need to be added to this set) always form closed
%loops.  Thus factors of $-1$ can again be counted by simply looking
%for caps in this set of closed loops.

\subsection{$\mathbb{Z}_2$ Frobenius-Schur Grading of the Fusion Algebra }
\label{sub:defininggrading}

In order to keep track of the particles with $\kappa=-1$, we introduce the notion of a $\mathbb{Z}_2$ Frobenius-Schur grading.  We will use this grading below in our bookkeeping schemes.
First, we explain what this grading is.  In Appendix \ref{sec:many} we explain why
such gradings usually exist, we give examples of the large families of
theories where such gradings exist and we explain some reasons why
exceptions are so rare.

We say that a theory can be given a $\mathbb{Z}_2$ Frobenius-Schur
grading if we can define indices $\tilde \kappa_a = \pm 1$ for all
particles $a$ such that $\tilde \kappa_a = \kappa_a$ for any self-dual
particle and 
\begin{equation}
  \tilde \kappa_a  \tilde \kappa_b = \tilde \kappa_c~~~~ \mbox{when} ~~~ N_{ab}^c > 0
  \label{eq:FSgrading}
\end{equation}
for any $a,b,c$ whether or not they are self-dual.  Here, $N_{ab}^c$
is the fusion multiplicity, so that $N_{ab}^c > 0$ means that $c$ is
among the possible fusion products of $a$ and $b$.

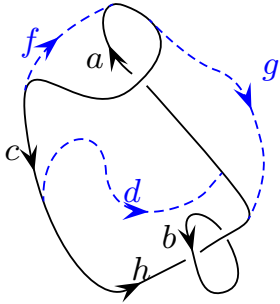
\begin{figure}
\scalebox{.7}{%LaTeX with PSTricks extensions
%%Creator: inkscape 0.91
%%Please note this file requires PSTricks extensions
\psset{xunit=.5pt,yunit=.5pt,runit=.5pt}
\begin{pspicture}(744.09448819,1052.36220472)(0,650)
{
\newrgbcolor{curcolor}{0 0 0}
\pscustom[linewidth=2,linecolor=curcolor]
{
\newpath
\psbezier[ArrowInside={->},ArrowInsidePos=.4,arrowsize=20](230,935.71429472)(190,965.71428572)(190,1019.99999972)(220,1009.99999972)
\curveto(250,999.99999972)(290,969.99999972)(230,920.00000472)
\curveto(170,870.00000472)(90,1009.99999972)(120,850.00000472)
\psbezier[ArrowInside={->},ArrowInsidePos=.1,arrowsize=20]{->}(138.24363,752.70062472)(191.97782,685.63054472)(240,710.00000472)
\psline(240,710.00000472)(287.85714,734.28571472)
}
}

\rput(190,950){\scalebox{2}{\color{black}$a$}}
\rput(100,850){\scalebox{2}{\color{black}$c$}}
\rput(240,725){\scalebox{2}{\color{black}$h$}}

{
\newrgbcolor{curcolor}{0 0 0}
\pscustom[linewidth=2,linecolor=curcolor]
{
\newpath
\moveto(303.57143,742.85714472)
\curveto(343.57143,772.85714472)(370,770.00000472)(350,800.00000472)
\curveto(330,830.00000472)(245.71429,922.85714472)(245.71429,922.85714472)
}
}
{
\newrgbcolor{curcolor}{0 0 0}
\pscustom[linewidth=2,linecolor=curcolor]
{
\newpath
\psbezier(326.42857,767.14286472)
(306.42857,791.42857472)(280.71429,790.00000472)(290.71429,760.00000472)
\psbezier[ArrowInside={->},ArrowInsidePos=.2,arrowsize=20](300.71429,730.00000472)(300.71429,690.00000472)(330.71429,700.00000472)
\curveto(360.71429,710.00000472)(332.85715,756.42857472)(332.85715,756.42857472)
}
}

\rput(270,760){\scalebox{2}{\color{black}$b$}}

{
\newrgbcolor{curcolor}{0 0 1}
\pscustom[linewidth=2,linecolor=curcolor,linestyle=dashed]
{
\newpath
\moveto(134.28571,797.85714472)
\curveto(130,870.00000472)(200,890.00000472)(201.78572,828.92857472)
\psbezier[ArrowInside={->},ArrowInsidePos=.5,arrowsize=20]
(210,760.00000472)(295.44786,791.59394472)(310,808.57143472)
\lineto(327.14286,828.57143472)
}
}

\rput(230,810){\scalebox{2}{\color{blue}$d$}}

{
\newrgbcolor{curcolor}{0 0 1}
\pscustom[linewidth=2,linecolor=curcolor,linestyle=dashed]
{
\newpath
\moveto(248.57142,994.28571172)
\curveto(248.57142,994.28571172)(290,950.00000472)(320,940.00000472)
\psbezier[ArrowInside={->},arrowsize=20,ArrowInsidePos=.6](350,930.00000472)(360,900.00000472)(370,860.00000472)
\curveto(380,820.00000472)(355.71429,778.57143472)(355.71429,778.57143472)
}
}

\rput(380,940){\scalebox{2}{\color{blue}$g$}}

{
\newrgbcolor{curcolor}{0 0 1}
\pscustom[linewidth=2,linecolor=curcolor,linestyle=dashed]
{
\newpath
\psbezier[ArrowInside={->},arrowsize=20](114.28571,917.85714472)(124.28571,977.85714272)(209.28571,1010.71428572)(209.28571,1010.71428572)
}
}

\rput(120,970){\scalebox{2}{\color{blue}$f$}}

\end{pspicture}}
\caption{A spacetime diagram for a $\mathbb{Z}_2$ graded theory.  Here
  particle types with $\tilde \kappa = -1$ are drawn black and solid.
  Particle types with $\tilde \kappa = +1$ are drawn blue dotted.  Note
  that the $\tilde \kappa =-1$ lines form closed loops. There are
  three caps of $\tilde \kappa=-1$ particles in this figure. }
\label{fig:generalz2}
\end{figure}

The point of the $\mathbb{Z}_2$ grading is that the $\tilde \kappa$
indices multiply at any vertex.  As a result, in any diagram, the
union of all the paths of particles having $\tilde \kappa = -1$ is a collection of
closed loops.  An example of a diagram in a $\mathbb{Z}_2$ graded
theory is shown in Fig.~\ref{fig:generalz2} where we can see the
closed loops explicitly.  In Convention 1, up-down zig-zags
(Fig.~\ref{fig:fig2}) in these closed loops will incur a minus sign
analogous to the case of the semion model.  Here we introduce an
analogous cap counting bookkeeping (Convention 3) which is equivalent
to Convention 1.

Our bookkeeping system will be a fairly simply generalization of the
bookkeeping for the semion model discussed above.  As in that case we
will arrange that all particles with $\tilde \kappa_a=-1$ have
$d_a < 0$ with $d_a$ the value of a loop.  Full evaluation of a
diagram will require (analogous to Rule C0) that we first count the
number of caps of particles with $\tilde \kappa_a =-1$ and call this
number $n$.  (An example of such cap counting is shown in
Fig.~\ref{fig:generalz2}).  We may then evaluate the diagram (making
our ``nonunitary evaluation'') using isotopy invariant rules (analogous
to Rules C1-C3) which allow us to straighten zig-zags and at the end we
multiply by $(-1)^n$ (analogous to rule C4) which breaks the isotopy
invariance again. As before, we can think of this procedure as simply working in a different gauge, either by applying a unitary basis transformation to a string net wave function, or by applying a non-unitary gauge transformation in our tensor category.

In counting the number of caps with negative $\tilde \kappa$, one has
to be a bit cautious, because one can run into vertices of the type
shown on the right of Fig.~\ref{fig:braandket}.  Such a vertex should
be counted as a cap if both $\tilde \kappa_b$ and $\tilde \kappa_c$
are negative.  A simple rule that makes counting easy is just to erase
all lines with positive $\tilde \kappa$ which leaves only loops, and
then we count the caps of these loops.

In this scheme we arrange that the sign of $d_a$ matches the sign of
$\tilde \kappa_a$.  Further since the $\tilde \kappa$ are
multiplicative as in Eq.~\ref{eq:FSgrading} we also have
${\rm sign}(d_a) {\rm sign}(d_b)= {\rm sign}(d_c)$ for any nonzero
$a,b,c$ vertex (such as shown in Fig.~\ref{fig:braandket}).  As a
result, the argument of the square roots in
Figs.~\ref{fig:bubblecollapse} and \ref{fig:completeness} are always
positive.  However we still have the freedom to choose the sign of the
square-root.  It is easy to establish that {\it for consistency we
  must choose an overall negative sign to the square root if and only
  if $d_a$ and $d_b$ are both negative}.  For example, in
Fig.~\ref{fig:bubblecollapse} if we consider the case of $a$ and $d$
being the vacuum, so that $b=\bar c$, the square root takes the sign
of $d_b$, so that a loop of $b$ has weight $d_b$ rather than $|d_b|$.
Another way to see that we should choose such a sign is that both the
moves shown in Figs.~\ref{fig:bubblecollapse} and
\ref{fig:completeness} change the parity of the number of caps exactly
when $d_a$ and $d_b$ are both negative.

This cap-counting technique allows us to work with diagrammatic rules
that allow straightening of zig-zags, so long as we have a
$\mathbb{Z}_2$ Frobenius-Schur grading.  As mentioned above, we
outline in Appendix \ref{sec:many} how many (but not
all) theories of interest do have such gradings.  Such gradings can be
subdivided into two types:

\begin{enumerate}

\item {\bf Simply Graded Theories}

  For a particularly simple subclass of theories, all
non-self-dual particles can be simply assigned $\kappa_a = +1$.  For
lack of a better word, we will call these theories ``simply
$\mathbb{Z}_2$ graded".

\item {\bf Non-Simply Graded Theories}

Although there are many theories that are simply graded, some are
graded, but are not simply graded.

The simplest example of a non-simply graded theory is
$SU(6)_1$. This is an abelian theory with six particle types (called
$\mathbb{Z}_6^{(3+1/2)}$ in the notation of
Ref.~\onlinecite{bondersonthesis}).  In this particular case, the
particle types can be written as $a^n$ for $n=0\ldots 5$ (with $a^0$
meaning the identity) with fusion rules
$a^n \times a^m = a^{(n + m) {\rm mod} 6}$. The particle type $a^3$ is self dual with $\kappa_3=-1$.  Hence we must take the grading
$\tilde \kappa_n = (-1)^n$.  We see that this type of theory allows a
$\mathbb{Z}_2$ grading so long as we allow some of the non-self-dual
particles ($a^1$ and $a^5$) to be assigned negative values of $\tilde{\kappa}$.

In a non-simply graded theory, we now have non-self-dual particles
which we have assigned $\tilde \kappa_a = -1$.  For our bookkeeping
scheme to work, we want the corresponding value $d_a$ of a
loop to be negative.  Fortunately, this can be arranged.  To do so,
we need only choose an appropriate gauge (See
Eq.~\ref{eq:gaugechange}).  In particular, for these particles we now
choose $[F^{a \bar a a}_a]_{00}$ negative and choose $d_a$ negative as
well so that we have isotopy invariance in Fig.~\ref{fig:Wig}.  Note
that the gauge choice can have an effect on other $F$ matrix elements
as well.

\end{enumerate}

In either case, we have a bookkeeping scheme that allows one to
straighten zig-zags freely and minus signs are re-inserted at the last
step.  In particular, it assures us that, if we are considering knots
(or links), the value of the diagram is unchanged under any isotopy of
the knot, with again minus signs being only re-inserted at the end.

% \subsection{Cap Counting for $\mathbb{Z}_2$ Graded Theories}

% \label{sub:simplyz2}

% the only particles with $\tilde \kappa = -1$ are self dual particles.
% In this case, our bookkeeping scheme is hardly more complicated than
% that for the semion model.  First, we assign $d_a$ to be negative for
% any particle with $\tilde \kappa = -1$.  Caps of particles with
% $\tilde \kappa = -1$ give a factor of $-1$ as in the semion case.
% Vertices are a bit more complicated.  Due to the $\mathbb{Z}_2$
% grading, each vertex must have an even number of particles with
% $\tilde \kappa = -1$.  The important case is for vertices of the form
% of the right of Fig.~\ref{fig:braandket}.  If the nontrivial
% $\mathbb{Z}_2$ charge forms a cap (i.e, goes in and out from the
% bottom) then we count this vertex also as a cap.  In other words, in
% the right of Fig.~\ref{fig:braandket}, we count a $-1$ factor if both
% $b$ and $c$ have $\tilde \kappa = -1$.

% \subsection{Cap Counting for Non-Simply $\mathbb{Z}_2$ Graded
%   Theories}

%  Once we have made this change, our bookkeeping
% scheme is exactly as in section \ref{sub:simplyz2} above: particles
% with $\tilde \kappa = -1$ have $d$ negative and incur a minus sign for
% each cap.  Similarly to above, for vertices caps are counted when the
% $\mathbb{Z}_2$ charge forms a cap.

\section{Isotopy of Diagrams with Vertices}
\label{sec:fullisotopy}

In studying topological theories, we would like to be able to deform
3D diagrams in any way, and still have the diagram correspond to the
same value.  I.e., we want our theories to be ``regular isotopic''
(The word ``regular'' here meaning that we must be careful not to
insert twists in strands.  I.e, the diagrams should be thought of as
being composed of ribbons rather than straight lines).  As discussed
above, negative Frobenius-Schur indicators present complications in
obtaining such isotopy invariance.

So long as we are considering knots and links, using Convention 2 from
section \ref{sec:flags} we are guaranteed to have a regular isotopic
theory --- i.e., a knot or link invariant.  If we use Convention 1
(which is often used by physicists and particularly useful for string net models) we have to worry
about Frobenius-Schur indicators, and straightening a zig-zag incurs a
minus sign.  However, so long as there is a $\mathbb{Z}_2$
Frobenius-Schur grading, our Convention 3 (cap counting) allows us to
work with isotopy invariant rules.  As described above, for particles
with $\tilde \kappa_a = -1$ we set $d_a$, the value of a loop
negative.  Before a computation we count caps having
$\tilde \kappa =-1$, and call this number $n$.  We may then apply
isotopy freely to our diagram to simplify it.  After fully evaluating
the diagram we multiply the final result by $(-1)^n$.  This
prescription is equivalent to Convention 1 where all flags are
right-pointing and we incur a minus sign for straightening each
zig-zag with $\tilde \kappa=-1$.

Despite these approaches to obtaining isotopy invariance for knots and
links, there is still an issue to be cautious of.  When we have
diagrams involving vertices --- fusions or splittings --- we are not
always guaranteed full isotopy invariance.  We have so far obtained
isotopy invariance for diagrams of knots and links, but not of graphs
with fusion.  This is somewhat curious: we have obtained isotopy
invariance for the value of a knot or link, but in order to actually
evaluate the knot or link we typically use $F$-moves to turn it into a
fusion diagram, which may then break the isotopy invariance.  Further
we certainly had to consider fusions and $F$-moves in order to even
define the Frobenius-Schur grading, just to obtain isotopy invariance
of diagrams that have no fusions or splitting.

To give an example of how fusion diagrams may lose isotopy invariance
we consider the diagrams shown in Fig.~\ref{fig:turningup}
% and Fig.~\ref{fig:slidingover}
where factors
of $F$ and $d$ are incurred in diagrammatic transformation that would
be allowed in a theory having full isotopy invariance.

\begin{figure}[h!]
	\begin{center}
\scalebox{1}{  \scalebox{.9}{\begin{pspicture}(6.5,-.3)(5,2.0)
      \psset{linewidth=.03,linecolor=black,arrowsize=.2}
   
%    \psellipticarc(1,2)(.7,1.5){180}{359}
%    \psellipticarc{->}(1,2)(.7,1.5){180}{340}
%    \psellipticarc{<-}(1,2)(.7,1.5){200}{340}
%    \psline[ArrowInside={->}](1.57,-.5)(1.57,1.2)
%    \rput[lB](1.7,0){\scalebox{1.5}{\color{blue}{$a$}}}
%    \rput[lB](.5,1.5){\scalebox{1.5}{\color{blue}{$c$}}}
%    \rput[lB](1.8,1.5){\scalebox{1.5}{\color{blue}{$b$}}}
%    \rput[lB](2,.75){\scalebox{1.5}{\color{blue}{$=$}}}
%   
    \rput[lB](2.1,0){    
   	\psellipticarc(1,2)(.7,1.5){180}{359}
   	\psellipticarc{->}(1,2)(.7,1.5){180}{340}
   	\psellipticarc{->}(1,2)(.7,1.5){180}{290}
   	\psellipticarc{<-}(1,2)(.7,1.5){200}{340}
   	\psline[ArrowInside={->}](1.57,-.5)(1.57,1.2)
%   	\psline[linestyle=dotted,linewidth=0.06](1,.5)(1.57,-.25)
%   	\rput[lB](.8,-.25){\scalebox{1.5}{\color{black}{$I$}}}
   	\rput[lB](1.7,0){\scalebox{1.5}{\color{black}{$a$}}}
   	\rput[lB](.5,1.5){\scalebox{1.5}{\color{black}{$c$}}}
   	\rput[lB](1.1,.8){\scalebox{1.5}{\color{black}{$\bar c$}}}
   	\rput[lB](1.8,1.5){\scalebox{1.5}{\color{black}{$b$}}}
%   	\rput[lB](2,.75){\scalebox{1.5}{\color{black}{$=[F^{c\bar c a}_a]_{Ib}$}}}
   		}
   	
  %  	  \rput[lB](5.5,0){    
  %  		\psline[ArrowInside={->},ArrowInsideNo=4](1.3,-.5)(2,1.8)
  %  		\psline[ArrowInside={->},ArrowInsidePos=0.7](1.5,.15)(.8,1.8)
  % \rput[lB](1.7,.8){	\rotatebox{-20}{\psellipticarc(0,0)(.4,.35){110}{250}
  % \psellipticarc{<-}(0,0)(.4,.35){170}{250}}}
   		
  %  		\rput[lB](1.5,-.4){\scalebox{1.5}{\color{blue}{$a$}}}
  %  		\rput[lB](1.9,.6){\scalebox{1.5}{\color{blue}{$a$}}}
  %  		\rput[lB](1.7,0){\scalebox{1.5}{\color{blue}{$b$}}}
  %  		\rput[lB](1,1.5){\scalebox{1.5}{\color{blue}{$c$}}}
  %  		\rput[lB](1.4,1.2){\scalebox{1.5}{\color{blue}{$\bar c$}}}
  %  		\rput[lB](2,1.4){\scalebox{1.5}{\color{blue}{$b$}}}
    		\rput[lB](4.7,.75){\scalebox{1.5}{\color{black}{$={\sqrt{\frac{d_a d_c}{d_b}}}[F^{c\bar c a}_a]_{Ib}$}}}
   
  %  	}
   	
  \rput[lB](8.0,0){ 
  	\psline[ArrowInside={->}](1,-.4)(1,1)
  	\psline[ArrowInside={->}](1,1)(0,2)
  	\psline[ArrowInside={->}](1,1)(2,2)
 \rput[lB](1.2,-.3){\scalebox{1.5}{\color{black}{$a$}}}
\rput[lB](2.1,2){\scalebox{1.5}{\color{black}{$b$}}}
\rput[lB](0.2,2){\scalebox{1.5}{\color{black}{$c$}}}
  } 
    \end{pspicture}}}

\vspace*{25pt} 

\scalebox{1}{\scalebox{.9}{\begin{pspicture}(6.5,-.3)(5,2.0)
	\psset{linewidth=.03,linecolor=black,arrowsize=.2}
	
	\rput[lB](2.1,0){    
  \rput(2,0){	\psscalebox{-1 1}{	\psellipticarc(1,2)(.7,1.5){180}{359}
		\psellipticarc{->}(1,2)(.7,1.5){180}{340}
		\psellipticarc{->}(1,2)(.7,1.5){180}{290}
		\psellipticarc{<-}(1,2)(.7,1.5){200}{340}
		\psline[ArrowInside={->}](1.57,-.5)(1.57,1.2)
%		\psline[linestyle=dotted,linewidth=0.06](1,.5)(1.57,-.25)
              }
%              \rput[lB](-1.2,-.25){\scalebox{1.5}{\color{black}{$I$}}}
              }
		\rput[lB](0,0){\scalebox{1.5}{\color{black}{$a$}}}
		\rput[lB](.5,1.5){\scalebox{1.5}{\color{black}{$b$}}}
		\rput[lB](.8,.8){\scalebox{1.5}{\color{black}{$\bar c$}}}
		\rput[lB](1.8,1.5){\scalebox{1.5}{\color{black}{$c$}}}
		\rput[lB](2.2,.75){\scalebox{1.5}{\color{black}{$={\sqrt{\frac{d_a d_c}{d_b}}}[F^{a \bar c c}_a]^*_{bI}$}}}
	}
	
%	\rput[lB](5.5,0){    
%		\psline[ArrowInside={->},ArrowInsideNo=4](1.3,-.5)(2,1.8)
%		\psline[ArrowInside={->},ArrowInsidePos=0.7](1.5,.15)(.8,1.8)
%		\rput[lB](1.7,.8){	\rotatebox{-20}{\psellipticarc(0,0)(.4,.35){110}{250}
%				\psellipticarc{<-}(0,0)(.4,.35){170}{250}}}
%		
%		\rput[lB](1.5,-.4){\scalebox{1.5}{\color{blue}{$a$}}}
%		\rput[lB](1.9,.6){\scalebox{1.5}{\color{blue}{$a$}}}
%		\rput[lB](1.7,0){\scalebox{1.5}{\color{blue}{$b$}}}
%		\rput[lB](1,1.5){\scalebox{1.5}{\color{blue}{$c$}}}
%		\rput[lB](1.4,1.2){\scalebox{1.5}{\color{blue}{$\bar c$}}}
%		\rput[lB](2,1.4){\scalebox{1.5}{\color{blue}{$b$}}}
%		\rput[lB](2.2,.75){\scalebox{1.5}{\color{black}{$={\sqrt{\frac{d_a d_c}{d_b}}}[F^{c\bar c a}_a]_{Ib}$}}}
%		
%	}
%	
	\rput[lB](8.0,0){ 
		\psline[ArrowInside={->}](1,-.4)(1,1)
		\psline[ArrowInside={->}](1,1)(0,2)
		\psline[ArrowInside={->}](1,1)(2,2)
		\rput[lB](1.2,-.3){\scalebox{1.5}{\color{black}{$a$}}}
		\rput[lB](2.1,2){\scalebox{1.5}{\color{black}{$c$}}}
		\rput[lB](0.2,2){\scalebox{1.5}{\color{black}{$b$}}}
	} 
	\end{pspicture}}}
	\end{center}
	\vspace*{-15pt}
	\caption{Turning-up and Turning-Down legs.  The factors of
          $d^{1/2}$ in these equations are due to the vertex
          renormalization factors in Eq.~\ref{eq:vertexfactor}.  For
          theories with full isotopy invariance the total factor out
          front on the right is unity. Note the case $b=1$ is just straightening out zigzags.}
	\label{fig:turningup}
\end{figure}

% \begin{figure}[h!]
% 	\begin{center}
% \vspace*{20pt}
% \hspace*{15pt}\scalebox{.15}{\input{Diagram14.tex}}
% 	\end{center}
% 	\vspace*{-15pt}
% 	\caption{For theories with full isotopy invariance, this
%           $F$-matrix element is unity. }
% 	\label{fig:slidingover}
% \end{figure}

Let us assume that we are working with Convention 3 (cap counting)
such that we can freely straighten zig-zags.  It turns out that very
often we can fix a gauge such that all factors from turning-up and
turning-down legs as in Fig.~\ref{fig:turningup} are trivial.  If we
can set these factors to unity, then we have full isotopy invariance
in the plane even with vertices.

The condition to have such full planar isotopy is a condition on a
quantity known as the {\it third Frobenius-Schur
  indicator}\cite{Ng,Ng2,bondersonthesis,Kitaev20062}$^,$\footnote{See for example, L. M. Isaacs.{\it Character Theory of Finite Groups}, Academic Press, 1976, p. 49 Lemma 4.4 for a discussion of how general $n^{th}$ Frobenius-Schur indicators of elements of a finite group can be used to determine how many $n^{th}$ roots an element of a group has.}.  This indicator
can be defined if the fusion multiplicity $N_{aa}^{\bar a}$ is
nonzero.  I.e., if $a$ and $a$ fused together has $\bar a$ as one of
the possible fusion products.  If this is the case we define the
operator $C_a$ as in Fig.~\ref{fig:rot120} to rotate a vertex by
$2\pi/3$.  This operator is an $N_{aa}^{\bar a}$ dimensional matrix
(and in particular is just a scalar if there is no fusion
multiplicity).

\begin{figure}[h]
 \begin{pspicture}(1,0)(4,2.5)
              \psset{arrowsize=6pt}
      \rput(-.5,0){ \scalebox{1}{

               \rput(1,1){\pscurve[ArrowInside=->,ArrowInsidePos=.7](0,0)(-.05,.3)(-.5,0)(0,-.8)(0,-1)}
          \rput{120}(1,1){\pscurve[ArrowInside=->,ArrowInsidePos=.7](0,0)(-.05,.3)(-.5,0)(0,-.8)(0,-1)}
          \rput{240}(1,1){\pscurve[ArrowInside=->,ArrowInsidePos=.7](0,0)(-.05,.3)(-.5,0)(0,-.8)(0,-1)}
          \pscircle[fillstyle=solid,fillcolor=black](1,1){.05}
          \psline{->}(.35,1.4)(.25,1.4)
          \psline{->}(.98,.3)(1.04,.2)
          \psline{->}(1.65,1.35)(1.68,1.4)
      \rput(.3,1.6){\scalebox{1.2}{\mbox{$a$}}}
      \rput(1.7,1.6){\scalebox{1.2}{\mbox{$a$}}}
      \rput(.8,.1){\scalebox{1.2}{\mbox{$a$}}}
      \rput(1.2,1.1){\scalebox{1.2}{\mbox{$\mu$}}}
      }  }
    \rput(3,1){\scalebox{1.5}{$=\sum_\nu [C_a]_{\mu \nu}$}}

    \rput(4,0){ \scalebox{1}{
      \pscircle[fillstyle=solid,fillcolor=black](1,1){.05}
      \psline[ArrowInside=->](1,1)(1,0)
      \psline[ArrowInside=->](1,1)(.2,2)
      \psline[ArrowInside=->](1,1)(1.8,2)
      \rput(1.2,1){\scalebox{1.2}{\mbox{$\nu$}}}
      \rput(.5,2){\scalebox{1.2}{\mbox{$a$}}}
      \rput(1.5,2){\scalebox{1.2}{\mbox{$a$}}}
      \rput(.8,.3){\scalebox{1.2}{\mbox{$a$}}}
    }  }    
\end{pspicture}
\caption{Rotating a vertex by $2\pi/3$. The indices $\mu, \nu$ are
  vertex indices which must be included if there is a fusion
  multiplicity $N_{aa}^{\bar a} > 1$.}
\label{fig:rot120}
\end{figure}

The third Frobenius-Schur indicator $\nu_3(a)$ is then defined as the trace of this matrix
\begin{equation}
 \nu_3(a) = {\rm Tr}[C_a] \label{eq:thirdfrobdef}
\end{equation}
We say that this indicator is trivial if $\nu_3(a) = N^{aa}_{\bar a}$,
i.e., if all $N^{aa}_{\bar a}$ of the eigenvalues of $C_a$ are unity.
Otherwise we say that $\nu_3(a)$ is nontrivial.  (A simple example of a
theory with a nontrivial $\nu_3(a)$ is the generating cocycle of the
group $\mathbb{Z}_3$.).

In appendix~\ref{app:turning}, we prove the following important theorem: 

\noindent \textbf{Theorem: } For a spherical tensor category with a $\mathbb{Z}_2$ Frobenius-Schur grading and trivial third Frobenius-Schur indicator, one
can always choose a gauge which realizes full planar isotopy
invariance, i.e., the prefactors in Fig.~\ref{fig:turningup} are all
unity.   In particular this means that one can always
obtain planar isotopy invariance if there is no particle such that
$N_{aa}^{\bar a} > 0$. 

Many theories we want to consider are also \emph{ribbon}, meaning that in
addition to having $F$ matrices satisfying the pentagon, we have $R$
matrices (see Fig.~\ref{fig:Rmatrix}) satisfying the hexagon equation
with these $F$'s, and we have a consistent set of twist factors
$\theta$ (See for example,
Refs.~\onlinecite{Kitaev20062,bondersonthesis}). In fact every braided
unitary theory has a unique ribbon structure (i.e., uniquely defined
consistent twists)\cite{Galindo1}.  If we have such a ribbon theory
then it is much harder to have nontrivial third Frobenius-Schur
indicators.  We study this case in detail in appendix
\ref{app:braided} with the following results: 
\begin{enumerate}
\item
In the case of a ribbon
theory we cannot have such a nontrivial $\nu_3(a)$ unless
$N_{aa}^{\bar a} > 1$ (i.e., unless there is a fusion multiplicity
such that $\bar a$ occurs $N_{aa}^{\bar a}>1$ times, in
the fusion product of $a$ with $a$).  
\item 
If we do have such a fusion
multiplicity, then we examine the $N_{aa}^{\bar a}$ dimensional matrix
matrix $[R^{aa}_{\bar a}]_{\mu \nu}$ and the $N_{aa}^{\bar a}$
dimensional matrix $[(F^{a \bar a \bar a}_{\bar a})_{I a}]_{\mu \nu}$.
If $R^{aa}_{\bar a}$ commutes with
$(F^{a \bar a \bar a}_{\bar a})_{I a}$, then the third Frobenius-Schur
indicator is trivial.  
\end{enumerate}
Obviously in the case where
$N_{aa}^{\bar a}=1$, these are scalars not matrices and therefore
commute.

An example of a modular (therefore braided) theory with nontrivial
third Frobenius-Schur indicator is given in Appendix \ref{app:modularZ3}. 

We note that there are other ``higher'' Frobenius-Schur
indicators\cite{Ng,Ng2,LakeWu} which we write as $\nu_p(a)$ with
$p > 3$.  These are defined analogous to Fig.~\ref{fig:rot120} except
that the vertex has $p$ lines all labeled $a$ coming into a single
point.  However, our diagrammatic algebra is defined only for
trivalent vertices so one can only define $\nu_p$ with $p>3$ by
resolving a $p$-valent vertex into multiple trivalent vertices.  Then
if the manipulation of the trivalent vertices is isotopy invariant, so
will be the full diagram.  Thus, $\nu_p$ for $p>3$ cannot 
present a further obstruction to obtaining isotopy invariance
for any theory which already has trivial third indicators and where
all vertices in diagrams are defined to be trivalent only (which we
generally assume).

\subsection{How Much Isotopy?}
\label{sub:howmuch}

Assuming we have a $\mathbb{Z}_2$ Frobenius-Schur grading, and we use
Convention 3, and further we do not have any nontrivial third
Frobenius-Schur indicators, then we have full isotopy for planar
diagrams --- i.e we can deform the diagrams in the plane, and turn up
and down the legs of the vertices freely.  However, for ribbon
theories (i.e, those with $R$ matrices satisfying the hexagon, and
consistent twist factors $\theta$) this does not necessarily translate
into full 3D isotopy.

Let us consider ribbon theories with a $\mathbb{Z}_2$ Frobenius-Schur
grading and with no nontrivial third Frobenius-Schur indicators.  We
use Convention 3 and consider the ``nonunitary'' evaluation of the
diagram --- i.e., the steps after the counting of caps and before the
re-introduction of minus signs at the end.  For this evaluation, when
we start with a diagram, we think of it as being made of blackboard
framed ribbons with branches of ribbons at vertices.  We paint the
front side (out of the blackboard) of the ribbon white and the
backside (which we do not see in the blackboard framing) black.  Any
deformation (isotopy) of the ribbons leaves the value of the diagram
unchanged if it again ends up with all the white sides facing
forwards~\cite{Reshetikhin1990ribbon}.  Any $F$-moves or $R$-moves
performed are defined to start with white sides facing forwards, and
end with white sides again facing forwards.  However, it is generally
not the case that the value of a diagram will be unchanged (and in
fact may not even be well defined) if the diagram is isotoped so that
any of the black sides are forward.  This appears to agree with the
discussion of Refs.~\onlinecite{Turaev,Kirillov,Reshetikhin1990ribbon}.

\begin{figure}
\hspace*{3cm} \scalebox{.5}{\begin{pspicture}(-5,-1)(10,3) 
\newcommand{\texttauhere}{2}		
\psset{linewidth=.04,linecolor=black,arrowsize=.3,xunit=.8cm,yunit=1cm}

\rput(-5.5,0){
\psline[ArrowInside={->}](-2.5,0)(2.5,0)
\psline[ArrowInside={->},ArrowInsidePos=0.6](0,3)(-2.5,0)
\psline[ArrowInside={->},ArrowInsidePos=0.6](0,3)(2.5,0)
\psline[ArrowInside={->},ArrowInsidePos=0.7](0,3)(0,1.1)
\psline[ArrowInside={->}](-2.5,0)(0,1.1)
\psline[ArrowInside={->}](2.5,0)(0,1.1)

\pscircle[fillcolor=black,fillstyle=solid](-2.5,0){.1}
\pscircle[fillcolor=black,fillstyle=solid](2.5,0){.1}
\pscircle[fillcolor=black,fillstyle=solid](0,3){.1}
\pscircle[fillcolor=black,fillstyle=solid](0,1.1){.1}

\rput[cB](-2.8,-.3){\scalebox{\texttauhere}{\color{black}{$\mu$}}}
\rput[cB](2.8,-.3){\scalebox{\texttauhere}{\color{black}{$\nu$}}}
\rput[cB](-.4,1.3){\scalebox{\texttauhere}{\color{black}{$\alpha$}}}
\rput[cB](0,3.4){\scalebox{\texttauhere}{\color{black}{$\beta$}}}

\rput[cB](-2,1.4){\scalebox{\texttauhere}{\color{black}{$f$}}}
\rput[cB](1.8,1.4){\scalebox{\texttauhere}{\color{black}{$c$}}}
\rput[cB](.5,1.7){\scalebox{\texttauhere}{\color{black}{$b$}}}
\rput[cB](-.7,.4){\scalebox{\texttauhere}{\color{black}{$a$}}}
\rput[cB](.6,.4){\scalebox{\texttauhere}{\color{black}{$d$}}}
\rput[cB](0,-.5){\scalebox{\texttauhere}{\color{black}{$e$}}}

}

\rput(-2.4,1.25){\scalebox{2.5}{$=$}}

\psset{linewidth=.04,linecolor=black,arrowsize=.3,xunit=.8cm,yunit=1cm}
\psline[ArrowInside={->}](2.5,1)(.9,1)

\pscircle[fillcolor=black,fillstyle=solid](2.5,1){.1}
\pscircle[fillcolor=black,fillstyle=solid](.9,1){.1}

\rput[cB](.4,1){\scalebox{\texttauhere}{\color{black}{$\alpha$}}}
\rput[cB](3,1){\scalebox{\texttauhere}{\color{black}{$\nu$}}}

\pscircle[fillcolor=black,fillstyle=solid](1.45,2.4){.1}
\pscircle[fillcolor=black,fillstyle=solid](1.45,-.4){.1}

\rput[cB](1.8,2.7){\scalebox{\texttauhere}{\color{black}{$\beta$}}}
\rput[cB](1.8,-.7){\scalebox{\texttauhere}{\color{black}{$\mu$}}}

\psarc[arrowscale=1]{-}(1.7,1.6){.85}{-45}{0}
\psarcn[arrowscale=1]{->}(1.7,1.6){.85}{90}{0}

\psarcn[arrowscale=1]{-}(1.7,1.6){.85}{225}{180}
\psarcn[arrowscale=1]{<-}(1.7,1.6){.85}{180}{90}

\psarc[arrowscale=1]{->}(1.7,.4){.85}{270}{0}
\psarc[arrowscale=1]{-}(1.7,.4){.85}{0}{45}

\psarcn[arrowscale=1]{->}(1.7,.4){.85}{270}{180}
\psarcn[arrowscale=1]{-}(1.7,.4){.85}{180}{135}

\psellipticarc[arrowscale=1]{->}(.6,1)(1.1,2.5){58}{180}
\psellipticarc[arrowscale=1]{-}(.6,1)(1.1,2.5){180}{-58}

\rput[cB](.3,2.1){\scalebox{\texttauhere}{\color{black}{$b$}}}
\rput[cB](.3,0){\scalebox{\texttauhere}{\color{black}{$a$}}}
\rput[cB](3.1,2.1){\scalebox{\texttauhere}{\color{black}{$c$}}}
\rput[cB](3.1,0){\scalebox{\texttauhere}{\color{black}{$e$}}}
\rput[cB](1.75,1.5){\scalebox{\texttauhere}{\color{black}{$d$}}}
\rput[cB](-1,1.5){\scalebox{\texttauhere}{\color{black}{$f$}}}

\end{pspicture}}
\caption{The Tetrahedral Diagram.  Variables are attached to vertices for the general case with fusion multiplicity.}
\label{fig:tetrahedral}
\end{figure}
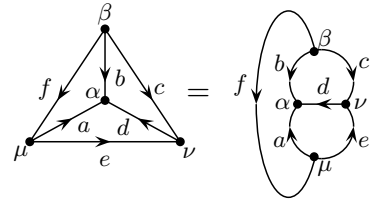

One often considers theories with a higher degree of isotopy known as
``tetrahedral symmetry''.  This is generally taken to mean that the
tetrahedral diagram in Fig.~\ref{fig:tetrahedral} is invariant under
all 24 symmetry operations of the tetrahedron --- 12 rotations and 12
inversion-rotations.  For ribbon theories with $\mathbb{Z}_2$
Frobenius-Schur grading, and with no nontrivial third
Frobenius-Schur indicators, if we use Convention 3 as described in the
previous paragraph, the tetrahedral diagram is invariant under all 12
rotations, but not necessarily under inversions.  It is worth noting
that, if we think of the tetrahedral diagram as being made out of
ribbons, one can smoothly deform the diagram into the inverted
tetrahedron, but if in the initial position if all the ribbons have
the white side facing forwards, when the tetrahedron is inverted all
of the black sides face forwards instead.  Thus invariance under
inversion is not something that our diagrammatic algebra generally
guarantees.  In appendix \ref{app:modularinv} an example is given of a
modular (therefore braided) theory where the tetrahedral diagram is
invariant under all rotations, but not inversions.

We note in passing that there have been a number of
attempts\cite{Barenz,Tingley1,Tingley2} to develop a diagrammatic
calculus which is able to more generally describe ribbon diagrams such
that one can properly give a value to half-twisting a ribbon ---
something that we cannot describe in the usual diagrammatic algebra.

Of course there certainly do exist many (ribbon) theories for which
the tetrahedral diagram can be rotated and inverted without changing
its value.  For example, any Chern-Simons theory $SU(N)_k$ without
fusion multiplicity\cite{Gu2015} has this full tetrahedral symmetry
including inversion.  %(However, for $SU(N)_k$ whenever there is fusion
%multiplicity, one can find situations where inverting the tetrahedron
%accumulates a sign).
However, in cases where there are fusion multiplicities it is possible
to find cases (indeed it may even be generic) where one can obtain
tetrahedral rotational symmetry but cannot obtain inversion symmetry
in any gauge.  For theories which enjoy full tetrahedral invariance,
including invariance under inversion, one does not need to keep track
of the front and back of ribbons, and we believe diagrams for such
theories are fully isotopy invariance in three-dimensional space.

\section{More General Transformations?}
\label{sec:moregeneral}

The cap-counting scheme described above allows us (usually) to work
with isotopy invariant diagram algebras, at the price of (at least
temporarily) working with a non-unitary theory (i.e., with negative
loop-weights).  The cap-counting signs added back in at the end fix
the result so that it is completely equivalent to a unitary theory.
In string-net models, as discussed briefly in section
\ref{sec:stringnets}, the Convention 3 ``nonunitary evaluation'' is
completely equivalent to making a well-chosen gauge transformation.
It is clear that we can do this not only for the theories discussed in 
\ref{sec:stringnets},  but in fact for in any theory with a
$\mathbb{Z}_2$ Frobenius-Schur grading.

Since we have accepted that we will be working with a non-unitary
theory, we might wonder if we can do something more general than just
pushing signs from the Frobenius-Schur indicators onto loop weights.
For the case of constructing string-nets this possibility has been
explored in detail by Refs.~\onlinecite{Burnell,LevinLin}.  There one
considers independent gauge transforms for ``bras'' and ``kets'' for
the planar diagram algebra, and these transformations can be more
general than just involving signs.  One might ask whether we can use a
similar generalized strategy for describing 2+1 dimensional (braided)
theories.  Of course, we give up unitarity to do this, and there may
not be any easy trick (like cap counting) that will recover unitarity
in the end.  We leave it as an open question as to whether there are
advantages to working with such more general non-unitary braided
theories.

% In defining a string-net model, we have taken advantage of the ability
% to separately gauge transform the two different vertex types shown in
% Fig.~\ref{fig:braandket} for our planar diagram algebra.  We have done
% this for the sake of keeping track of minus signs associated with
% Frobenius-Schur indicators.  However, more generally one can gauge
% transform the two vertex types separately, and the gauge
% transformations can be more general than just involving signs.  This
% allows one to construct some string net models that cannot be
% described if the two types of vertices have their gauges tied to each
% other.  Such a strategy is explored in detail in
% Ref.~\onlinecite{Burnell}.

\section{Conclusion}
\label{sec:conclusions}

The Frobenius-Schur indicator has been a source of a great deal of
confusion in both the physics and mathematics literature.  In this
paper we have elucidated the meaning of this quantity.  We showed that
it is intimately related to the need to frame particle world lines.
We show that different conventions for interpreting diagrams ends up
treating signs associated with the Frobenius-Schur indicator
differently, and it is this distinction that is the source of much of
the confusion.  One method of interpretation is isotopy invariant
(``Convention 2'') applicable to framed world lines in Chern-Simons
theories, whereas the other convention (``Convention 1'') is not
isotopy invariant on account of signs associated with zig-zags.  We
discuss an alternate convention (``Convention 3'') which, while
equivalent to the non-isotopy invariant conventions, allows one to count
caps of the diagram then work with an isotopy invariant theory
thereafter, adding back in the non-isotopy invariant signs at the end.
This method works so long as the theory has a $\mathbb{Z}_2$
Frobenius-Schur grading, which we argue is quite general.  We point
out that in the construction of string nets, the non-isotopy invariant
signs can be removed by gauge transformation.  

Assuming that we have a $\mathbb{Z}_2$ Frobenius-Schur grading, using
Convention 3, we are able to straighten zig-zags freely.  One can
obtain full isotopy of (fusion) diagrams in a plane, unless there is an
obstruction caused by a nontrivial third Frobenius-Schur indicator.
In the absence of this obstruction, we can freely deform diagrams in
the plane.  For ribbon theories (where nontrivial third
Frobenius-Schur indicators are very rare) we can freely deform
diagrams in three dimensions but we must treat the diagram as
being made of ribbons and generally we must keep track of which side
of the ribbon is facing forward.

Finally we would like to remark that any non-invariance of diagrams
under isotopy that occurs in the topological formalism is always due
to a change in the order of events. If we consider all fusion and
splitting vertices in a diagram (including the creations and
annihilations) to be the events of the spacetime history, then these
events naturally form a partially ordered set, with the ordering
induced by the flow of the particles from one vertex to another. Any
deformation of the history which changes this partial order requires
either bending the legs of one or more vertices from forward to
backward in time or vice versa, and/or adding new creation and
annihilation events (caps and cups). Isotopies which do not do this
always leave the amplitudes invariant. We might call these ``causal
isotopies", since partially ordered sets essentially store the causal
information in a history while forgetting, as much as possible, any
distance information. Because of this, they have inspired a number of
approaches to the study of quantum
gravity\cite{Bombelli1987space,Ambjorn1998non}, where the distance
information must be somehow added back in to reproduce the classical
limit.  In describing systems of particles with only topological
interactions, one might consider starting from the idea that
amplitudes are allowed to depend on the partial order of the events as
long as they are invariant under causal isotopy. Naively this could
lead to richer theories. However the Hamiltonian approach to
histories, essentially interpreting every timeslice as a state, seems
to naturally allow only the limited non-invariance provided by the
presence of nontrivial Frobenius-Schur indicators, at least in the planar case.  With
braiding, we gain non-invariance under inversion of the tetrahedron
and one may wonder if there is more. Starting from the action
formalism of Chern-Simons theory, one would have expected complete
invariance, since the field theory is Euclidean and does not support
any intrinsic order of events. It turns out the requirement of framing
of particle world lines, which is a global requirement, not local in
any ``timeslice", actually allows for the expected full isotopy
invariance at least for histories which are knots or links, resolving
the non-invariance which is found in the Hamiltonian approach when there
are nontrivial Frobenius-Schur indicators. With vertices, some non-invariance can actuallly
remain.  It would be interesting to know if there are any other global
requirements on spacetimes histories, beyond framing, which can be
imposed to enrich the symmetries of Hamiltonian models.

We hope that this paper goes a long way towards clarifying the physics
of the Frobenius-Schur indicators and will be useful to those studying
topological models in both the physics and mathematics communities.

\begin{acknowledgments}
  The authors are grateful to Andre Henriques and Parsa Bonderson for
  helpful conversations.  We are particularly grateful to Eddy
  Ardonne, both for giving comments on a draft of this work and also
  for helping us establish the $SU(3)_3$ example in Appendix C3. 
  Statement of compliance with EPSRC policy framework on
  research data: This publication is theoretical work that does not
  require supporting research data.  SHS has been supported by EPSRC
  Grants EP/I031014/1, EP/N01930X/1, and EP/S020527/1.  JKS has been
  supported by Science Foundation Ireland through Principal
  Investigator Awards 12/IA/1697 and 16/IA/4524
\end{acknowledgments}

\appendix

\section{The Many Theories With $\mathbb{Z}_2$ Frobenius-Schur Grading}
\label{sec:many}

The purpose of this appendix is to explore how common it is to be
able to give a theory a $\mathbb{Z}_2$ Frobenius-Schur grading.

There appear to be a great many theories that admit a $\mathbb{Z}_2$
grading (simple or non-simple in the language of section \ref{sub:defininggrading}) --- and indeed while exceptions
(non-gradeable theories) do exist, they are a bit unusual. To examine
this further, let us consider a few common types of theories of
interest which all have $\mathbb{Z}_2$ Frobenius-Schur gradings.

\begin{enumerate}[(i)]

\item {\bf Products (and Possibly Condensations, and Cosets)}

  In
  appendix \ref{sec:Cosets} we discuss how, given theories that have a
  $\mathbb{Z}_2$ Frobenius-Schur grading, additional theories also
  having such a $\mathbb{Z}_2$ Frobenius-Schur grading may be
  constructed by several procedures. These procedures include taking
  the product of two theories, taking the quotient of two theories
  (i.e., forming a coset theory), and condensing a boson from a
  theory.  It is easy to show that if two theories both admit a
  $\mathbb{Z}_2$ Frobenius-Schur grading then their product will also.
  The situation is less clear for condensations and cosets, but we
  conjecture that if we start with theories having $\mathbb{Z}_2$
  Frobenius-Schur gradings, under fairly general conditions their
  condensations and cosets will too.

\item {\bf Chern-Simons Theories}

  Consider Chern-Simons theories $G_k$ where $G$ is a Lie-group and
  $k$ is the level.  Such Chern-Simons theories are examined in detail
  in Appendix \ref{app:ChernSimons}.  All of these theories admit a
  $\mathbb{Z}_2$ Frobenius-Schur grading.  Almost all of these
  theories can obtain a $\mathbb{Z}_2$ Frobenius-Schur grading by
  assigning $\tilde \kappa_a = 1$ for any non-self-dual particles
  (i.e., they are ``simply $\mathbb{Z}_2$ graded" in the language we
  introduced in section \ref{sub:defininggrading}, item 1).  The exception are
  theories of the form $SU(6+4n)_k$ where $n\ge 0$.  These cases can
  also be given a $\mathbb{Z}_2$ grading by assigning some of the
  non-self-dual particles $\tilde \kappa_a = -1$.  A detailed
  discussion of Chern-Simons theories is given in Appendix
  \ref{app:ChernSimons}.  
 
\item {\bf All particles self-dual}

  In many theories, all particles
  are self-dual (such as, $SU(2)_k$).  As long as the so-called
  positivity conjecture holds (See appendix \ref{sec:positivity}),
  then the Frobenius-Schur indicators immediately give the theory a
  $\mathbb{Z}_2$ grading.  As mentioned in appendix
  \ref{sec:positivity}, exceptions to the positivity conjecture are
  extremely rare\cite{Mason,Stack}, and cannot occur in a braided theory
  without fusion multiplicity $N_{ab}^c > 1$.

\item {\bf Braided abelian Theories}

  It is easy to show that any braided abelian theories will
  admit a $\mathbb{Z}_2$ Frobenius-Schur grading (See Appendix
  \ref{app:abelian}). 

\item {\bf ``Small'' Discrete (twisted and untwisted) Gauge Theories}

  As detailed in section \ref{sub:exceptions} discrete gauge theories
  (twisted or untwisted) of groups of order 15 and less all admit a
  $\mathbb{Z}_2$ Frobenius-Schur grading.  \label{thisitem}

\end{enumerate}

In appendix \ref{sub:exceptions} we give some theories which we know
{\it do not} admit $\mathbb{Z}_2$ Frobenius-Schur gradings.

\subsection{Products (and Possibly Condensations, and Cosets)}
\label{sec:Cosets}

Given two theories $G$ and $H$ having a $\mathbb{Z}_2$ Frobenius-Schur
grading, it is trivial to show that the product theory $G\times H$
will also have a $\mathbb{Z}_2$ Frobenius-Schur grading with
$\tilde \kappa_{a_G \times b_H} = \tilde \kappa_{a_G} \tilde
\kappa_{b_H}$ where $a_G$ is a particle type from theory $G$, and
$b_H$ is a particle type from theory $H$.

What is more interesting is the possibility of taking
condensations\cite{Bais} of theories with $\mathbb{Z}_2$
Frobenius-Schur gradings.  Bosons can only condense if they have
trivial Frobenius-Schur indicators\cite{Eliens}.  Further we can think
of a condensed particle as being the fusion of a particle in the
uncondensed theory with the vacuum --- thus having the same
$\mathbb{Z}_2$ Frobenius-Schur index as the uncondensed particle.  It
is also possible that under condensation a particle may split into
multiple species.  Physically we can think of this as the creation of
new conserved quantities that can be assigned to a particle in the
condensed phase.  However, such splitting does not change the
$\mathbb{Z}_2$ index.  Under condensation, the fusion rules are
preserved, and this is entirely consistent with the $\mathbb{Z}_2$
index being inherited from the uncondensed theory.

What is nontrvial here is the possibility that a new self-dual
particle may emerge in the condensed theory that was not there in the
uncondensed theory.  In particular, if we condense a particle $b$, if
there is a particle $a$ in the uncondensed theory such that
$a \times a = b + \ldots$, then while $a$ is not self-dual in the
uncondensed theory, it becomes self-dual in the condensed theory and
its Frobenius-Schur indicator seems as if it could be arbitrary, and
this may not match the value of its Frobenius-Schur grading when it
was uncondensed, thus breaking the idea of the condensed theory
inheriting its grading from the uncondensed theory.

That said, if the uncondensed theory has a $\mathbb{Z}_2$
Frobenius-Schur grading and if it also has no nontrivial third
Frobenius-Schur indicators, then the uncondensed theory can be put in
a form with isotopy invariance for its diagrammatic algebra.  Since
every process in the condensed theory can be described as a process in
the uncondensed theory, along with creation and annihilation of
bosons, it then seems as if the diagrammatic algebra for the condensed
theory must also have isotopy invariance, suggesting that it also has a
$\mathbb{Z}_2$ Frobenius-Schur grading and no nontrivial third
Frobenius-Schur indicators.  However, this is a conjecture, not a proof. 

Finally we consider the possibility of cosets of two theories. This
case is simply a special case of condensations.  Here we will use the
statement that we can write a coset $G/H$ as $G \times \bar H$ (with
$\bar H$ being the mirror image theory of $H$) where in the product
theory we condense all bosons\cite{Bais} (or fully extend the chiral
algebra, in another language).  We can then invoke the discussion above regarding condensations.

\subsection{Chern-Simons Theories}

\label{app:ChernSimons}

Here we discuss simple Chern-Simons theories for compact Lie group $G$
at level $k$ and we show that all can be given a $\mathbb{Z}_2$
Frobenius-Schur grading.

First we note that since the Frobenius-Schur indicator of a self-dual
object can only take the values $\pm 1$, it should be constant as a
function of a parameter $q$ when the representation theory is
$q$-deformed even when we choose $q$ to be root of unity despite the
fact that the tensor product becomes truncated.  Thus we expect that
the (second) Frobenius-Schur indicators are the same for all levels $k$,
and are the same as the corresponding classical Lie groups as well.
It is crucial here that the truncation does not affect the channel
where $a$ and $a$ fuse to the identity, so long as the particle $a$
exists at the given level.  As a result, we can focus only on the
corresponding classical group.

We thus want to find a ${\mathbb{Z}_2}$ Frobenius-Schur grading for
the classical Lie groups
($A_l, B_l, C_l, D_l, F_4, G_2, E_6, E_7, E_8$).  A tremendously
elegant way to do this is to invoke results given in
Ref.~\onlinecite{adams2014} (See also the discussion of
Ref.~\onlinecite{Overflow}).  Given a representation $a$, let $\chi_a$
be the so-called central character of $a$ (i.e., the rep $a$
restricted to the center of the group is $\chi_a$ times the identity).
For any irrep we set
$$
\tilde \kappa_a = \chi_a(\exp(2 \pi i \rho^\vee))
$$
where $\rho^\vee$ is half the sum of the positive coroots.  Here
$\exp(2 \pi i \rho^\vee)$ is necessarily an element of order 2 in the
center of the group.  Due to the multiplicative property of characters
of Abelian groups, this means that the $\tilde \kappa$'s form a
$\mathbb{Z}_2$ grading.  Further, as shown in
Ref.~\onlinecite{adams2014}, $\tilde \kappa_a$ is in fact $\kappa_a$,
the Frobenius-Schur indicator when $a$ is self-dual.  This then
confirms the existence of a $\mathbb{Z}_2$ Frobenius-Schur grading.

For completeness we mention some properties of the $\mathbb{Z}_2$
Frobenius-Schur gradings for the classical Lie groups. 

For the Lie groups $E_8$, $F_4$, $G_2$, and $D_{n}$ (or $SO(2n)$) with
$n=4m$, and $B_n$ (or $SO(2n+1)$) with $n =4m$ or $4m +3$ all irreps
are self-dual, and all Frobenius-Schur indicators are $+1$, so these
cases have trivial $\mathbb{Z}_2$ Frobenius-Schur gradings.

For $E_6$ and $D_n$ (or $SO(2n)$) with $n$ odd, and $A_n$ ($SU(n+1)$)
with $n \neq 4m + 1$ not all irreps are self-dual, but all self-dual
irreps have Frobenius-Schur indicator +1, so again we can trivially
assign $\tilde \kappa=+1$ for all irreps.

For $E_7$, $A_1$ (or $SU(2)$), $C_n$ (or $Sp(2n)$), $B_n$ (or $SO(2n+1)$)
with $n=4m+1$ or $4m+2$, and $D_n$ (or $SO(2n)$) with $n=4m+2$, all
irreps are self-dual and the Frobenius-Schur indicator is
multiplicative under fusion as required.

The most interesting case is $A_{n}$ or $SU(n+1)$ with $n={4m+1}$ and
$m \geq 1$.  In this case not all irreps are self-dual and for those
irreps which are self-dual not all have positive Frobenius-Schur
indicators.  For this case we must nontrivially assign
$\tilde \kappa = \pm 1$ to the non-self-dual irreps.  I.e., this case is ``non-simply graded''. 
In this case it
is well known that one can assign an index to each irrep which is
conserved under fusion modulo $(n+1)$ (this is sometimes known as
$(n+1)$-ality or the congruence class\cite{Congruence}).  Further, for
self-dual irreps this index is even for $\kappa_a = 1$ and odd for
$\kappa_a = -1$.  Assigning $\tilde \kappa_a$ to be the parity of this
index then gives the Frobenius-Schur grading.

\subsection{Positivity Conjecture}
\label{sec:positivity}

In 2003, Bantay proposed\cite{Bantay,Mason} a positivity conjecture
that states that for any category with fusion rules
$$
a \times b = \sum_{c} N_{ab}^c \,\, c
$$ 
and corresponding Frobenius-Schur indicators $\kappa_a$ for particle
of type $a$, we should have 
\begin{equation}
  \label{eq:positivity}
\kappa_a \kappa_b =  \kappa_c ~~~ \mbox{when} ~~~~ N_{ab}^c > 0 ~~~\mbox{and $a,b,c$ all self-dual}
\end{equation}
Note that this conjecture applies to cases
where $a$, $b$ and $c$ are all self-dual (in fact we do not even
define $\kappa$ for non-self dual particles here).

While it turns out that this conjecture is not actually true in all
cases, it is quite challenging to find cases where the conjecture
fails, the first one being published only in 2017 by Mason\cite{Mason}
being based on a group of order 128 (Another exception based on a
group of order 64 was informally discussed on a website\cite{Stack} in
2011).  Further, there does exist a proof that the conjecture must be
true whenever $N_{ab}^c$ is odd (for any braided theory and indeed,
many other theories that do not actually require a braiding, see
Refs~\onlinecite{Fuchs,Mason}).

\subsection{Braided Abelian  Theories}
\label{app:abelian}

Here we briefly show that any braided abelian theory, can be given a
$\mathbb{Z}_2$ Frobenius-Schur grading and further cannot have a
nontrivial third Frobenius-Schur indicator.

A beautiful theorem by Galindo and Jaramillo\cite{Galindo,WangPrime} reduces any
modular abelian category to a product of so-called {\it prime} modular abelian
categories.  Of the prime categories only the right- and left-handed
semion theories have a nontrivial Frobenius-Schur indicator (!).
Since we know these two admit a $\mathbb{Z}_2$ Frobenius-Schur
grading, and taking products of theories with gradings gives a theory
that allows a grading (see section \ref{sec:Cosets}), this implies
that all modular abelian categories admit a $\mathbb{Z}_2$
Frobenius-Schur grading.

This result can also be extended to non-modular but braided abelian
categories.  To do so, we simply use a result from Ref.~\onlinecite{Cano} that
any non-modular but braided abelian theory can be written as a product of a
modular abelian theory along with some number of fermions which also
have trivial Frobenius-Schur indicators.  Thus our result applies to
all braided abelian theories.

Further, abelian theories have no fusion multiplicities, and as
mentioned in the main text (and shown in appendix \ref{app:braided})
there can be no nontrivial third Frobenius-Schur indicator without
fusion multiplicity.  This means that it is always possible to put
abelian theories into isotopy invariant form.

\section{Turning-Up/Turning-Down and Gauge Transformation}
\label{app:turning}

In this appendix we consider the transformations of
``turning-up/down'' edges from vertices (as in
Fig.~\ref{fig:turningup}).  We will assume a $\mathbb{Z}_2$
Frobenius-Schur grading.  We will assume we have handled the minus
signs from Frobenius-Schur indicators using Convention 3.  We now want
to know whether we can further choose a gauge such that the
turning-up/down transformations are trivial --- i.e., in the
diagrammatic algebra, one can turn up and down edges for free.  While
it is not always possible to choose such a gauge, we will isolate a
single possible ``obstruction'' which is the
third Frobenius-Schur
  indicator\cite{bondersonthesis,Kitaev20062,Ng,Ng2,Kitaev20062}. 

We will not assume any braiding for now.  We introduce some notation
that we will use to simplify the discussion later.
 \begin{equation}
   \scalebox{.7}{\begin{pspicture}(0,0)(2,2)
       \psset{arrowsize=6pt}
      \psline[ArrowInside=->](1,0)(1,1)
       \psline[ArrowInside=->](1,1)(.2,2)
       \psline[ArrowInside=->](1,1)(1.8,2)
       \pscircle[fillstyle=solid,fillcolor=black](1,1){.1}
       \rput(1.4,1){\scalebox{1.2}{\mbox{$\mu$}}}
       \rput(.5,2){\scalebox{1.2}{\mbox{$a$}}}
       \rput(1.5,2){\scalebox{1.2}{\mbox{$b$}}}
       \rput(.8,.3){\scalebox{1.2}{\mbox{$c$}}}
     \end{pspicture}} \mbox{\raisebox{20pt}{$= V^{ab}_{c;\mu}$}} 
\end{equation}
The index $\mu$ is included when there is a fusion multiplicity
$N_{ab}^c > 1$.  For simplicity of notation we may supress these vertex indices except when they become important.  Similarly we have
\begin{equation}
  \scalebox{.7}{\begin{pspicture}(0,0)(2,2)
    \psset{arrowsize=6pt}
       \psline[ArrowInside=->](1,1)(1,2)
      \psline[ArrowInside=->](.2,0)(1,1)
      \psline[ArrowInside=->](1.8,0)(1,1)
      \pscircle[fillstyle=solid,fillcolor=black](1,1){.1}
      \rput(1.4,1){\scalebox{1.2}{\mbox{$\mu$}}}
      \rput(.5,0){\scalebox{1.2}{\mbox{$a$}}}
      \rput(1.5,0){\scalebox{1.2}{\mbox{$b$}}}
      \rput(.8,2){\scalebox{1.2}{\mbox{$c$}}}
    \end{pspicture}} \mbox{\raisebox{20pt}{$= V_{ab;\mu}^{c}$}} 
  \end{equation}

 We will assume we are working with ``Convention 3'' so that we can
 straighten zig-zags freely.  Hermitian conjugation of a diagram is
 achieved by reflecting the diagram across a horizontal line and
 reversing all arrows, for example as in Eq.~\ref{eq:hermco}
\begin{equation}
  \left[\scalebox{.7}{\begin{pspicture}(0,1)(2,2)
      \psset{arrowsize=6pt}
      \psline[ArrowInside=->](1,0)(1,1)
      \psline[ArrowInside=->](1,1)(.2,2)
      \psline[ArrowInside=->](1,1)(1.8,2)
      \rput(.5,2){\scalebox{1.2}{\mbox{$a$}}}
      \rput(1.5,2){\scalebox{1.2}{\mbox{$b$}}}
      \rput(.8,.3){\scalebox{1.2}{\mbox{$c$}}}
    \end{pspicture}} \right]^\dagger =
  \scalebox{.7}{\begin{pspicture}(0,1)(2,2)
    \psset{arrowsize=6pt}
       \psline[ArrowInside=->](1,1)(1,2)
      \psline[ArrowInside=->](.2,0)(1,1)
      \psline[ArrowInside=->](1.8,0)(1,1)
      \rput(.5,0){\scalebox{1.2}{\mbox{$a$}}}
      \rput(1.5,0){\scalebox{1.2}{\mbox{$b$}}}
      \rput(.8,2){\scalebox{1.2}{\mbox{$c$}}}
    \end{pspicture}}
    \label{eq:hermco}
  \end{equation}
or equivalently $[V^{ab}_c]^\dagger = V_{ab}^c$.  Note however, we
will generally have a non-positive-definite inner product because we
may have chosen $d_a < 0$ for some particle types.   

\begin{figure}[h]
\scalebox{.35}{
		\newcommand{\texttauhere}{3}
		\newcommand{\arrowsizehere}{0.5}
		\begin{pspicture}(15,-5)(6,3)
                  
		\psset{linewidth=.04,linecolor=black,arrowsize=\arrowsizehere,xunit=.8cm,yunit=1cm}

                \rput(1,-1){
		\psline[ArrowInside={->},ArrowInsidePos=.7](5,1)(3,3)
		\psline[ArrowInside={->},ArrowInsidePos=.7](5,1)(7,3)
		\psline[ArrowInside={->},ArrowInsidePos=.4](5,1)(5,-1)
		
		\rput[cB](5,-1.5){\scalebox{\texttauhere}{\color{black}{$c$}}}
		\rput[cB](2.9, 3.3){\scalebox{\texttauhere}{\color{black}{$a$}}}
		\rput[cB](7.4,3.3){\scalebox{\texttauhere}{\color{black}{$b$}}}
		\rput[cB](9,1.0){\scalebox{5}{\color{black}{$=$}}}
              }

         \rput(12,0){
              
                \rput(6,0){
                \rput(0,0){\pscurve(0,0)(0,-1)(-2,1)(0,2)(4,0)(0,-4)(0,-4.5)
                            \psline[ArrowInside={->},ArrowInsidePos=.1](0,-4.5)(0,-5.5)}
                  \rput{120}(0,0){\pscurve(0,0)(0,-1)(-2,1)(0,2)(4,0)(0,-4)(0,-4.5)
                            \psline[ArrowInside={->},ArrowInsidePos=.1](0,-4.5)(0,-5.5)}
                \rput{240}(0,0){\pscurve(0,0)(0,-1)(-2,1)(0,2)(4,0)(0,-4)(0,-4.5)
                            \psline[ArrowInside={->},ArrowInsidePos=.1](0,-4.5)(0,-5.5)}
                          \rput[cB](.1,-5.9){\scalebox{\texttauhere}{\color{black}{$c$}}}
                          \rput[cB](-6.5, 3){\scalebox{\texttauhere}{\color{black}{$a$}}}
                           \rput[cB](6.5,3){\scalebox{\texttauhere}{\color{black}{$b$}}}
                        }
}
              \end{pspicture}
            }
            \caption{The pivotal property is rotation of a vertex by
              $2 \pi$}
	\label{fig:pivotalidentity}
\end{figure}
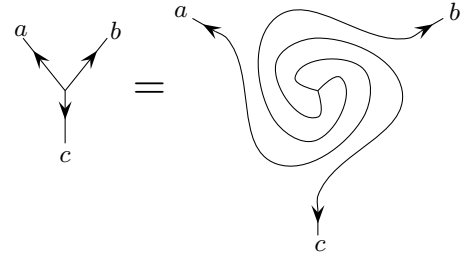

We will make use of the so-called pivotal property, which we write diagrammatically as shown in Fig.~\ref{fig:pivotalidentity}. This identity is proven very generally in Ref.~\onlinecite{Kitaev20062}. 

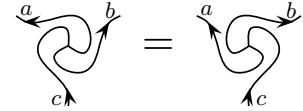
\begin{figure}[h]
  \begin{pspicture}(0,0)(4,1.5)
           \psset{arrowsize=6pt}
      \rput(0,0){ \scalebox{.8}{

               \rput(1,1){\pscurve[ArrowInside=->,ArrowInsidePos=.7](0,0)(-.05,.3)(-.5,0)(0,-.8)(0,-1)}
          \rput{120}(1,1){\pscurve[ArrowInside=->,ArrowInsidePos=.7](0,0)(-.05,.3)(-.5,0)(0,-.8)(0,-1)}
          \rput{240}(1,1){\pscurve[ArrowInside=->,ArrowInsidePos=.7](0,0)(-.05,.3)(-.5,0)(0,-.8)(0,-1)} 
          \psline{->}(.35,1.4)(.25,1.4)
          \psline{->}(1,.2)(.98,.3)
          \psline{->}(1.65,1.35)(1.68,1.4)
          
      \rput(.3,1.6){\scalebox{1.2}{\mbox{$a$}}}
      \rput(1.7,1.6){\scalebox{1.2}{\mbox{$b$}}}
      \rput(.8,.1){\scalebox{1.2}{\mbox{$c$}}}
      }  }

    \rput(2,.8){\scalebox{1.7}{$=$}}
    \rput(4,0)
    { \psscalebox{-.8 .8}{

               \rput(1,1){\pscurve[ArrowInside=->,ArrowInsidePos=.7](0,0)(-.05,.3)(-.5,0)(0,-.8)(0,-1)}
          \rput{120}(1,1){\pscurve[ArrowInside=->,ArrowInsidePos=.7](0,0)(-.05,.3)(-.5,0)(0,-.8)(0,-1)}
          \rput{240}(1,1){\pscurve[ArrowInside=->,ArrowInsidePos=.7](0,0)(-.05,.3)(-.5,0)(0,-.8)(0,-1)} 
          \psline{->}(.35,1.4)(.25,1.4)
          \psline{->}(1,.2)(.98,.3)
          \psline{->}(1.65,1.35)(1.68,1.4)
          
      \rput(.3,1.6){\psscalebox{-1.2 1.2}{\mbox{$b$}}}
      \rput(1.7,1.6){\psscalebox{-1.2 1.2}{\mbox{$a$}}}
      \rput(.8,.1){\psscalebox{-1.2 1.2}{\mbox{$c$}}}
      }  }
    
  \end{pspicture}
  \caption{Another version of the pivotal property.  Clockwise rotation of a vertex by $\pi$ is equal to counterclockwise rotation by $\pi$.}
  \label{fig:pivotal2}
\end{figure}

This property is equivalent to Fig.~\ref{fig:pivotal2}\cite{Kitaev20062}.

Since the pivotal property is derived in Ref.~\onlinecite{Kitaev20062}
using Convention 1, one might worry that a sign could be introduced
when moving to Convention 3.  However, since there are three caps on the right hand side of Fig.
\ref{fig:pivotalidentity} (one
of each type) the assumed $\mathbb{Z}_2$ grading assures that signs cancel and the
pivotal property holds in this Convention 3 as well.

We now define the turning-up and turning-down operators
\begin{eqnarray}
T^{ab}_c  \left[\scalebox{.7}{\begin{pspicture}(0,1)(2,2)
      \psset{arrowsize=6pt}
      \psline[ArrowInside=->](1,0)(1,1)
      \psline[ArrowInside=->](1,1)(.2,2)
      \psline[ArrowInside=->](1,1)(1.8,2)
      \rput(.5,2){\scalebox{1.2}{\mbox{$a$}}}
      \rput(1.5,2){\scalebox{1.2}{\mbox{$b$}}}
      \rput(.8,.3){\scalebox{1.2}{\mbox{$c$}}}
    \end{pspicture}} \right] &=&
  \scalebox{.7}{\begin{pspicture}(0,1)(2,2)
    \psset{arrowsize=6pt}
       \psline[ArrowInside=->](1,1)(1,2)
      \psline[ArrowInside=->](.7,0)(1,1)
      \psbezier[ArrowInside=->,ArrowInsidePos=.7](1,1)(1.5,-.5)(1.9,1)(2,2)
      \rput(.5,0){\scalebox{1.2}{\mbox{$c$}}}
      \rput(1.8,2){\scalebox{1.2}{\mbox{$b$}}}
      \rput(.8,2){\scalebox{1.2}{\mbox{$a$}}}
    \end{pspicture}} \\
\rule{0pt}{35pt}  T^{a}_{bc}  \left[ \scalebox{.7}{\begin{pspicture}(0,1)(2,2)
    \psset{arrowsize=6pt}
       \psline[ArrowInside=->](1,1)(1,2)
      \psline[ArrowInside=->](.2,0)(1,1)
      \psline[ArrowInside=->](1.8,0)(1,1)
      \rput(.5,0){\scalebox{1.2}{\mbox{$b$}}}
      \rput(1.5,0){\scalebox{1.2}{\mbox{$c$}}}
      \rput(.8,2){\scalebox{1.2}{\mbox{$a$}}}
    \end{pspicture}}  \right] &=&\,\,\,\,\, \scalebox{.7}{\begin{pspicture}(0,1)(2,2)
      \psset{arrowsize=6pt}
      \psline[ArrowInside=->](1,0)(1,1)
      \psbezier[ArrowInside=->,ArrowInsidePos=.3](0,0)(.1,1)(.5,2.5)(1,1)
      \psline[ArrowInside=->](1,1)(1.3,2)
      \rput(.2,.3){\scalebox{1.2}{\mbox{$b$}}}
      \rput(1.5,2){\scalebox{1.2}{\mbox{$a$}}}
      \rput(.8,.3){\scalebox{1.2}{\mbox{$c$}}}
    \end{pspicture}}  \\ \nonumber
   \end{eqnarray}
Explicit expressions for these $T$ operators in terms of $F$-matrices are given in 
Fig.~\ref{fig:turningup}.

 If there are no fusion multiplicities these $T$ operators are just complex phases. 
 More generally, in cases where there are fusion multiplicities, the
 vertices need indices as well, and the raising and lowering
 operators become unitary matrices, as shown here:
\begin{equation}
\sum_\mu [T^{ab}_c]_{\nu \mu}  \left[\scalebox{.7}{\begin{pspicture}(0,1)(2,2)
      \psset{arrowsize=6pt}
      \psline[ArrowInside=->](1,0)(1,1)
      \psline[ArrowInside=->](1,1)(.2,2)
      \psline[ArrowInside=->](1,1)(1.8,2)
      \pscircle[fillstyle=solid,fillcolor=black](1,1){.1}
      \rput(1.3,1){\scalebox{1.2}{\mbox{$\mu$}}}
      \rput(.5,2){\scalebox{1.2}{\mbox{$a$}}}
      \rput(1.5,2){\scalebox{1.2}{\mbox{$b$}}}
      \rput(.8,.3){\scalebox{1.2}{\mbox{$c$}}}
    \end{pspicture}} \right] =
  \scalebox{.7}{\begin{pspicture}(0,1)(2,2)
    \psset{arrowsize=6pt}
       \psline[ArrowInside=->](1,1)(1,2)
      \psline[ArrowInside=->](.7,0)(1,1)
      \psbezier[ArrowInside=->,ArrowInsidePos=.7](1,1)(1.5,-.5)(1.9,1)(2,2)
            \pscircle[fillstyle=solid,fillcolor=black](1,1){.1}
      \rput(1.3,1){\scalebox{1.2}{\mbox{$\nu$}}}
      \rput(.5,0){\scalebox{1.2}{\mbox{$c$}}}
      \rput(1.8,2){\scalebox{1.2}{\mbox{$b$}}}
      \rput(.8,2){\scalebox{1.2}{\mbox{$a$}}}
    \end{pspicture}} 
\end{equation}
For simplicity of notation, we will typically not write these matrix
indices, but we should remember that they are implied.  We will insert
them explicitly when they become important.

Note that these turning-up and turning-down operators are gauge
dependent.  Under gauge transformation the vertices transform as 
% \begin{eqnarray}
%   \scalebox{.7}{\begin{pspicture}(0,1)(2,2)
%       \psset{arrowsize=6pt}
%       \psline[ArrowInside=->](1,0)(1,1)
%       \psline[ArrowInside=->](1,1)(.2,2)
%       \psline[ArrowInside=->](1,1)(1.8,2)
%       \rput(.5,2){\scalebox{1.2}{\mbox{$a$}}}
%       \rput(1.5,2){\scalebox{1.2}{\mbox{$b$}}}
%       \rput(.8,.3){\scalebox{1.2}{\mbox{$c$}}}
%     \end{pspicture}} \longrightarrow  u^{ab}_c
%   \scalebox{.7}{\begin{pspicture}(0,1)(2,2)
%       \psset{arrowsize=6pt}
%       \psline[ArrowInside=->](1,0)(1,1)
%       \psline[ArrowInside=->](1,1)(.2,2)
%       \psline[ArrowInside=->](1,1)(1.8,2)
%       \rput(.5,2){\scalebox{1.2}{\mbox{$a$}}}
%       \rput(1.5,2){\scalebox{1.2}{\mbox{$b$}}}
%       \rput(.8,.3){\scalebox{1.2}{\mbox{$c$}}}
%     \end{pspicture}} \\
% \rule{0pt}{45pt}  \scalebox{.7}{\begin{pspicture}(0,1)(2,2)
%     \psset{arrowsize=6pt}
%        \psline[ArrowInside=->](1,1)(1,2)
%       \psline[ArrowInside=->](.2,0)(1,1)
%       \psline[ArrowInside=->](1.8,0)(1,1)
%       \rput(.5,0){\scalebox{1.2}{\mbox{$a$}}}
%       \rput(1.5,0){\scalebox{1.2}{\mbox{$b$}}}
%       \rput(.8,2){\scalebox{1.2}{\mbox{$c$}}}
%     \end{pspicture}}
%   \longrightarrow
%   u^c_{a b}   \scalebox{.7}{\begin{pspicture}(0,1)(2,2)
%     \psset{arrowsize=6pt}
%        \psline[ArrowInside=->](1,1)(1,2)
%       \psline[ArrowInside=->](.2,0)(1,1)
%       \psline[ArrowInside=->](1.8,0)(1,1)
%       \rput(.5,0){\scalebox{1.2}{\mbox{$a$}}}
%       \rput(1.5,0){\scalebox{1.2}{\mbox{$b$}}}
%       \rput(.8,2){\scalebox{1.2}{\mbox{$c$}}}
%     \end{pspicture}} \\ \nonumber
    %   \end{eqnarray}
\begin{eqnarray}
  V^{ab}_c  & \longrightarrow & u^{ab}_c  V^{ab}_c \\
  V_{ab}^c & \longrightarrow & u_{ab}^c V_{ab}^c
\end{eqnarray}
Where again the $u$ factors become unitary matrices in cases where
there are fusion multiplicities.  Note that due to the Hermitian
conjugation principle (Eq.~\ref{eq:hermco}) we have
$$
  u^{ab}_c = [u^c_{ab}]^\dagger
$$

Under gauge transformation the $T$ operators transform as
\begin{eqnarray}
  T^{ab}_c &\longrightarrow& [u^a_{c\bar b}]  \, T^{ab}_c \, [u^{ab}_c]^\dagger  \label{eq:Ttransform1} \\
  T_{bc}^a &\longrightarrow&  [u^{\bar b a}_c] \, T_{bc}^a \, [u_{bc}^a]^\dagger  \label{eq:Ttransform2} 
\end{eqnarray}
Again, if there are fusion multiplicities then the $u$'s and $T$'s are matrices in the $\mu$ vertex variables. 

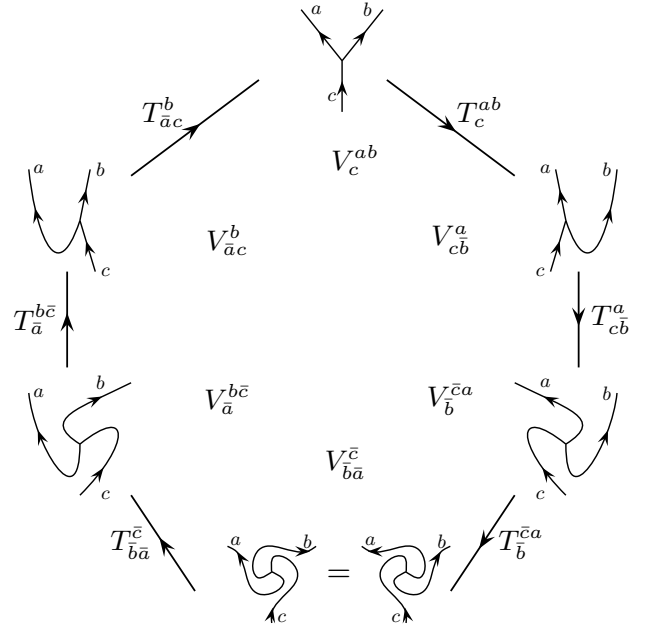
\begin{figure}[h]
    \hspace*{-.5cm}\scalebox{.85}{    \begin{pspicture}(0,-2)(8,7.5)
           \psset{arrowsize=6pt}
      \rput(4,6){ \scalebox{.8}{
      \psline[ArrowInside=->](1,0)(1,1)
      \psline[ArrowInside=->](1,1)(.2,2)
      \psline[ArrowInside=->](1,1)(1.8,2)
      \rput(.5,2){\scalebox{1.2}{\mbox{$a$}}}
      \rput(1.5,2){\scalebox{1.2}{\mbox{$b$}}}
      \rput(.8,.3){\scalebox{1.2}{\mbox{$c$}}}
    }  }

  \rput(5,5.25){\scalebox{1.4}{$V^{ab}_{c}$}}
  \rput(4.85,.5){\scalebox{1.4}{$V^{\bar c}_{\bar b \bar a}$}}
  
    \psline[ArrowInside=->](5.5,6.5)(7.5,5)
    \rput(6.95,6){\scalebox{1.4}{$T^{ab}_c$}}
    \psline[ArrowInside=->](1.5,5)(3.5,6.5)
    \rput(2,6){\scalebox{1.4}{$T^{b}_{\bar a c}$}}

    \rput(7.5,3.5){\scalebox{.8}{
     \psline[ArrowInside=->](1,1)(.8,2)
      \psline[ArrowInside=->](.7,0)(1,1)
      \psbezier[ArrowInside=->,ArrowInsidePos=.7](1,1)(1.5,-.5)(1.9,1)(2,2)
%      \psbezier[ArrowInside=->](1,0)(.5 
        
      \rput(.5,0){\scalebox{1.2}{\mbox{$c$}}}
      \rput(1.8,2){\scalebox{1.2}{\mbox{$b$}}}
      \rput(.6,2){\scalebox{1.2}{\mbox{$a$}}}
}}

\rput(6.5,4){\scalebox{1.4}{$V^{a}_{c\bar b}$}}

    \rput(1.5,3.5){\psscalebox{-.8 .8}{
     \psline[ArrowInside=->](1,1)(.8,2)
      \psline[ArrowInside=->](.7,0)(1,1)
      \psbezier[ArrowInside=->,ArrowInsidePos=.7](1,1)(1.5,-.5)(1.9,1)(2,2)
%      \psbezier[ArrowInside=->](1,0)(.5 
        
      \rput(.5,0){\psscalebox{-1.2 1.2}{\mbox{$c$}}}
      \rput(1.8,2){\psscalebox{-1.2 1.2}{\mbox{$a$}}}
      \rput(.6,2){\psscalebox{-1.2 1.2}{\mbox{$b$}}}
}}

\rput(3,4){\scalebox{1.4}{$V^{b}_{\bar a c}$}}

\psline[ArrowInside=->](8.5,3.5)(8.5,2)
\rput(9,2.75){\scalebox{1.4}{$T^{a}_{c\bar b}$}}
\psline[ArrowInside=->](.5,2)(.5,3.5)
\rput(0,2.75){\scalebox{1.4}{$T^{b\bar c}_{\bar a}$}}

    \rput(7.5,0){\scalebox{.8}{
      \psbezier[ArrowInside=->,ArrowInsidePos=.3](1,0)(0,1)(0,1.8)(1,1)
      \psbezier[ArrowInside=->,ArrowInsidePos=.7](1,1)(1.8,1.5)(1,1.7)(0,2.2)
      \psbezier[ArrowInside=->,ArrowInsidePos=.7](1,1)(1,-.5)(1.9,1)(2,2)
      \rput(.5,0){\scalebox{1.2}{\mbox{$c$}}}
      \rput(1.8,2){\scalebox{1.2}{\mbox{$b$}}}
      \rput(.6,2.2){\scalebox{1.2}{\mbox{$a$}}}
}}

\rput(6.5,1.5){\scalebox{1.4}{$V^{\bar c a}_{\bar b}$}}

    \rput(1.5,0){\psscalebox{-.8 .8}{
      \psbezier[ArrowInside=->,ArrowInsidePos=.3](1,0)(0,1)(0,1.8)(1,1)
      \psbezier[ArrowInside=->,ArrowInsidePos=.7](1,1)(1.8,1.5)(1,1.7)(0,2.2)
      \psbezier[ArrowInside=->,ArrowInsidePos=.7](1,1)(1,-.5)(1.9,1)(2,2)
      \rput(.5,0){\psscalebox{-1.2 1.2}{\mbox{$c$}}}
      \rput(1.8,2){\psscalebox{-1.2 1.2}{\mbox{$a$}}}
      \rput(.6,2.2){\psscalebox{-1.2 1.2}{\mbox{$b$}}}
}}

\rput(3,1.5){\scalebox{1.4}{$V^{b \bar c}_{\bar a}$}}

\psline[ArrowInside=->](7.5,0)(6.5,-1.5)
\rput(7.6,-.75){\scalebox{1.4}{$T^{\bar c a}_{\bar b}$}}
\psline[ArrowInside=->](2.5,-1.5)(1.5,0)
\rput(1.5,-.75){\scalebox{1.4}{$T^{\bar c}_{\bar b \bar a}$}}

      \rput(5,-2){ \scalebox{.8}{

               \rput(1,1){\pscurve[ArrowInside=->,ArrowInsidePos=.7](0,0)(-.05,.3)(-.5,0)(0,-.8)(0,-1)}
          \rput{120}(1,1){\pscurve[ArrowInside=->,ArrowInsidePos=.7](0,0)(-.05,.3)(-.5,0)(0,-.8)(0,-1)}
          \rput{240}(1,1){\pscurve[ArrowInside=->,ArrowInsidePos=.7](0,0)(-.05,.3)(-.5,0)(0,-.8)(0,-1)} 
          \psline{->}(.35,1.4)(.25,1.4)
          \psline{->}(1,.2)(.98,.3)
          \psline{->}(1.65,1.35)(1.68,1.4)
          
      \rput(.3,1.6){\scalebox{1.2}{\mbox{$a$}}}
      \rput(1.7,1.6){\scalebox{1.2}{\mbox{$b$}}}
      \rput(.8,.1){\scalebox{1.2}{\mbox{$c$}}}
      }  }

    \rput(4.75,-1.25){\scalebox{1.5}{$=$}}
    
   \rput(4.5,-2){ \psscalebox{-.8 .8}{

               \rput(1,1){\pscurve[ArrowInside=->,ArrowInsidePos=.7](0,0)(-.05,.3)(-.5,0)(0,-.8)(0,-1)}
          \rput{120}(1,1){\pscurve[ArrowInside=->,ArrowInsidePos=.7](0,0)(-.05,.3)(-.5,0)(0,-.8)(0,-1)}
          \rput{240}(1,1){\pscurve[ArrowInside=->,ArrowInsidePos=.7](0,0)(-.05,.3)(-.5,0)(0,-.8)(0,-1)} 
          \psline{->}(.35,1.4)(.25,1.4)
          \psline{->}(1,.2)(.98,.3)
          \psline{->}(1.65,1.35)(1.68,1.4)
          
      \rput(.3,1.6){\psscalebox{-1.2 1.2}{\mbox{$b$}}}
      \rput(1.7,1.6){\psscalebox{-1.2 1.2}{\mbox{$a$}}}
      \rput(.8,.1){\psscalebox{-1.2 1.2}{\mbox{$c$}}}
      }  }

  \end{pspicture}}
  \caption{Rotating a vertex by $2\pi$ with turning-up and turning-down operators. The equality at the bottom is assured by the pivotal property, Fig.~\ref{fig:pivotal2}}
  \label{fig:hexagonal}
\end{figure}

The key question here is whether we have enough gauge freedom to set
all of the turning-up and turning-down factors $T$ to unity.  To answer this question, we refer to the hexagonal diagram in Fig.~\ref{fig:hexagonal}.

%\begin{widetext}

%\end{widetext}

In this diagram, the equality at the bottom is assured by the pivotal property, Fig.~\ref{fig:pivotal2}.   Thus, following the arrows, all the way around the hexagon, we have the following identity.
 \begin{equation}
   T^{b}_{\bar a c} \, T^{b \bar c}_{\bar a} \, T^{\bar c}_{\bar b \bar a} \, T^{\bar c a}_{\bar b} \, 
   T^a_{c \bar b} \, T^{ab}_c = 1  \label{eq:sixTs}
  \end{equation}
where again, with fusion multiplicities, this is a matrix equation in the $\mu$ variables and the right hand side is the identity matrix.  Note that we can start the circle around the hexagon at any point on the circle and we will still get the identity.  For example, we also have 
 \begin{equation}
    T^{b \bar c}_{\bar a} \, T^{\bar c}_{\bar b \bar a} \, T^{\bar c a}_{\bar b} \, 
   T^a_{c \bar b} \, T^{ab}_c \, T^{b}_{\bar a c} = 1  \label{eq:sixTs2}
  \end{equation}

What we would like to know here is whether we have enough gauge
freedom such that we can always choose a gauge such that all of the
$T$ operators are unity --- i.e., so that one can turn up and turn
down legs at vertices freely.

Let us first consider the case where all three quantum numbers $a,b,c$
are distinct, meaning they are not the same and they are not related
to each other by duality.  In this case it is clear that it is always
possible to set all of the $T$'s to unity.  To see this let us start
at the top of the hexagon with $V^{ab}_c$.  We pick an arbitrary gauge
for this vertex (i.e., we fix $u^{ab}_c$).  We then want to gauge
transform $V^{a}_{c\bar b}$ such that $T^{ab}_c$ becomes the identity.
To do this, using Eq.~\ref{eq:Ttransform1} we may choose
$$
u^a_{c\bar b}  =  ([T^{ab}_c]_{old-gauge})^\dagger  [u^{ab}_c]
$$
and we use the fact that the $u$'s and $T$'s are always unitary. In
the new gauge $T^{ab}_c$ is unity.  Even in the case where there is
fusion multiplicity, this strategy successfully trivializes
$T^{ab}_c$.

Having set this $T$ to unity, we continue around the hexagon clockwise
and attempt to set $T^a_{c \bar b}$ to unity.  To do this, we
similarly gauge transform $V^{\bar c a}_{\bar b}$ (Using
Eq. ~\ref{eq:Ttransform2} in this case).  We can continue all the way
around the hexagon setting each $T$ to unity.  When we have set five
of these $T$'s to unity, the last one is guaranteed to be unity also
by Eq.~\ref{eq:sixTs}.  Note that there is still one overall gauge
freedom which is given by how we set the gauge of the first vertex at
the start of the process.

\subsection{Obstructions}

The procedure described for gauge-fixing $T$'s to unity can run into
obstructions if not all of the three quantum numbers $a,b,c$ are distinct. The
problem one can run into is if two of the vertices $V$ around the
hexagon are the same (or are related by complex conjugation) then one
cannot independently choose the gauge of each $V$.

With some thought it is clear that there are only two situations that
may cause trouble.  We will take these cases one at a time.

\subsubsection{Case 1: $a=\bar b$, and $c$ self-dual. Not an obstruction}

To avoid notational confusion, let us set $a=x$, $b=\bar x$ and we
assume $c=y$ is self-dual.    We may also assume that $c$ is not the
vacuum field since we have already assumed that we can freely add and
remove cups and caps, which can be thought of as a vertex with $c=I$.

At the top of the hexagon we have $V^{ab}_c = V^{x \bar x}_y$ whereas
at the bottom we have $V^{\bar c}_{\bar b \bar a} = V^y_{x\bar
  x}$. These two are Hermitian conjugates of each other and cannot be
independently gauge fixed.  Similarly on the right hand side of the
hexagon we have $V^a_{c\bar b} = V^x_{y x}$ and $V^{\bar c a}_{\bar b} = V^{y x}_x$ which
are also Hermitian conjugates of each other; and on the left hand side
of the hexagon we have $V^b_{\bar a c} = V^{\bar x}_{\bar x y}$ and
$V^{b \bar c}_{\bar a} = V^{\bar x y}_{\bar x}$ which are again
conjugate of each other.  We thus have only three gauge freedoms in
the hexagon instead of six.  We might wonder if we still have enough freedom
to set all of the $T$'s to unity.

First, we claim that $T^{ab}_c = T^{x \bar x}_y$ is related to
$T^{\bar c a}_{\bar b} = T^{y x}_{x}$ and
$T^{\bar c}_{\bar b\bar a} = T^y_{x \bar x}$ is related to
$T^b_{\bar a c} = T^{\bar x}_{\bar x y}$ via the equations
\begin{eqnarray}
  T^{x \bar x}_y  &=& \left(\left[ T^{y x}_{x} \right]^\dagger\right)^{-1} \label{eq:TTr} \\
  T^y_{x \bar x} &=&  \left(\left[ T^{x}_{\bar x y} \right]^\dagger\right)^{-1}  \label{eq:TTs}
\end{eqnarray}
Let us examine Eq.~\ref{eq:TTr} in a bit of depth.  The left hand side
is the $T^{ab}_c$ connecting two diagrams at the top right of the
hexagon.  However, if we flip the diagrams over a horizontal (i.e.,
Hermitian conjugate) we will find exactly the vertices at the bottom
right of the hexagon which are connected by $T^{\bar c a}_{\bar b}$.
This is on the right hand side of Eq.~\ref{eq:TTr} and the fact that we
flipped the diagrams accounts for the Hermitian conjugation.  However,
the arrow on the operator $T^{\bar c a}_{\bar b}$ needs to be
reversed, which accounts for the inverse sign.  The argument is similar for Eq.~\ref{eq:TTs}.

Now let us try to set all of the $T$'s to the identity.  First, as
above, by choosing a gauge for $V^a_{c\bar b}$ we can set the first
link $T^{ab}_c = T^{x\bar x}_y$ to unity.  However, given
Eq.~\ref{eq:TTr} this also sets $T^{\bar c a}_{\bar b} = T^{y x}_x$ to
unity.  Now let us think about the first three steps of the hexagon together, which are
$$
  T^{\bar c a}_{\bar b} T^a_{c \bar b} T^{ab}_c  = T^{y x}_x  T^x_{y x} T^{x \bar x}_y
$$
If we consider gauge transforming the top of the hexagon
$V^{ab}_c = V^{x\bar x}_y$ with $u^{x\bar x}_y$ this also transforms
the bottom $V^{\bar c}_{\bar b \bar a} = V^y_{x \bar x}$ as the
Hermitian conjugate.   The result will be
$$
  [u_{x \bar x}^y] T^{y x}_x  T^x_{y x} T^{x \bar x}_y [u^{x \bar x}_y]^\dagger =     [u_{x \bar x}^y] T^{y x}_x  T^x_{y x} T^{x \bar x}_y [u_{x \bar x}^y]
$$
We can choose the gauge so that this is unity.  Since $T^{y x}_x$ and
$T^{x \bar x}_y$ can both be set to unity by choosing the gauge of
$V^a_{c\bar b}$, this means we must have also set the remaining
$T^x_{yx}$ to unity.

For the second half of the hexagon, we proceed similarly.  By choosing
the gauge of $V^{b \bar c}_{\bar a}$ we can set both
$T^{\bar c}_{\bar b\bar a}= T^y_{x \bar x}$ and
$T^b_{\bar a c} = T^{\bar x}_{\bar x y}$ to unity.  However, now the
product of $T$'s all the way around the hexagon must be unity
(Eq.~\ref{eq:sixTs}) so the one remaining $T^{b \bar c}_{\bar a}=T^{\bar x y}_{\bar x}$ is also unity.

Thus we conclude that we still have enough gauge freedom to set all of
the $T$'s to unity and there is no obstruction.

\subsubsection{Case 2: $a=b=\bar c$.   Possible Obstruction}
  
In the case where $a=b=\bar c$ we do have a possible obstruction.
Again for clarity of notation we set $a=b=x$ and $c=\bar x$.

In the hexagon diagram, there are now only two different operators
$T^{xx}_{\bar x}$ and $T^x_{\bar x \bar x}$ which alternate around the hexagon.
Taking any two consecutive steps along the hexagon, the vertex one
ends up with is identical to the vertex one starts with.  We can thus
construct two  gauge invariant quantities.
\begin{eqnarray}
  C_x &=&  T^{xx}_{\bar x} T^x_{\bar x \bar x} \\
  C'_x &=&  T^x_{\bar x \bar x} T^{xx}_{\bar x} 
\end{eqnarray}
Note that if $N^{xx}_{\bar x} > 1$ then these are matrix equations
(with indices $\mu, \nu$ not written).

From Eq.~\ref{eq:sixTs}, i.e., going all the way around the hexagon, we have
$$
[C_x]^3 = [C'_x]^3 = 1
$$
so that the eigenvalues of $C$ and $C'$ must be cube roots of unity.
We can then define the so-called {\it third Frobenius-Schur
indicator}\cite{Ng}
$$
 \nu_3(x) = {\rm Tr}[C] = {\rm Tr}[C']
$$
We say the that this indicator is trivial if
$$
 \nu_3(x) = N_{xx}^{\bar x}
$$
i.e., if $C_x$ (or equivalently $C'_x$) is the unit matrix.  If this
is not the case, and there is an eigenvalue which is not unity, then
there is no way to choose a gauge such that turning up and down legs
does not incur any phases.  (Indeed, this eigenvalue tells us that we
have a gauge invariant nontrivial phase associated with taking two
steps around the hexagon, or twisting a vertex by 120 degrees).  A
crucial result here is that such a nontrivial third Frobenius-Schur
indicator is in fact the {\it only} possible obstruction to obtaining
isotopy of planar diagrams (given that we have a $\mathbb{Z}_2$
Frobenius-Schur grading and we have used our Convention 3 to account
for signs associated with zig-zags).  For planar diagram algebras
there are simple cases of theories having nontrivial third
Frobenius-Schur indicators --- for example, the generating cocycle of
the group $\mathbb{Z}_3$.

\subsection{$\mathbb{Z}_3$ Frobenius-Schur in Ribbon Theories}

\label{app:braided}

It is rather difficult for ribbon theories to have nontrivial
third Frobenius-Schur indicators.  To see why this is, we
will use the $R$ matrices to explicitly calculate $\nu_3(x)$.  We will
consider taking two steps around the hexagon in
Fig.~\ref{fig:hexagonal} starting at the upper left and going to the
upper right.  I.e., $ C_x = T^{xx}_{\bar x} T^{x}_{\bar x \bar x} $.
Our plan will be to evaluate $T^{xx}_{\bar x}$ in terms of
$T^x_{\bar x \bar x}$ by using the $R$ matrix.  Recall that we are
using Convention 3 so that we may add and remove cups and caps freely. 
We start with

\scalebox{.8}{\begin{pspicture}(-2,0)(2,2.5)
	\psset{linewidth=.04,arrowsize=.2}

\rput(4,0){
        \psline[ArrowInside=->](1,.4)(1,1)
	\psbezier[ArrowInside=->,ArrowInsidePos=.85]
	(1,1)(1.8,1.4)(.6,1.4)(.4,2)
	\psbezier[border=.1,bordercolor=white]
	(1,1)(.2,1.4)(1.4,1.4)(1.6,2)
	
	\psbezier[ArrowInside=->,ArrowInsidePos=.85]
	(1,1)(.2,1.4)(1.4,1.4)(1.6,2)

	\rput(.2,1.7){\scalebox{1.5}{\color{black}$x$}}
	
	\rput(1.75,1.65){\scalebox{1.5}{\color{black}$x$}}
	
	\rput(.7,.4){\scalebox{1.5}{\color{black}$\bar x$}}
      }

    	\rput(3,1){\scalebox{1.5}{\color{black}$=[R^{xx}_{\bar x}]^{-1}$}}
      
	\rput(-3,0){

		\psline[ArrowInside=->](3.5,.4)(3.5,1.2)
		
		\psline[ArrowInside=->](3.5,1.2)(4.2,2)
		
		\psline[ArrowInside=->](3.5,1.2)(2.8,2)
		
		\rput(2.8,1.7){\scalebox{1.5}{\color{black}$x$}}
		
		\rput(4.15,1.65){\scalebox{1.5}{\color{black}$x$}}
		
		\rput(3.2,.6){\scalebox{1.5}{\color{black}$\bar x$}}
	}
	
      \end{pspicture}}

      Note that with fusion multiplicity $N_{xx}^{\bar x} > 1$, here
$R^{xx}_{\bar x}$ is a matrix in these internal indices which we suppress for simplicity of notation. 
    
We then lower the left leg by using $[T^x_{\bar x \bar x}]^{-1}$ (i.e., walking from the top of the hexagon, one step to the left.)   We then have

\scalebox{.8}{\begin{pspicture}(-2,0)(2,2.5)
	\psset{linewidth=.04,arrowsize=.2}
\rput(0,0){
        \psline[ArrowInside=->](1,.4)(1,1)
	\psbezier[ArrowInside=->,ArrowInsidePos=.85]
	(1,1)(1.8,1.4)(.6,1.4)(.4,2)
	\psbezier[border=.1,bordercolor=white]
	(1,1)(.2,1.4)(1.4,1.4)(1.6,2)
	
	\psbezier[ArrowInside=->,ArrowInsidePos=.85]
	(1,1)(.2,1.4)(1.4,1.4)(1.6,2)

	\rput(.2,1.7){\scalebox{1.5}{\color{black}$x$}}
	
	\rput(1.75,1.65){\scalebox{1.5}{\color{black}$x$}}
	
	\rput(.7,.4){\scalebox{1.5}{\color{black}$\bar x$}}
      }

     	\rput(3,1){\scalebox{1.5}{\color{black}$=[T^x_{\bar x \bar x}]^{-1}$}}

        \rput(4,0){

	\psbezier[ArrowInside=->,ArrowInsidePos=.85]
	(1,1)(1.8,1.4)(.6,1.4)(.4,2)

	\psbezier[border=.1,bordercolor=white]
	(1,1)(.3,0)(.6,1.4)(1.6,2)
	\psbezier[ArrowInside=->,ArrowInsidePos=.85]
	(1,1)(.3,0)(.6,1.4)(1.6,2)
	
        \psline[ArrowInside=->](1.2,.4)(1,1)
	
	\rput(.2,1.7){\scalebox{1.5}{\color{black}$x$}}
	
	\rput(1.75,1.65){\scalebox{1.5}{\color{black}$x$}}
	
	\rput(1.5,.4){\scalebox{1.5}{\color{black}$\bar x$}}
      }
\end{pspicture}  
}

The diagram on the right can be deformed into the following diagram and untwisted as shown to yield $\theta_a^*$.  Note that 

\scalebox{.8}{
\begin{pspicture}(-2,-1)(2,2.5)
	\psset{linewidth=.04,arrowsize=.2}

     	\rput(3,1){\scalebox{1.5}{\color{black}$=\theta_x^*$}}

        \rput(0,0){

%	\psbezier[ArrowInside=->,ArrowInsidePos=.85]
%	(1,1)(1.8,1.4)(.6,1.4)(.4,2)

          \psline[ArrowInside=->,ArrowInsidePos=.85](1,1)(.5,2)

	\psbezier[border=.1,bordercolor=white]
	(1,1)(.5,-.5)(0,.5)(.5,.5)

          \psline[ArrowInside=->,ArrowInsidePos=.3](1.4,.2)(1,1)
        
        \psbezier[border=.1,bordercolor=white]
        (.5,.5)(1.5,.5)(1.5,1)(1.5,2)

        \psbezier[ArrowInside=->,ArrowInsidePos=.7]
        (.5,.5)(1.5,.5)(1.5,1)(1.5,2)

	%\psbezier[ArrowInside=->,ArrowInsidePos=.85]
	%(1,1)(.3,0)(.6,1.4)(1.6,2)

	\rput(.2,1.7){\scalebox{1.5}{\color{black}$x$}}
	
	\rput(1.75,1.65){\scalebox{1.5}{\color{black}$x$}}
	
	\rput(1.7,.3){\scalebox{1.5}{\color{black}$\bar x$}}
      }

        \rput(4,0){

          \psline[ArrowInside=->,ArrowInsidePos=.85](1,1)(.5,2)

          \psbezier[ArrowInside=->,ArrowInsidePos=.3](.7,-0.5)(1,0)(1,.5)(1,1)

	\psbezier[border=.1,bordercolor=white]
	(1,1)(0,.5)(2,-1.5)(1.5,2)

        \psbezier[ArrowInside=->,ArrowInsidePos=.8]
	(1,1)(0,.5)(2,-1.5)(1.5,2)

          \psline(1,.8)(1,1)

	\rput(.2,1.7){\scalebox{1.5}{\color{black}$x$}}
	
	\rput(1.75,1.65){\scalebox{1.5}{\color{black}$x$}}
	
	\rput(.3,-.3){\scalebox{1.5}{\color{black}$\bar x$}}
      }
      
    \end{pspicture}}
    
This diagram can then be untwisted with $[R^{xx}_{\bar x}]^{-1}$ to obtain
the diagram on the upper right of the hexagon.  Putting these pieces
together we have
$$
C_x =  \theta_x^* [R^{xx}_{\bar x}]^{-1} [T^x_{\bar x \bar x}]^{-1} [R^{xx}_{\bar x}]^{-1}  T^x_{\bar x \bar x}
$$
Generally each term on the right (except $\theta_x$) is a
$N_{x x}^{\bar x}$ dimensional matrix.  If $T^{x}_{\bar x \bar x}$ and
$R^{xx}_{\bar x}$ commute then we can bring the two $T$ terms together
and they will cancel.  We will then be left with
$C_x = \theta_x^* [R^{xx}_{\bar x}]^{-2}$.  We then use the ribbon identity
$[R^{ab}_c R^{ba}_c]_{\mu \nu} = \delta_{\mu \nu} \theta_c/(\theta_a \theta_b)$ (see Ref.~\onlinecite{Kitaev20062}) to give $C_x$ equal to the identity matrix. 
  
We conclude that the third Frobenius-Schur indicator is
trivial if $R^{xx}_{\bar x}$ commutes with
$T^x_{\bar x \bar x}$.  This is obviously satisfied if these
quantities are scalars, i.e., if $N_{xx}^{\bar x} = 1$.  (In the main
text we use $F$ instead of $T$, but these are equivalent up to
constant factors).

If $N_{xx}^{\bar x} > 1$, the commutation of these two matrices may
seem like a rather strong condition.  However due to the so-called
ribbon identity we
must have the eigenvalues of $R$ given by
$$
  {\rm eigs}[R^{xx}_{\bar x}] = \pm \frac{1}{\sqrt{\theta_x}}
$$
where $\theta_x$ is the twist factor for $x$.  If all of the $\pm$
happen to be the same, then this matrix is proportional to the
identity and it commutes with $T^x_{\bar x \bar x}$.

\section{Unusual Examples}
\label{app:examples}

\subsection{Theories without $\mathbb{Z}_2$ Frobenius-Schur Gradings}
\label{sub:exceptions}

There are theories that do not admit $\mathbb{Z}_2$
Frobenius-Schur grading.  For planar algebras (i.e., solutions of the
pentagon without solution of the hexagon) it is fairly easy to find
such exceptions.

A simple example is the generating cocycle of the group
$\mathbb{Z}_4$. This theory has four objects $a=0,1,2,3$ with fusion
rules $a \times b = (a+b) \, {\rm mod} \, 4$.  For the case of the
generating cocycle, the Frobenius-Schur indicator for the second object
is $\kappa_2=-1$, but $1 \times 1 =2$ so it is impossible to have a
$\mathbb{Z}_2$ Frobenius-Schur grading.  There are obvious
generalizations to cocycles of the group $\mathbb{Z}_{4n}$.

However, for braided theories exceptions are much harder to find.  By
searching a database\cite{Gruen} of discrete (twisted and untwisted)
gauge theories (i.e., Dijkgraaf-Witten theories\cite{Dijkgraaf90}
theories) we have found examples of modular theories which do not have
$\mathbb{Z}_2$ Frobenius-Schur grading.  Using this database we
generate fusion relations using the Verlinde formula and
Frobenius-Schur indicators using the Bantay formula\cite{BANTAY1997}
and then determine whether a grading is possible.  As mentioned in
item (\ref{thisitem}) of section \ref{sec:many} above, all gauge
theories (twisted or untwisted) for groups of order 15 and less do
have $\mathbb{Z}_2$ gradings.  For groups of order 16, there are
several exceptions, the simplest being the untwisted quantum double of
the quasi-dihedral group of order 16 (group [16,8] in GAP
notation\cite{GAP}).  This modular theory has 46 simple objects in it.
In addition certain twisted doubles of the groups [16,3],[16,4],
[16,6], [16,10], and [16,11] also fail to have gradings (none of these
have fewer than 46 objects).  We note, however, that the group
$Z_5 \rtimes Z_4$ (or [20,3]) which is a group of 20 elements, has
some twisted quantum doubles with only 22 simple objects which also
fails to have a $\mathbb{Z}_2$ gradings.  This modular theory with 22
simple objects is the smallest modular (or braided) theory we have
found which fails to admit a $\mathbb{Z}_2$ Frobenius-Schur grading.

\subsection{Some modular theories with nontrivial third Frobenius-Schur Indicator}

\label{app:modularZ3}

An example of a braided (and modular) theory with a nontrivial third
Frobenius-Schur indicator is the (untwisted) quantum double of the
group $Z_5 \rtimes Z_4$ (or [20,3] in GAP notation\cite{GAP}). The
group has 20 elements, and its quantum double has only 22 elements.
In fact, this group (before taking the quantum double) is the smallest
group where a representation has a nontrivial third Frobenius-Schur
indicator $\frac{1}{|G|} \sum_g \chi(g^3)$ with $\chi$ the character
of the rep.  All of the Frobenius-Schur indicators of this quantum
double are either 1 or 0 so the theory has a trivial $\mathbb{Z}_2$
Frobenius-Schur grading and yet we cannot put it into a form where
turning-up and turning-down does not incur a phase.  By searching the
database (Ref.~\onlinecite{Gruen}) we have found that among discrete
gauge theories (twisted or untwisted) this is the smallest example
with nontrivial third Frobenius-Schur indicator (There are twisted
quantum doubles of the same group which also have nontrivial third
Frobenius-Schur indicators).  The method of calculation is similar to
that of section \ref{sub:exceptions} above.  Using the database we
generate fusion relations using the Verlinde formula and third
Frobenius-Schur indicators using a generalization of the Bantay
formula\cite{BANTAY1997,Ng2}.

There are also examples of Chern-Simons theories with nontrivial third
Frobenius-Schur indicators.  The simplest few are $SU(4)_5$,
$SO(5)_4$, $SO(8)_4$, $(E_6)_4$, $(E_7)_4$, $(E_8)_5$, $(F_4)_4$,
$(G_2)_9$. As mentioned in section \ref{app:ChernSimons} all of these
have $\mathbb{Z}_2$ Frobenius-Schur gradings.  Among these examples
$SO(5)_4$ and $(E_8)_5$ both have only 15 particle types, and
$(F_4)_4$ has 16 (And further, the cases $(E_8)_5$ and $(F_4)_4$ have all particles
with $\kappa_a = +1$).  The method of finding these is similar to the
previous paragraph: we use the program Kac\cite{Kac} to generate $S$
matrices and twist factors for a given Chern-Simons theory then
generate fusion relations using the Verlinde formula and the third
Frobenius-Schur indicators using a generalization of the Bantay
formula\cite{BANTAY1997,Ng2}.

While in section \ref{app:braided} we showed that $N_{aa}^{\bar a}$
must be greater than 1 in order to have a nontrivial third
Frobenius-Schur indicator.  In fact, we have not found any case of a
modular theory where $N_{aa}^{\bar a} = 2$ with nontrivial third
Frobenius-Schur indicator.  We conjecture that this can never happen. 

\vspace*{10pt}

\subsection{Example of a modular theory with tetrahedral rotation, but
  not inversion}

\label{app:modularinv}

As an example of a modular theory which has full planar isotopy, and
allows rotation of the tetrahedral diagram, but not inversion, we
consider the example of $SU(3)_3$.  The $F$-matrices for this are
calculated explicitly in Ref.~\onlinecite{Ardonne2010} section B.3
(Note that actually what is shown in this reference is the category
$SU(3)_3/{\mathbb{Z}_3}$ but this is a subcategory of $SU(3)_3$).
This theory has no negative $\mathbb{Z}_2$ Frobenius-Schur indicators,
and no nontrivial third Frobenius-Schur indicators (The third
Frobenius-Schur indicator is calculated using the formula in
Ref.~\onlinecite{Ng2} using modular data for $SU(3)_3$ obtained from
Ref.~\onlinecite{Kac}.)  Because there are no nontrivial
Frobenius-Schur indicators, we can put the theory in a form so that
diagrams can be deformed in the plane as discussed above and
tetrahedral diagrams can be rotated freely.  However, we can give an
example of a tetrahedron that is not invariant under inversion.  We
consider labeling every edge of the tetrahedron with the self-dual
quantum number $8$ except one edge which we label with $10$ (these are
names of the particle types in $SU(3)$ notation).  There are two
vertices where three 8's meet, and since $N_{88}^8 = 2$, we must label
each of these vertices with additional quantum numbers, $\mu$ at one
vertex and $\nu$ at the other.  Up to positive constants (square-roots
of $d$'s) the value of this tetrahedron diagram is given by an
$F$-matrix symbol $([F^{8,8,8}_{10}]_{8,8})_{\mu \nu}$.  Inverting the
tetrahedron exchanges $\mu$ and $\nu$ or equivalently flips $10$ to
its dual $\overline{10}$.  If $\mu$ and $\nu$ are different then this
inversion changes the sign of the result.  One might wonder if one can
choose a different gauge for the indices $\mu,\nu$ so that this sign
goes away.  In fact, one cannot.  A gauge transform would be a unitary
matrix $U_{\alpha \nu}$ and would result in the $F$-matrix changed to
$\sum_{\alpha, \beta} U_{\mu \alpha} U_{\nu \beta}
([F^{8,8,8}_{10}]_{8,8})_{\alpha \beta} $ which one can show cannot be
made symmetric in $\mu,\nu$.

% In fact something similar happens for all theories $SU(N)_k$ when
% there is fusion multiplicity.  As outlined in Ref.~\onlinecite{Gu2015}
% inversion of the 6j symbol (the tetrahedral diagram) generally
% accumulates $(-1)^r$ where $r$ is the sum of the multiplicity indices
% at all the vertices (given that one is working in the right basis).

\bibliographystyle{apsrev4-1}
\bibliography{Frob}

\end{document}